\documentclass[preprint1]{aastex631}
\usepackage{graphicx}
\usepackage{epstopdf}
\usepackage{natbib}
\usepackage{amsmath}
\bibpunct{(}{)}{;}{a}{}{,}

\newcommand{\ee}[1]{\mbox{${} \times 10^{#1}$}}
\newcommand{\eten}[1]{\mbox{$10^{#1}$}}

\newcommand{\kms}{\mbox{km s$^{-1}$}}
\newcommand\cmv{\mbox{cm$^{-3}$}}
\newcommand\cmc{\mbox{cm$^{-2}$}}

\newcommand{\x}{\mbox{${}\times{}$}}

\newcommand\smm{submillimeter}

\newcommand{\sfrth }{\mbox{$SFR_{\rm th}$}}

\newcommand{\msun}{\mbox{M$_\odot$}}

\newcommand{\tk}{\mbox{$T_K$}}
\newcommand{\td}{\mbox{$T_d$}}

\newcommand{\dv}{\mbox{$\Delta v$}}
\newcommand{\vlsr}{\mbox{$v_{LSR}$}}
\newcommand{\n}{\mbox{$n$}}

\newcommand{\mvir}{\mbox{$M_{\rm vir}$}} 
\newcommand{\mco}{\mbox{$M_{\rm CO}$}} 
\newcommand{\mrat}{\mbox{$M_{\rm 13}/M_{\rm CO}$}} 
\newcommand{\mdust}{\mbox{$M_{\rm dust}$}} 
\newcommand{\lco}{\mbox{$L_{\rm CO}$}} 
\newcommand{\epsff}{\mbox{$\epsilon_{\rm ff}$}} 
\newcommand{\epssf}{\mbox{$\epsilon_{\rm sf}$}} 
\newcommand{\tff}{\mbox{$t_{\rm ff}$}} 
\newcommand{\tdep}{\mbox{$t_{\rm dep}$}} 
\newcommand{\tcloud}{\mbox{$t_{\rm cloud}$}} 
\newcommand{\alphavir}{\mbox{$\alpha_{\rm vir}$}} 
\newcommand{\alphavirz}{\mbox{$\alpha_{\rm vir, 0}$}} 
\newcommand{\alphaco}{\mbox{$\alpha_{\rm CO}$}} 
\newcommand{\muhh}{\mbox{$\mu_{\rm H_{2}}$}} 

\newcommand{\mean}[1]{\mbox{$\langle#1\rangle$}} 
\newcommand{\av}{\mbox{$A_V$}} 

\newcommand{\hh}{\mbox{{\rm H}$_2$}}

\newcommand{\ammonia}{\mbox{{\rm NH}$_3$}}
\newcommand{\coo}{\mbox{$^{13}$CO}}

\newcommand{\jj}[2]{\mbox{$J = #1\rightarrow#2$}}

\newcommand{\amax}{\mbox{$\alpha_{\rm max}$}}

\newcommand{\sigmav}{\mbox{$\sigma_v$}}
\newcommand{\sigmam}{\mbox{$\Sigma_{\rm M}$}}
\newcommand{\go}{\mbox{$G_0$}}

\newcommand{\fn}{\mbox{$f(N)$}}
\newcommand{\fcn}{\mbox{$f_{\rm c}(N)$}}
\newcommand{\fmass}{\mbox{$f(M)$}}
\newcommand{\fcmass}{\mbox{$f_{\rm c}(M)$}}

\newcommand{\kkms}{\mbox{K\ \kms}}
\newcommand{\alphacounit}{\mbox{\msun (K\ \kms\ pc$^2$)$^{-1}$}}
\newcommand{\neff}{\mbox{$n_{\rm eff}$}}
\newcommand{\rgal}{\mbox{$R_{\rm gal}$}}

\newcommand{\isorat}{\mbox{${\rm CO}/{\rm ^{13}CO}$}}

\newcommand{\msunyr}{\mbox{M$_\odot$ yr$^{-1}$}}
\newcommand{\msunpc}{\mbox{M$_\odot$ pc$^{-2}$}}

\shorttitle {Boundedness}
\shortauthors{Evans et al.}

\begin{document}

\setcounter{table}{0}

\title{Which Molecular Cloud Structures Are Bound?}

\correspondingauthor{Neal J. Evans II}
\email{nje@astro.as.utexas.edu}

\author{Neal J. Evans II}
\affiliation{Department of Astronomy, The University of Texas at Austin,
2515 Speedway, Stop C1400, Austin, Texas 78712-1205, U.S.A.}
\affiliation{Korea Astronomy and Space Science Institute, 776 Daedeokdaero, Daejeon 305-348, Korea}
\affiliation{Humanitas College, Global Campus, Kyung Hee University, Yongin-shi 17104, Korea}
\author{Mark Heyer}
\affiliation{Department of Astronomy, University of Massachusetts, Amherst, MA 01003, USA}
\author{Marc-Antoine Miville-Desch{\^e}nes}
\affiliation{Universit{\'e} Paris Saclay and Universit{\'e} de Paris, CEA, CNRS, AIM, F-91190 Gif-sur-Yvette, France}
\author{Quang Nguyen-Luong}
\affiliation{The American University of Paris, 2bis, Passage Landrieu 75007 Paris, France}
\author{Manuel Merello}
\affiliation{Departamento de Astronom\'ia, Universidad de Chile, Casilla 36-D, Santiago, Chile}

\begin{abstract}
We analyze surveys of molecular cloud structures defined by tracers
ranging from CO \jj10\ through \coo\ \jj10\ to dust emission together with
NH$_3$ data. The mean value of the virial parameter and the fraction of mass in bound structures depends on the method used to identify structures. Generally,
the virial parameter decreases and the fraction of mass in bound structures
increases with the effective density of the tracer, the surface density and
mass of the structures, and the distance from the center of a galaxy.
For the most complete surveys of structures \added{in the Galaxy} defined by CO \jj10, the fraction of mass that is in bound
structures is 0.19. \deleted{for the Galaxy} 
\added{For catalogs of other galaxies based on CO \jj21, the fraction is 0.35.}
\deleted{and 0.32 in a sample of other galaxies, providing}
\added{These results offer} substantial alleviation of the fundamental
problem of slow star formation. If only clouds found to be bound are counted and they are assumed to collapse in a free-fall time at their mean cloud density, the sum over all clouds in a complete survey of the Galaxy yields a predicted star formation rate of 46 \msunyr, a factor of 6.5 less than if all clouds are bound.
\end{abstract}

\keywords{interstellar medium, molecular clouds, star formation}


\section{Introduction}\label{intro}

The fundamental problem of star formation was first stated by
\citet{1974ARA&A..12..279Z}
and
\citet{1974ApJ...192L.149Z}.
Simply stated, a theoretical estimate of the star formation rate
in the Milky Way exceeded by a factor of about 100 the rate averaged
over the last Gyr or so, as inferred from observations. 
Those early estimates were based on preliminary
surveys of the Galaxy in CO and estimates of the average star formation rate,
 but the situation has only worsened after
nearly a half century of improved surveys.
\citet{2015ARA&A..53..583H} \replaced{estimate}{estimated} a mass of $(1.0 \pm 0.3) \times 10^9$ \msun\
for the total molecular gas in the Galaxy.
The free-fall time,
\begin{equation}
\tff = \bigg(\frac{3\pi}{32G\rho}\bigg)^{0.5} = 3.34 \ee7  n^{-0.5} {\rm yr}  = \\ 1.66\ee7  r_{\rm pc}^{1.5} M_{\msun}^{-0.5} {\rm yr} ,
\end{equation}
where $n$ is the particle density 
\added{in \cmv}, 
related to $\rho$ by the mean mass per
particle of 2.37
\citep{2008A&A...487..993K}.
For a characteristic density of 100 \cmv, 
$\tff = 3.34\ee6$ yr, predicting a star formation rate of 300 \msunyr.
In contrast, observational data indicate much lower rates.
Counting young stars from the GLIMPSE 
\citep{2003PASP..115..953B, 2009PASP..121..213C} survey,
\citet{2010ApJ...710L..11R}
constrained the star formation rate to 0.68-1.45 \msunyr, averaged over the lifetime
of infrared excesses around young stars, roughly the last few Myr.
Using a wider range of indicators,
\citet{2011AJ....142..197C} 
derived a star formation rate of $1.9\pm 0.4$ \msunyr, which we adopt.
The discrepancy has, if anything, worsened with further research to a factor of 160.

This problem arises because the molecular clouds identified
by CO emission are assumed to be gravitationally bound and collapsing on a free-fall
timescale. This problem is properly called the {\it slowness} of
star formation
\citep{2014prpl.conf..243K, 2014prpl.conf...77P}
 and parameterized by the efficiency per free-fall time,
\epsff. which is taken to be 0.01, reflecting the factor of 100
problem identified by the first references, but the latest numbers suggest $\epsff = 0.006$. Another expression of the slowness is that the depletion time, given by the molecular mass divided by the star formation rate, is $\tdep = 5\ee8$ yr, much larger than \tff. The depletion time is even longer in other galaxies, $\tdep \approx 2$ Gyr 
\citep{2008AJ....136.2846B}.

A somewhat distinct
problem is the final efficiency of star formation, \epssf, defined by
\begin{equation}
{\epssf = M_*/M_{\rm gas},}
\end{equation}
where $M_*$ is the mass in stars at the end of the event and $M_{\rm gas}$
is the mass of gas at the beginning. While we cannot measure these for
an individual cloud, star clusters that have formed reasonably recently
(i.e., leaving aside globular clusters)
are much less massive than the mass in clouds. We can compare the 
mass function of molecular clouds to that of clusters. The mass functions,
expressed per logarithmic unit of mass are expressed as
\begin{equation}
\frac{dN}{d \log M} \propto  (M/M_{\rm u, x})^{\Gamma_{\rm x}}
\end{equation}
where $M$ is the mass of entity $x$, up to a maximum mass
$M_{\rm u, x}$.
For Milky Way  OB assocations and clusters, $\Gamma_{\rm cluster}$ 
is $-1.0$ and $M_{\rm u, cluster}$ is about 6\ee4 \msun\
\citep{1995ApJ...454..151M,1997ApJ...476..144M,2010ARA&A..48..431P}.
For molecular clouds, the values are $\Gamma_{\rm cloud} = -0.6$
and $M_{\rm u, cloud} = 6\ee6$ \msun\
\citep{1997ApJ...476..166W}.
See also 
\citet{2005PASP..117.1403R}, who give results in terms of
$\gamma = \Gamma_{\rm cloud} + 1$,
with our definition of $\Gamma_{\rm cloud}$.
The ratio of upper mass limits suggests that $\epssf = 0.01$, but the
differences in the slopes suggests that molecular clouds are not the immediate
precursors of clusters and associations. 
Dense clumps, in contrast, have a similar slope and upper mass limit
to those of clusters and associations.
\citet{2003ApJS..149..375S} found $\Gamma_{\rm clump} = -1.0$ for
dense clumps defined by strong CS \jj54 emission. 
\citet{2017MNRAS.471..100E}
found $\Gamma_{\rm clump} = -1.05 \pm 0.08$ in a sample of \eten4
clumps defined from submillimeter continuum emission. The
upper mass limit for clumps, $M_{\rm u, clump}$ is about  1\ee5 \msun\
based on various studies
\citep{2003ApJS..149..375S,2016ApJ...822...59S,2018MNRAS.473.1059U}.

Much theoretical ingenuity has been expended to solve the problem of slow
star formation.
Magnetic fields may help to ``support'' clouds, and turbulence can
slow collapse. However, numerical simulations generally found that
supersonic turbulence should decay rapidly, with or without magnetic
fields 
\citep{1998ApJ...508L..99S, 1999ApJ...524..169M}.
\deleted{Theorists have turned to feedback for relief;
the idea is that star formation is self-limiting as some star formation
either regenerates turbulence or dissipates the cloud
\citep{2018NatAs...2..896O,2011ApJ...729..133M}. 
However, the low efficiency and slowness of star formation 
in clouds forming only low mass stars are similar to those
in clouds forming massive stars, where the feedback is much
stronger
\citep{2016ApJ...831...73V}.}
Also, turbulence appears to be driven on scales larger than the cloud, rather
than from within the cloud 
\citep{1981MNRAS.194..809L, 2004ApJ...615L..45H, 2009A&A...504..883B}.

The kind of solution that invokes turbulence to support a cloud has
the problem of the inherent instability of a gravitationally bound entity.
It is hard to imagine an individual cloud avoiding gravitational collapse
for a depletion time that is over 200 times the free-fall time. Indeed,
\citet{2015ARA&A..53..583H}
placed a strong upper limit on cloud lifetime (\tcloud) of \eten8 yr and
preferred a lifetime of $3\ee7$ yr, based on
\citet{2009ApJS..184....1K},
closer to 10 times the canonical value for \tff. 
Clearly most of the mass in clouds 
must disperse after a star forming event. The low value of \epssf\ suggests
that 99.4\% is dispersed. Another approach has been to question the use of the current density to estimate the free-fall time \citep{2019MNRAS.490.3061V}. A lower initial density would slow the initial rate of collapse, but the conflict with the prediction based on current properties of molecular clouds remains.

Recent theoretical work has concentrated on feedback to remove most of the
gas before it forms stars 
\added{as some star formation
either regenerates turbulence or dissipates the cloud
\citep{2018NatAs...2..896O,2011ApJ...729..133M}.
However, the low efficiency and slowness of star formation 
in clouds forming only low mass stars are similar to those
in clouds forming massive stars, where the feedback is much
stronger
\citep{2016ApJ...831...73V}.
Also, most simulations find that feedback has difficulty accounting
for the very low efficiencies
(e.g., \citealt{2019MNRAS.488.1501G,2018ApJ...859...68K}).
}
For example,
\citet{2018ApJ...859...68K}
compared the effectiveness of radiation pressure and photoionization
in disrupting clouds. Even though they started with marginally bound
clouds, the final efficiencies, \epssf,
increased strongly with increasing initial cloud surface density (\sigmam), exceeding 5\% by initial $\sigmam = 20$ \msunpc\ (figure 7 of their paper).
Bound clouds overproduce stars even at very low initial \sigmam, even with
strong radiative feedback.  Another interesting feature of figure 7 of
\citet{2018ApJ...859...68K} is that \epssf\ reaches about 0.5 at 
$\sigmam \sim \eten3$ \msunpc, similar to the surface density of the dense clumps
studied by
\citet{2003ApJS..149..375S}.
If these were the bound entities, the ratio of maximum cluster mass to
initial gas mass would come out about right.
Simulations of unbound clouds look more promising.
\citet{2021ApJ...911..128K}
have performed radiative magnetohydrodynamical 
simulations (RMHD) for clouds with starting $\sigmam = 80$ \msunpc,
reasonably appropriate to the structures traced by CO. The star formation
is limited by feedback from massive stars.
They found that both \epsff\ and \epssf\ decrease with 
increasing \added{initial} virial parameter (\alphavirz), 
reaching values similar to the observed values for
$\alphavirz \sim 5$. 

In the current situation, it seems prudent to reexamine the
assumptions underlying the fundamental problem:
that all molecular clouds are gravitationally bound and
that they are collapsing at free-fall, with the free-fall
time calculated from the mean cloud density of 100 \cmv.
The question of whether the cloud is bound
 is usually addressed by calculating the virial
parameter, \alphavir\
\citep{1992ApJ...395..140B}. 
The definition of this parameter
varies. The most common definition assumes virial equilibrium.
Ignoring surface terms and magnetic fields,
\begin{equation}
\alphavir = \frac{2E_{kin}}{|E_g|},
\end{equation}
where $E_g$ is the gravitational potential energy and $E_{kin}$
is the kinetic energy. With this definition, the dividing line
between bound and unbound is $\alphavir = 2$, where $E_{kin} = |E_g|$.
Surface pressure from the surrounding medium would increase 
\alphavir\ while magnetic fields would decrease \alphavir\
\citep{2013IAUS..292...19T}. 
\replaced{Because we are interested in whether attractive (gravity) or
dispersive (turbulence) motions dominate,}
{Because we cannot measure them,} 
we neglect the largely
unknown pressure and magnetic terms.
Then a cloud
with $E_{kin} > E_g$ would tend to expand or dissipate, so we take
$\alphavir = 2$ as our critical value to define a cloud as bound
if $\alphavir < 2$ and unbound if $\alphavir > 2$. 

The calculation
of $E_g$ depends on cloud shape and structure 
\citep{1992ApJ...395..140B, 2013ApJ...779..185K}
but these
effects can be considered as correction factors for the expression
for a uniform density sphere:
\begin{equation}
E_g = \frac{-3 GM^2}{5R}.
\end{equation}
where M and R are the cloud mass and radius.
The kinetic energy is estimated from
\begin{equation}
E_{kin} = \frac{3 \sigmav^2 M}{2}
\end{equation}
where $\sigmav$ is the one-dimensional velocity dispersion averaged over the cloud.

Unbound clouds may of course contain bound structures.
Measuring \alphavir\ in the complex structures of the ISM is more problematic,
as surface terms are likely to be important
\citep{2020ApJ...898...52M}.
\citet{2021ApJ...911..128K}
\replaced{compare}{compared} the virial parameters from their model clouds
to those that would be derived from observations. They found that the
methods used by observers are more likely to overestimate the boundedness
of the cloud, but generally by factors of 2 or less. They caution that 
neglect of tidal effects in crowded regions would make the overestimation 
worse. This effect was indeed apparent in larger-scale simulations
\citep{2020ApJ...898...52M},
which found that many regions that were apparently bound, based on the
criteria that $\alphavir <2$ were in fact unbound. 
In particular, the boundedness of
large massive structures was more likely to be overestimated using the
simpler estimations available to observers.
\added{If they are correct, 
our neglect of surface terms for practical reasons is unlikely to
result in systematic underestimations of \alphavir\ for clouds.
Of course, there are circumstances where surface pressure may play 
an important role, 
especially on smaller scales (e.g., \citealt{2018MNRAS.481L...1H}).
}

Given our simple approach to the virial parameter, ignoring
external pressure, magnetic fields, and tides, we use the term ``bound"
as shorthand for gravitationally bound. A cloud could be unbound
in our definition, but contained or ``bound" by external pressure.
Our focus is on whether gravity or turbulence is dominant. If the results
of the simulations discussed above apply, our criteria for boundedness
will be more generous than reality.
We calculate virial parameters from
\begin{equation}\label{avireq}
\alphavir = \frac{2E_{kin}}{|E_g|} = 1160  \sigmav^2 R_{pc}/M_{\msun}
= \\ 209 \dv^2 R_{pc}/M_{\msun},
\end{equation}
where $M_{\msun}$ is the actual mass of the cloud in solar masses.

Estimating the actual mass of the cloud can be done in a number of ways, but
the original source of the fundamental problem relied on observations of
the lowest rotational transition of CO. The problem is that this line
is usually very optically thick. Many words have been expended explaining
why this can work, including an implicit assumption that the clouds {\it are}
in virial equilibrium (e.g.,
\citealt{2013ARA&A..51..207B}).
If this were the only argument, the whole discussion would be circular, but
we will take the CO-derived masses as independent of the assumption of
virial equilibrium.
The correlation of CO emission with dust extinction via the $X({\rm CO})$
factor provides an independent 
basis for its use in determining column density
\citep{1978ApJS...37..407D, 2010ApJ...721..686P, 2010ApJ...723.1019H}. 
While column density may be underestimated at high column density 
and overestimated at low column density, both by factors of 2 or more,
\citet{2013ARA&A..51..207B} 
\replaced{argues}{argued} that variations in $X$(CO) average out, so
$L$(CO) may reflect mass better than the intensity traces column density for
a particular line of sight.
The relation is $\mco = \alphaco \lco$.
We follow \citet{2013ARA&A..51..207B}  in taking \alphaco\ of 4.3 \alphacounit.
\added{\citet{2020ApJ...898....3L} have recently suggested that \alphaco\ varies
strongly with Galactocentric radius and is higher than the canonical value at
almost all radii. We do not adopt this approach, but we will discuss the issue
of possible variations in \alphaco\ in \S \ref{caveats}.
}

Samples using \coo\ \jj10\ to get mass usually combine the \coo\ data
with CO data to derive the optical depth and then column density of \coo,
$N_{\rm 13}$, assuming that the excitation
temperatures of CO and \coo\ are equal and that the partition function can
be approximated by assuming that the excitation temperature applies to
all levels (commonly referred to as the ``LTE" assumption). The excitation
temperature can be found from the CO brightness temperature. Because
we use CO data with low spatial resolution, we instead assume an
excitation temperature of 
8~K for all lines of sight \citep{2010ApJ...723..492R}. 
We follow the description by \citet{2013MNRAS.431.1296R} but note 
the error in their equation~4, in which the exponential factor should be in the denominator.
Conversion to the column density of \hh\ follows
\begin{equation}
N({\rm H_2}) = N_{\rm 13} \Bigl[\frac{{\rm H_2}}{\rm CO}\Bigr] \Bigl[\frac{\rm CO}{\coo}\Bigr].
\end{equation} 
We have used the latest number for \hh/CO of 6000
\citep{2017ApJ...838...66L}.
Because the new measurement directly relates the abundance of \hh, not including
helium, to that of
CO through infrared absorption observations, we use the mean molecular weight of
$\muhh m_{\rm H}$, with $\muhh = 2.8$ to convert to mass 
\citep{2008A&A...487..993K}.
We have corrected data to use these values. 
Unless otherwise noted, when analyzing the \coo\ \jj10\ data,
we account for a Galactic gradient in \isorat. 
We apply the elemental isotope gradient in the Galaxy most recently derived by \citet{2020A&A...640A.125J},
\begin{equation}
\frac{^{12}{\rm C}}{^{13}{\rm C}} = 5.87 \rgal + 13.25,
\end{equation}
where \rgal\ is the distance of the cloud from the Galactic Center in kpc.


\section{Samples}\label{samples}

The sets of data used in this paper are listed in Table \ref{tabsample}. 
We consider structures defined by CO emission, \coo\ emission,
and dust emission, supplemented by molecular line data.
From the catalogs in the papers, we extracted the information on size,
mass, mass surface density (\sigmam) and virial parameter (\alphavir),
or the parameters needed to calculate them.
In most cases, some pruning of the catalog entries was necessary.
The cuts used are explained below.

\subsection{Structures Defined by CO \jj10\ Emission}

The samples in this category rely on masses measured from the luminosity
of the \jj10\ transition of CO. The conversion from luminosity to mass
(\alphaco)
differed somewhat among the studies; to standardize them, 
the value of \mco\ in the Milky Way samples was calculated using
$\alphaco = 4.3$ \alphacounit\
\citep{2013ARA&A..51..207B}.

\begin{table*}[h]
\caption{Sample List} \label{tabsample} 
\vspace {3mm} 
\begin{tabular}{l r r r  } 
\tableline 
\tableline 
Name & Tracer & Reference & Notes  \cr  
\tableline 
CO-Rice & CO \jj10 & \citet{2016ApJ...822...52R} & 1 \cr 
CO-MD & CO \jj10 &  \citet{2017ApJ...834...57M} & 2 \cr
CO-CMZ & CO \jj10  & \citet{2001ApJ...562..348O} & 3  \cr
CO-OGS  & CO \jj10 & \citet{2001ApJ...551..852H} & 4 \cr
CO-Sun  & CO \jj10 & \citet{2020ApJ...901L...8S} & 5 \cr
\coo-GRS & \coo\ \jj10  & \cite{2010ApJ...723..492R} & 6 \cr
\coo-EXFC55-100 & \coo\ \jj10  & \cite{2016ApJ...818..144R} & 7 \cr
\coo-EXFC135-195 & \coo\ \jj10  & \cite{2016ApJ...818..144R} & 8 \cr
\coo-SEDIGISM & \coo\ \jj21 & \citet{2021MNRAS.500.3027D} & 9 \cr
Herschel & \smm $+$ \ammonia & \citet{2019MNRAS.483.5355M} & 10 \cr

\tableline 
\end{tabular} 

Notes: 1. Recovers 25\% of emission in the survey by 
\citet{2001ApJ...547..792D}. \\
2. Recovers 98\% of emission in the survey by 
\citet{2001ApJ...547..792D}. \\  
3. Attempted to restrict to CMZ. \\   
4. Outer Galaxy Survey: $\mean{\rgal} = 11.8 \pm 1.3$ \\
5. Survey of 66 galaxies.  \\
6. Galactic Ring Survey: $\mean{\rgal} = 5.2 \pm 1.0$ \\
7. Exeter-FCRAO Survey for $55< l <100$: $\mean{\rgal} = 10.3 \pm 1.9$ \\
8. Exeter-FCRAO Survey for $135< l <195$: $\mean{\rgal} = 12.9 \pm 2.1$ \\
9. Covers \rgal\ from 0.6 to 15.7 kpc: $\mean{\rgal} = 5.5 \pm 1.8$ \\
10. Based on the Herschel Infrared Galactic Plane Survey \\
\end{table*} 

New catalogs of clouds based on decomposition of the CO surveys are
now available
\citep{2016ApJ...822...52R, 2017ApJ...834...57M}.
These are particularly interesting as they used the data from 
\citet{2001ApJ...547..792D}
that was featured by 
\citet{2015ARA&A..53..583H}
but analyzed it in two different ways.
These two studies, labeled CO-Rice and CO-MD in Table \ref{tabsample}, take different approaches to breaking the large
scale CO emission in the Galaxy into clouds, and we will discuss
the effect of those differences in \S \ref{disc:method}.

\citet{2016ApJ...822...52R} used a dendrogram analysis to identify
1064 clouds. They recovered 
2.5\ee8 \msun, about 25\% of the total molecular mass of the Galaxy. 
Table 3 of
\citet{2016ApJ...822...52R}
provides a radius, \sigmav, and \mco, the mass determined from
the CO luminosity, which we use directly. 
The mass surface density was determined from \mco\ and the radius.
No mass or size information was given for clouds without clear resolution
 of the kinematic distance ambiguity, and one cloud was assigned zero size and mass in Table 3 of
\citet{2016ApJ...822...52R}; after removing these, 1037 clouds remained
for our analysis.

\citet{2017ApJ...834...57M} 
used a Gaussian decomposition and hierarchical cluster analysis
 to assign essentially all the CO emission in the Dame survey
to 8107 clouds comprising $1.2\ee9$ \msun.
Their online table 
provides radii and masses for both near and far kinematic distances,
along with a code for resolution of the kinematic distance ambiguity, 
which we use to select the best radius and mass. 
\citet{2017ApJ...834...57M} calculated a radius from the two projected radii by
$r = (r_{\rm max} r_{\rm min} r_{\rm min})^{1/3}$, arguing that
the depth was more likely equal to the smaller size projected on
the sky. They also
provide \sigmav, which we use in calculating \alphavir.
The masses were based on summing column densities, using
$X_{\rm CO} = 2\ee{20}$ \cmc\ (\kkms)$^{-1}$, which corresponds
to $\alphaco = 4.3$ \alphacounit\
\citep{2013ARA&A..51..207B}.
The catalog
contains a small number of clouds with nominal Galactocentric radii greater
than 30 kpc and a large number of low-mass objects with very high \alphavir.
To select against these, we require that $R_{\rm gal} < 30$ kpc,
 $\mco > 1$ \msun\ and 
$\alphavir < \amax$, where \amax\ is an arbitrarily chosen maximum value.
Entry 2 in Table \ref{tabstats} required $\alphavir < 100$, while entry 3 used
$\alphavir < 20$.

The sample labeled CO-CMZ in Table \ref{tabsample} is from
\citet{2001ApJ...562..348O}
and focuses on clouds in the Central Molecular Zone (CMZ) using
data obtained with the Nobeyama Radio Observatory 45-m telescope.
Clouds were defined by topologically closed surfaces in $(l, b, v)$ space.
Three different thresholds were used and some clouds appear multiple
times in the catalog, providing a range of surface densities. The
dispersions in all three coordinates were used to define sizes and 
linewidths. A distance of 8.5 kpc was used for all clouds. They
accounted for truncation by the threshold by comparing to simulated
clouds and provided corrected values for sizes, velocity dispersions,
and CO luminosities. These are what we use.  To calculate the 
mass surface density, we take the size as a radius. They also computed virial
masses, but we do not use these. We eliminate clouds determined
to be outside the CMZ (labeled D in their table 1) or confused
(labeled with C in their table 1).
The use of $\alphaco = 4.3$ \alphacounit\ in the CMZ is dubious.
\citet{1998ApJ...493..730O} suggested a value 10 times smaller.
Elemental abundances certainly are higher in the CMZ and a 
correction for that might be appropriate, as suggested by
\cite{2020ApJ...901L...8S}
for other galaxies.

The sample labeled CO-OGS in Table \ref{tabsample}
is a sample of clouds in the outer Galaxy mapped by the FCRAO 14-m telescope 
\citep{1998ApJS..115..241H}
and tabulated by 
\citet{2001ApJ...551..852H}.
The Table provides \lco, \dv\ (FWHM), and diameter of the structure.
\citet{2001ApJ...551..852H}
 used $\alphaco = 4.1$ to convert to mass,
but we changed to 4.3 for consistency.
We convert diameter to radius for consistency.
The linewidth used to calculate
\mvir\ was that of a spectrum averaged over the cloud.
These were almost always larger than an alternative estimate using the dispersion
of line centroids over the cloud, indicating that the clouds are generally
not dominated by systematic motions such as shear.
The catalog contains 10156 structures, but many are local. To select
outer Galaxy sources, 
\citet{2001ApJ...551..852H}
required $\vlsr < -25$ \kms.
To select against small structures, they also required
$\mco > 2.5\ee3$ \msun. We give values for this cut, but we relax it to
$\mco > 10$ \msun\ for
the figures and most analysis because we are more interested in completeness
in structures than in defining a mass function.

\subsection{Structures Defined by CO \jj21\ Emission in Other Galaxies}

Studies of other galaxies are beginning to provide similar 
information without the line-of-sight confusion of Milky Way
studies. 
For other galaxies, we use the remarkable compilation by
\citet{2020ApJ...901L...8S}
of the PHANGS-ALMA study of the CO \jj21\ transition, labeled
CO-Sun in Table \ref{tabsample}, which is part of a larger program
including HST observations
\citep{2021arXiv210102855L}.
They collected data on over \eten5 sightlines toward
66 galaxies, resampled to a uniform spatial resolution of 150 pc,
by far the largest data base available, albeit with very poor
spatial resolution by the standards of studies in our Galaxy.
They used a metallicity dependent value for \alphaco, and we
use their values. Their table 2 provides \sigmam\ and \alphavir.
We compute \mco\ from \sigmam\ and the fixed size of 150 pc and
use their \alphavir.

\subsection{Structures Defined by \coo\ \jj10\ Emission}

\citet{2010ApJ...723..492R}
provided the sample defined by \coo\ \jj10\ emission from the Galactic
Ring Survey (GRS) by 
\citet{2006ApJS..163..145J}. 
The \coo\ data were
 combined with the  CO  data from the 
University of Massachusetts-Stony Brook (UMSB) survey 
\citep{1985ApJ...289..373S, 1986ApJS...60..297C}
to determine optical depth and \coo\ column density.
The clouds in the 
\citet{2010ApJ...723..492R}
study were actually identified by
\citet{2009ApJS..182..131R},
who presented a detailed description of the method, using CLUMPFIND
\citep{1994ApJ...428..693W}
and a number of tests to be able to identify both bright, compact structures
and fainter, more diffuse structures. The effective lowest brightness level
was 0.2 K, so they should not have selected regions of abnormally high 
density.

For the original analysis by
\citet{2010ApJ...723..492R}
of the GRS survey,
 we consider these to be structures
defined by \coo.
Radius, mass, surface density, and \alphavir\ are provided by their
table 1, with 749 entries. Requiring $\mco > 1$ \msun, $\sigmam > 10$ \msunpc,
and $\alphavir < 20$ pruned the sample to 737.
The condition on \alphavir\ removed a few extreme outliers.
170 clouds in the sample  were not covered by the UMSB CO survey,
so the usual method to determine the column density of \coo\ was 
impractical. If we exclude those, the results do not differ substantially.
\citet{2010ApJ...723..492R}
used a different definition of \alphavir, assuming a correction for the
density distribution. For consistency, we recalculated \alphavir\
from Equation \ref{avireq}. These values were larger by a factor of 1.3
on average than those given in their paper when the same assumptions about
abundances were made.

For the \coo\ \jj10\ samples from other parts of the Galaxy, 
both \coo\ and CO \jj10\
were observed simultaneously. These are the Exeter-FCRAO surveys,
which  are divided into EXFC55-100 and EXFC135-195, but our analysis excludes the range of
longitude where kinematic distances are unreliable ($165 < l < 195$),
as described by
\citet{2016ApJ...818..144R}.
For the EXFC surveys, 
\citet{2016ApJ...818..144R}
attempted to identify all voxels with emission to distinguish diffuse
(CO only) from ``dense" gas (\coo\ and CO).  Structures were not identified
and no catalogs were provided.
We have re-analyzed all the \coo\ \jj10\ surveys in a uniform way, using
the CO-defined structures from 
\citet{2017ApJ...834...57M},
with the procedure detailed in the Appendix. Because the masses are based
on all the \coo\ data within the CO-defined structures, we refer to these
results as \coo-CO, followed by the survey name, such as GRS.

The GRS sample is concentrated in the inner Galaxy; the sub-sample that we
analyze has median(\rgal)$ = 5.2$ kpc, $\mean{\rgal} = 5.2 \pm 1.0$ kpc.
The EXFC55-100 sample is concentrated closer to the solar circle, with 
$ 7.2 < \rgal < 14.5$, median(\rgal)$ = 10.7$ kpc, $\mean{\rgal} = 10.3 \pm 1.9$ kpc.
The EXFC135-195 data is concentrated in the outer Galaxy, with 
$ 10.3 < \rgal < 20.4$, median(\rgal)$ = 12.5$ kpc, $\mean{\rgal} = 12.9 \pm 2.1$ kpc.
They thus provide convenient samples for examining the effects of \rgal.

\subsection{Structures Defined by \coo\ \jj21\ Emission }

The SEDIGISM survey 
\citep{2021MNRAS.500.3027D}
has provided a catalog of structures defined by emission from the \jj21\
transition of \coo.  They used the SCIMES algorithm described by
\citet{2019MNRAS.483.4291C}
to define the structures.
They determined column densities by assuming a constant ratio of \hh\  column density to integrated intensity of \coo\ \jj21\ of $X(\coo\ \jj21) = 1\ee{21} \cmc (\kkms)^{-1}$ and a mean molecular weight per \hh\ 
of 2.8. Because their choice of $X(\coo\ \jj21)$ was empirically based
\citep{2017A&A...601A.124S},
we did no re-scaling.
The  sample was pruned to remove those with ambiguous
distances, near  edges of maps, or smaller than 114 pixels, roughly re-creating
their ``science sample.'' The masses and virial parameters in their tables
were used directly, but \alphavir\ was checked and agreed with a calculation
from their mass, size, and linewidth information.

\subsection{Structures Defined by Herschel and \ammonia}

Structures identified from the Herschel
and ATLASGAL maps of the Milky Way were presented by 
\citet{2019MNRAS.483.5355M}.
They meet a number of requirements, but most relevant to us, they have \ammonia\
observations, which are used to obtain a linewidth from the inversion
transition in the $(J,K) = (1,1)$ level and a temperature from the ratio
of (1,1) and (2,2) lines. We use the sample of 1068,
called the ``Final Sample" by
\citet{2019MNRAS.483.5355M}.
They provide mass, \dv\ (FWHM), and radius, from which we compute 
\alphavir\ and \sigmam.
The masses were computed from a fit to the SED, the resulting optical depth
at 300 \micron, and an opacity of 0.1 cm$^2$ g$^{-1}$ of gas plus dust.
The radius was half the FWHM of the emission at 250 \micron.


\section{Results}\label{results}

For each of the samples, four plots are provided:
(1) the mean, median, and standard deviation of \alphavir\ versus mass;
(2) the mean, median, and standard deviation of \alphavir\ versus surface density (\sigmam);
(3) a histogram of $\log \sigmam$;
(4) a histogram of $\log \alphavir$.
Statistics of
the relevant properties are summarized in Table \ref{tabstats}.
For each sample,
Table \ref{tabstats} has the median, mean, and standard deviation of \alphavir\ 
and the median, mean, and standard deviation of \sigmam.
The table also includes the fraction of the total number of structures 
that satisfy  $\alphavir < 2$.  
The fraction of the number (\fn), and the fraction of the total mass
(\fmass) satisfying these conditions are both tabulated.

\begin{table*}[h] 
\begin{centering} 
\caption{Statistics} \label{tabstats} 
\vspace {3mm} 
\begin{tabular}{l r r r r r r r r r r r r} 
\tableline 
\tableline 
Sample & Number & \multicolumn{3}{c}{\alphavir} & \multicolumn{3}{c}{\sigmam\ (\msunpc)} & \fn & \fmass & \fcn & \fcmass & Note \cr 
         &        & Med & Mean & Std. & Med & Mean       & Std.        &  &  & & &  \cr 
\tableline 
CO-Rice & 1037  & 2.44  & 3.45  & 3.34  & 21.2 & 38.3 & 43.4 &  0.38 & 0.73 & 0.33 & 0.41 & 1 \cr 
CO-MD & 7516  & 7.58  & 15.20  & 18.53  & 21.9 & 41.1 & 53.9 &  0.07 & 0.19 & 0.06 & 0.09 & 2 \cr 
CO-MD & 5776  & 5.49  & 6.91  & 4.70  & 31.4 & 49.4 & 58.1 &  0.09 & 0.19 & 0.08 & 0.10 & 3 \cr 
CO-CMZ & 110  & 3.26  & 4.45  & 3.46  & 1311.3 & 1703.2 & 1271.0 &  0.29 & 0.71 & 0.00 & 0.00 & 4 \cr 
CO-OGS & 380  & 2.27  & 2.51  & 1.25  & 29.3 & 35.4 & 20.2 &  0.41 & 0.72 & 0.40 & 0.56 & 5 \cr 
CO-OGS & 3714  & 6.15  & 7.22  & 4.79  & 17.7 & 21.0 & 11.7 &  0.06 & 0.58 & 0.06 & 0.46 & 6 \cr 
CO-Sun & 102788  & 3.47  & 4.57  & 4.10  & 21.6 & 49.9 & 144.3 &  0.21 & 0.35 & 0.18 & 0.15 & 7 \cr
\coo(1-0) & 737  & 1.03  & 1.95  & 2.76  & 82.0 & 85.5 & 39.8 &  0.72 & 0.95 & 0.58 & 0.55 & 8 \cr 
$^{13}$CO-CO-GRS & 289  & 2.86  & 3.83  & 3.02  & 80.7 & 103.7 & 84.0 &  0.31 & 0.46 & 0.17 & 0.19 & 9 \cr 
$^{13}$CO-CO-EXFC55-100 & 105  & 0.99  & 1.92  & 2.24  & 25.9 & 36.1 & 32.8 &  0.62 & 0.73 & 0.58 & 0.70 & 10 \cr 
$^{13}$CO-CO-EXFC135-195 & 98  & 1.37  & 2.30  & 3.00  & 21.1 & 24.2 & 13.0 &  0.63 & 0.75 & 0.63 & 0.75 & 11 \cr 
\coo(2-1) & 6658  & 1.25  & 1.94  & 2.66  & 74.7 & 87.7 & 49.0 &  0.73 & 0.79 & 0.62 & 0.41 & 12 \cr 
Herschel & 1067  & 0.29  & 0.58  & 1.08  & 2497.7 & 3249.8 & 3032.2 &  0.96 & 1.00 & 0.00 & 0.00 & 13 \cr 
\tableline 
\end{tabular} 

\end{centering} 
Notes: 1. Data from \citet{2016ApJ...822...52R}. Clouds with ambiguous distance assignments and one cloud with zero mass were eliminated. \\        2. Data from \citet{2017ApJ...834...57M} using criteria $\mco > 1$ \msun, $R_{\rm gal} < 30$ kpc,  and $\alphavir < 100.$ \\         3. Data from \citet{2017ApJ...834...57M} using criteria $\mco > 1$ \msun, $R_{\rm gal} < 30$ kpc, and $\alphavir < 20.$ \\         4. Data from \citet{2001ApJ...562..348O}, considering only clouds in CMZ \\        5. Data from \cite{2001ApJ...551..852H}, using criterion $\mco > 2.5\ee3$ \msun  \\        6. Data from \cite{2001ApJ...551..852H}, using criterion $\mco > 1.0\ee1$ \msun  \\        7. Data from \citet{2020ApJ...901L...8S} \\        8. GRS Data from \citet{2010ApJ...723..492R}, updated abundance, some selection \\        9. Re-analysis of GRS \coo\ data within CO clouds \\         10. Re-analysis of EXFC 55-100 \coo\ data within CO clouds \\         11. Re-analysis of EXFC 135-195  \coo\ data within CO clouds \\         12. Data from \citet{2021MNRAS.500.3027D}  \\        13. Data from \citet{2019MNRAS.483.5355M}, using criteria $\mdust > 1$ \msun \\        
\end{table*}

We will compare the criterion for the virial parameter to 
the surface
density criterion for star formation. In local clouds, star formation
is strongly concentrated in  regions with $\sigmam \geq 120$ \msunpc\
\citep{2010ApJ...723.1019H,2010ApJ...724..687L, 2012ApJ...745..190L, 2014ApJ...782..114E}. 
 For simplicity, we refer to structures
with $\sigmam < 120$ \msunpc\ as clouds and those with
$\sigmam \ge 120$ \msunpc\ as clumps and the plots of
\alphavir\ versus \sigmam\ include a line demarcating that boundary.
We emphasize that this definition of ``clump" does not include constraints
often used to define clumps, such as size information, and we use the 
terminology only for simplicity. This usage of ``clump" is also distinguished
from ``dense clumps" as identified in lines of much higher effective density,
as in the Herschel sample or samples studied by
\citet{2010ApJS..188..313W}
and others.
We use bins of 0.3 in log \sigmam, on either side of 2.08, corresponding
to $\sigmam = 120$ \msunpc.
Similarly, we summarize the boundedness with labels of ``bound" or 
``unbound" on either side of a line at $\alphavir = 2$ in the plots.
Finally, the fraction of the number (\fcn) and mass (\fcmass) that
satisfy both $\alphavir < 2$ and $\sigmam < 120$ \msunpc\ are tabulated,
the subscript (c) indicating that they apply only to structures we have called
clouds.

\subsection{Structures Defined by CO \jj10\ Emission}

\begin{figure*}
\center
\includegraphics[scale=0.3]{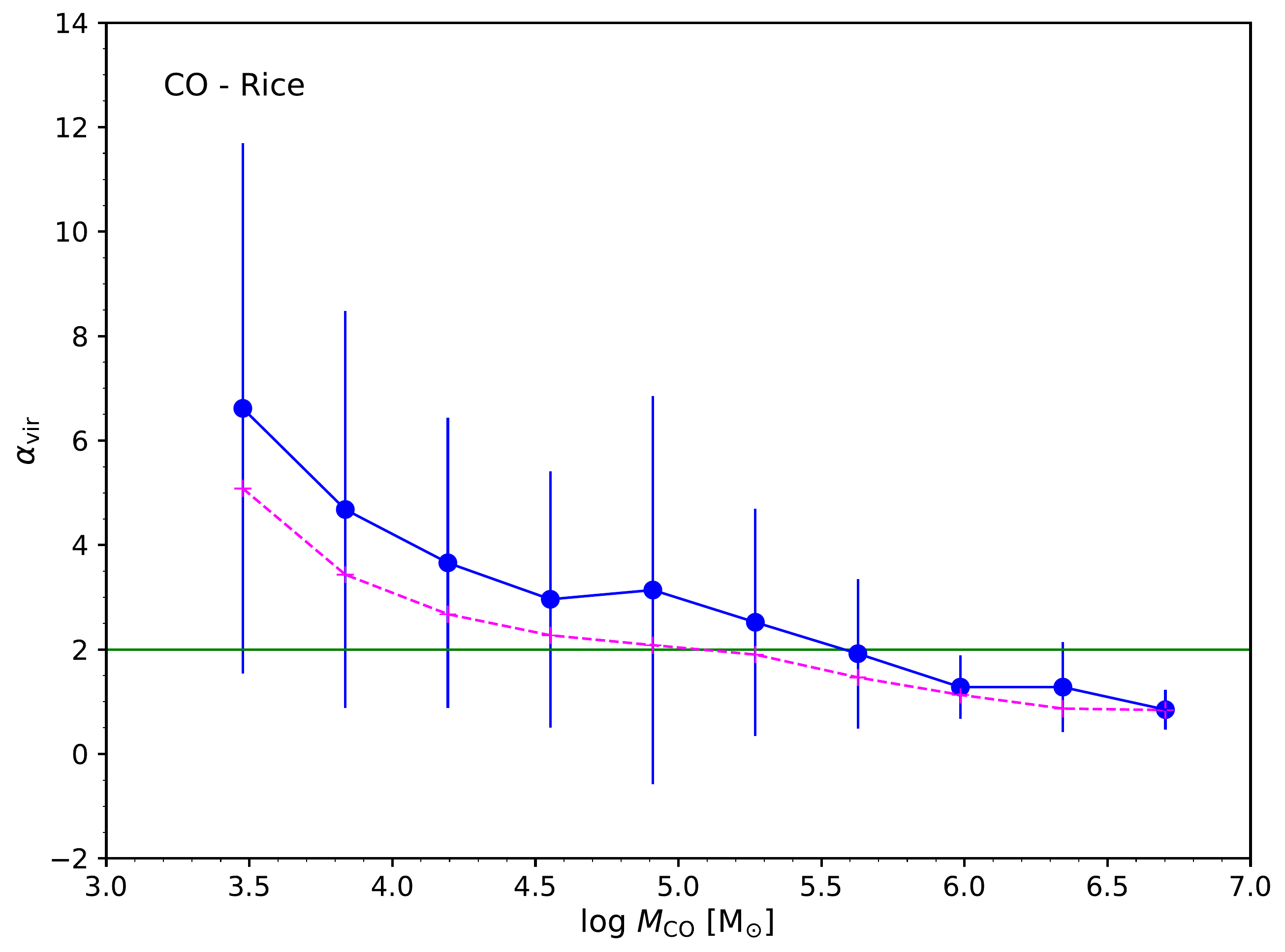}
\includegraphics[scale=0.3]{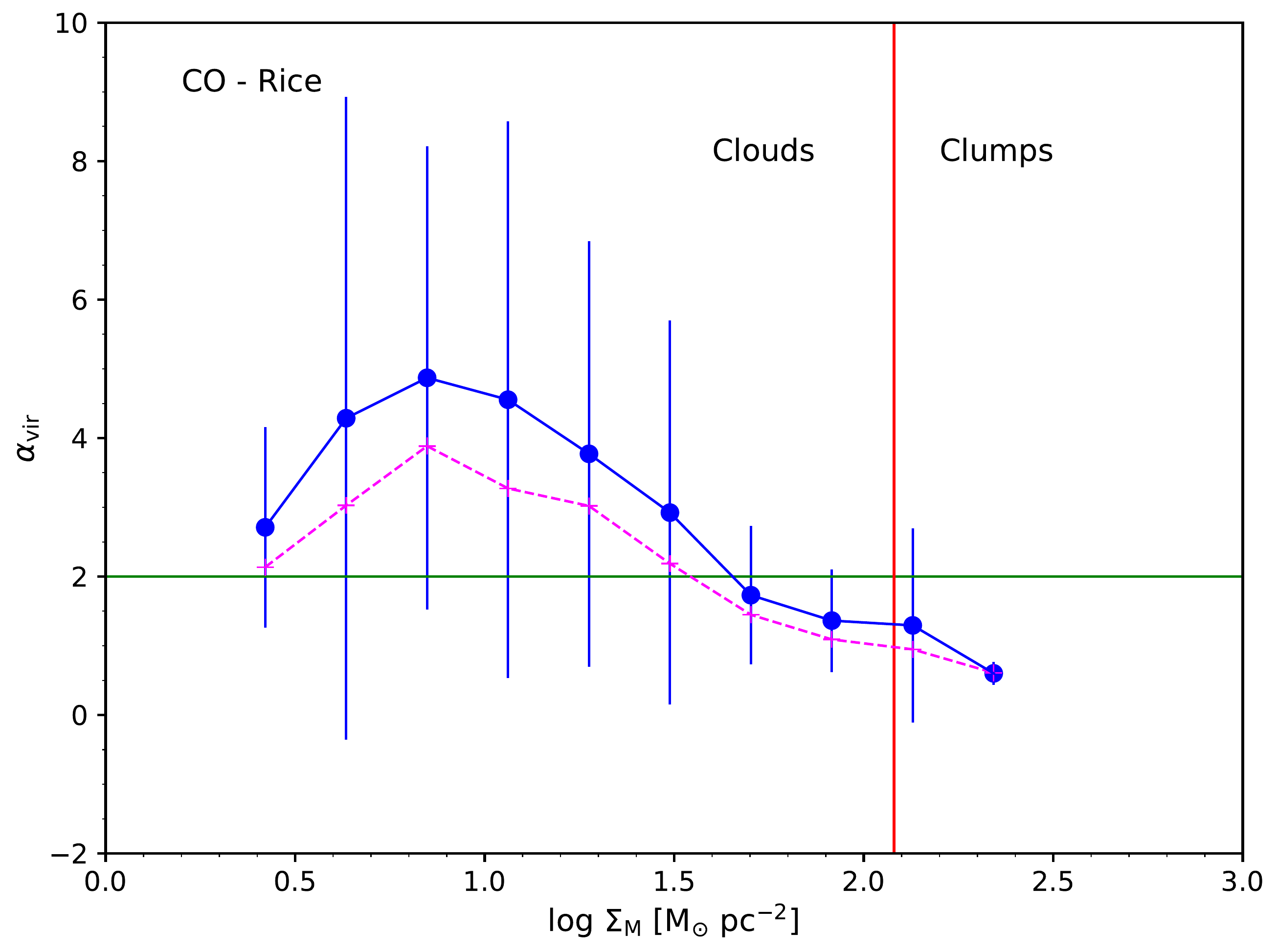}
\includegraphics[scale=0.3]{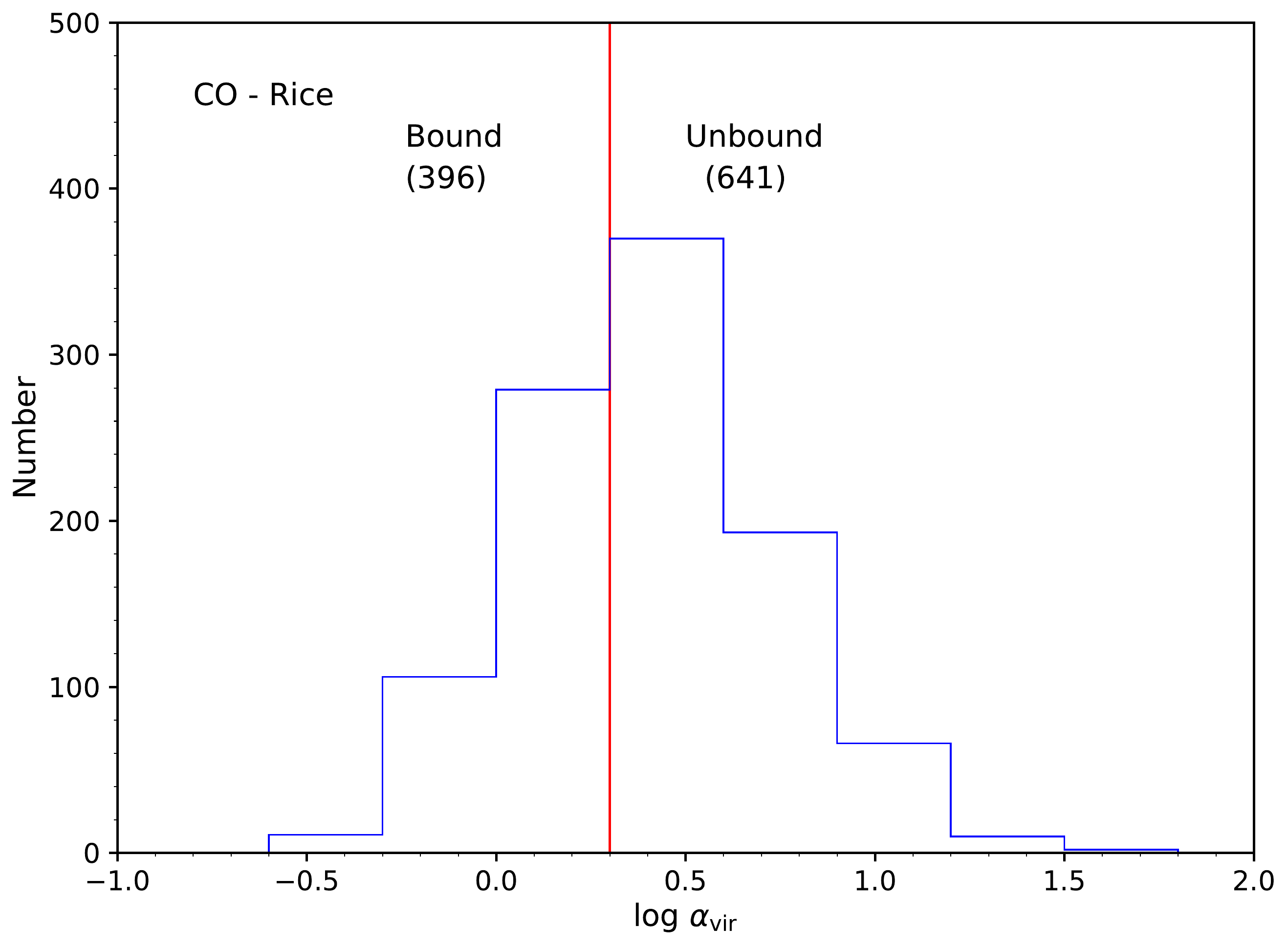}
\includegraphics[scale=0.3]{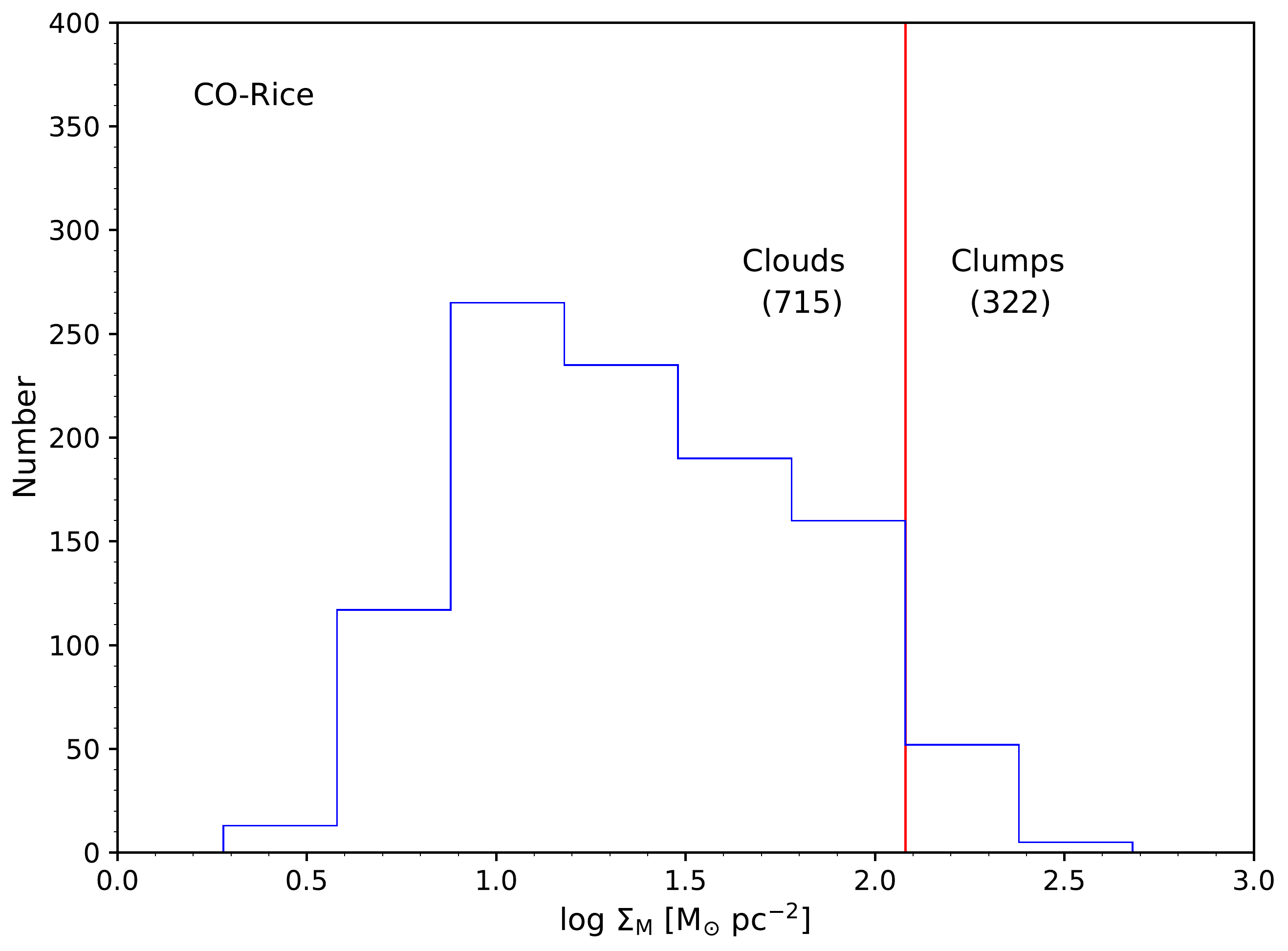}
\caption{
(Upper Left)
The mean and standard deviation of the virial ratio is plotted versus the logarithm of the clump mass, for the catalog of 
\citet{2016ApJ...822...52R}.
Median values are plotted in a magenta dashed line.
(Upper Right) The same quantities are plotted versus the mass surface density.
The horizontal line at $\alphavir = 2$ demarcates nominally unbound clouds above the
line from nominally bound clouds below the line. 
The vertical line in the right panel indicates $\sigmam = 120$ \msunpc.
(Lower left) The histogram of values of log \alphavir, with a vertical red line
at $\alphavir = 2.0$. 
The number of structures in each category are given in parentheses.
(Lower right) The histogram of values of log \sigmam, with a vertical red line at
$\sigmam = 120$ \msunpc.
}
\label{rice1}
\end{figure*}

We plot \alphavir\ calculated from the data in
\citet{2016ApJ...822...52R}
in Figure \ref{rice1}.
The mean \alphavir\ declines with mass, so that 
$\mean{\alphavir} < 2$ for log$\mco > 5.7$. 
The mean \alphavir\ also declines with \sigmam, so that 
$\mean{\alphavir} < 2$ for log$\sigmam > 1.7$. 
The histograms indicate that most structures are clouds and 
most are \added{unbound (\fn = 0.38), but the fraction of mass in bound 
structures is 0.73, reflecting the tendency for more 
massive structures to be bound.
}

\begin{figure*}
\center
\includegraphics[scale=0.3]{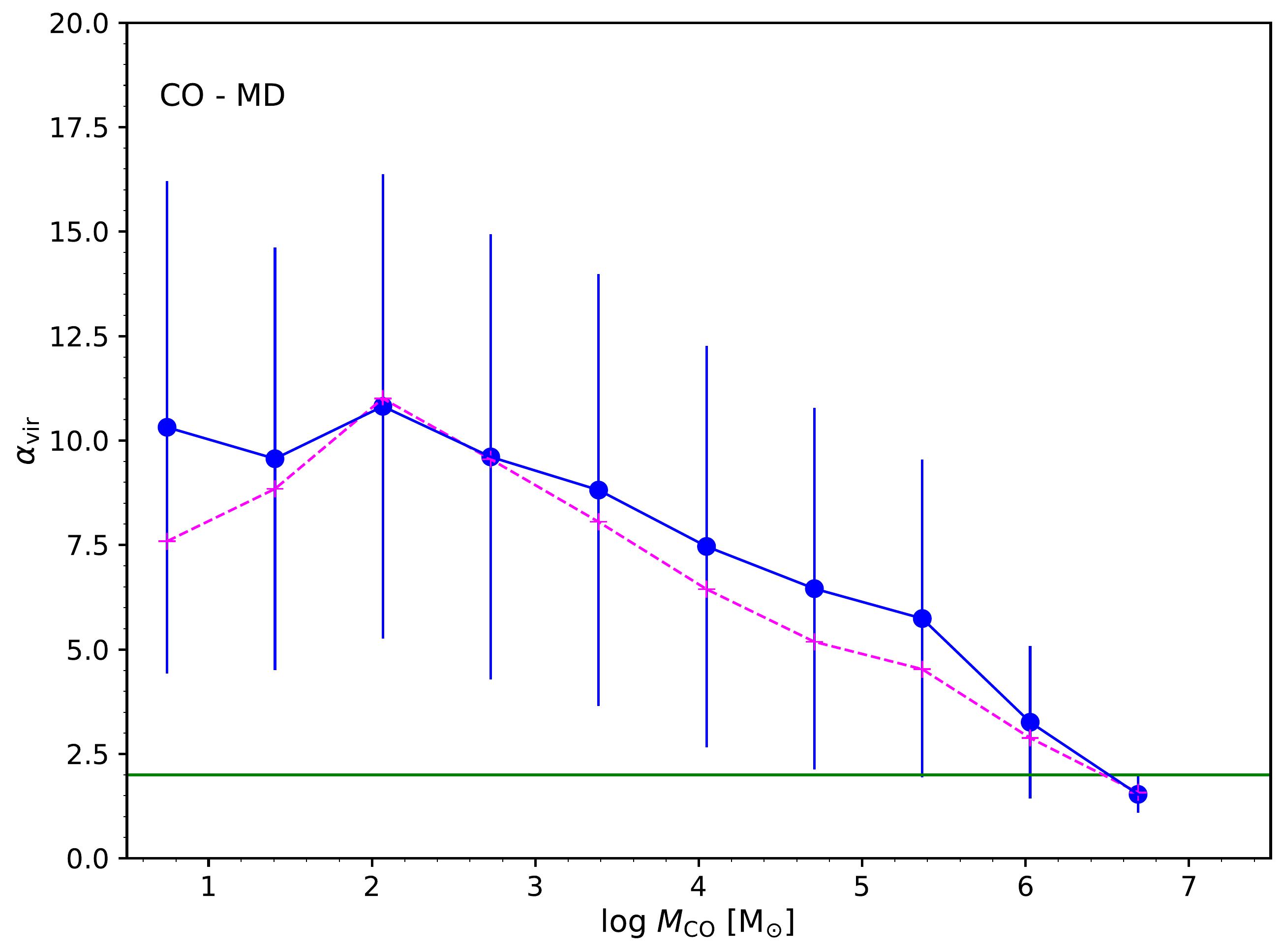}
\includegraphics[scale=0.3]{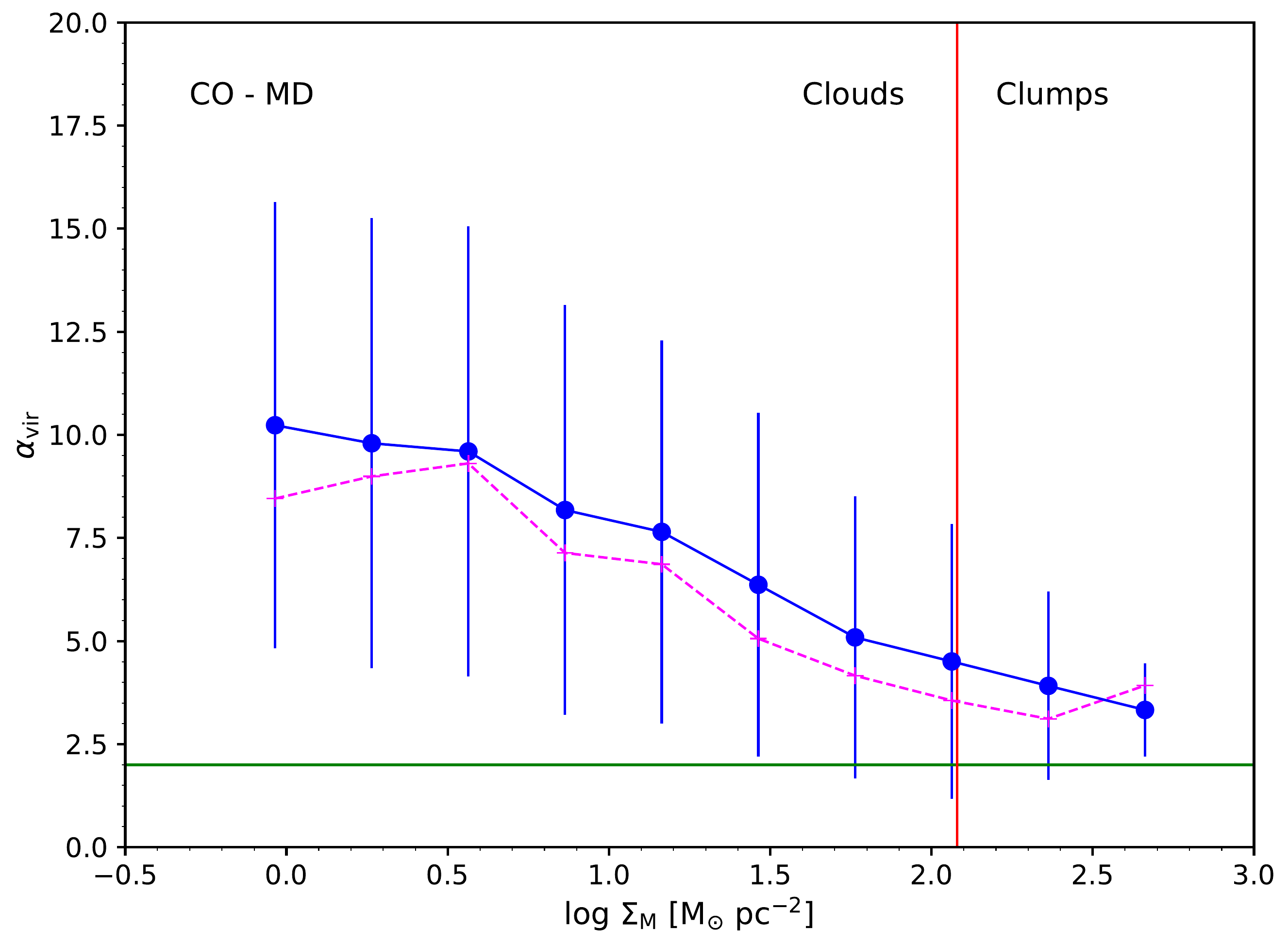}
\includegraphics[scale=0.3]{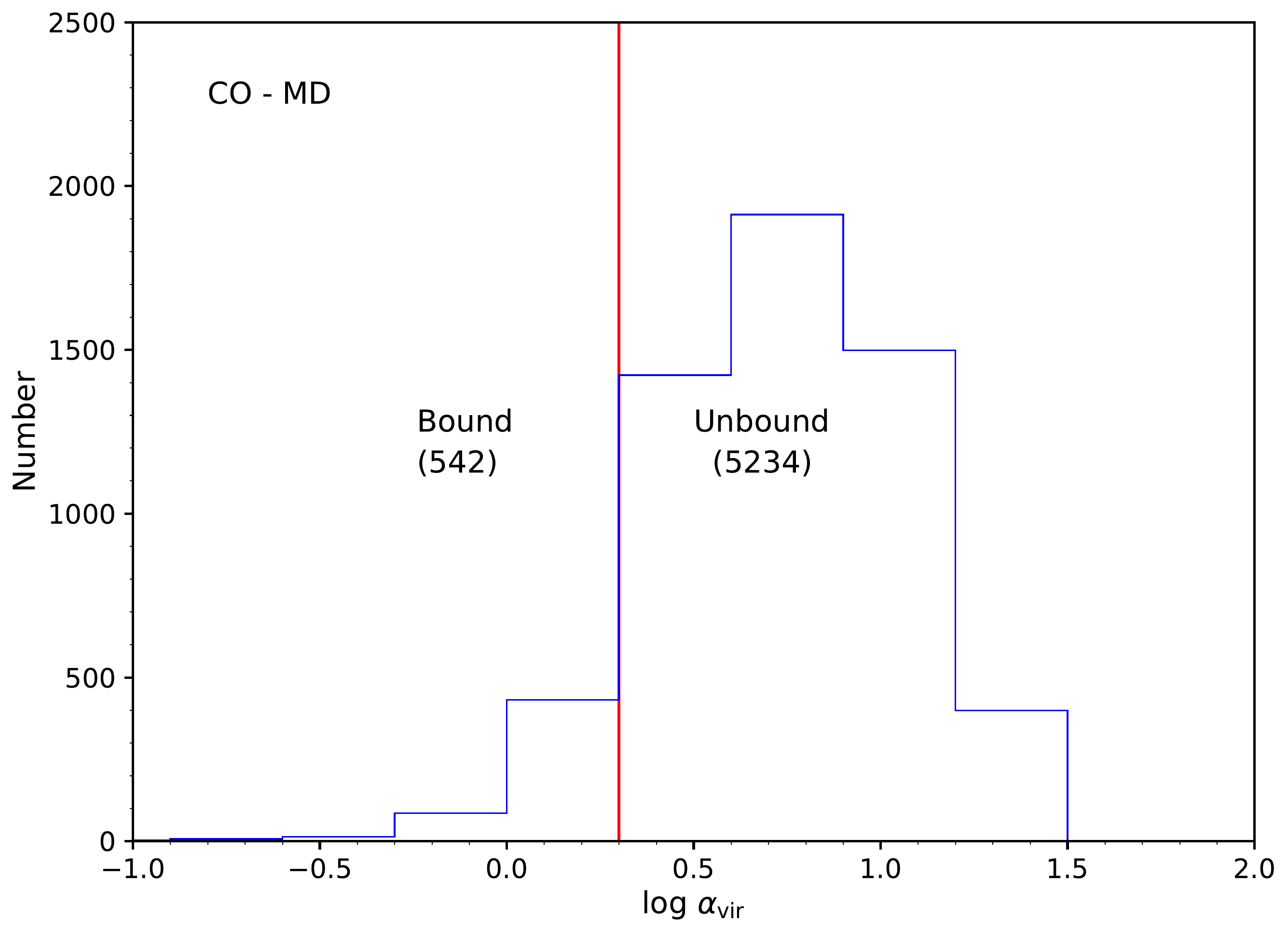}
\includegraphics[scale=0.3]{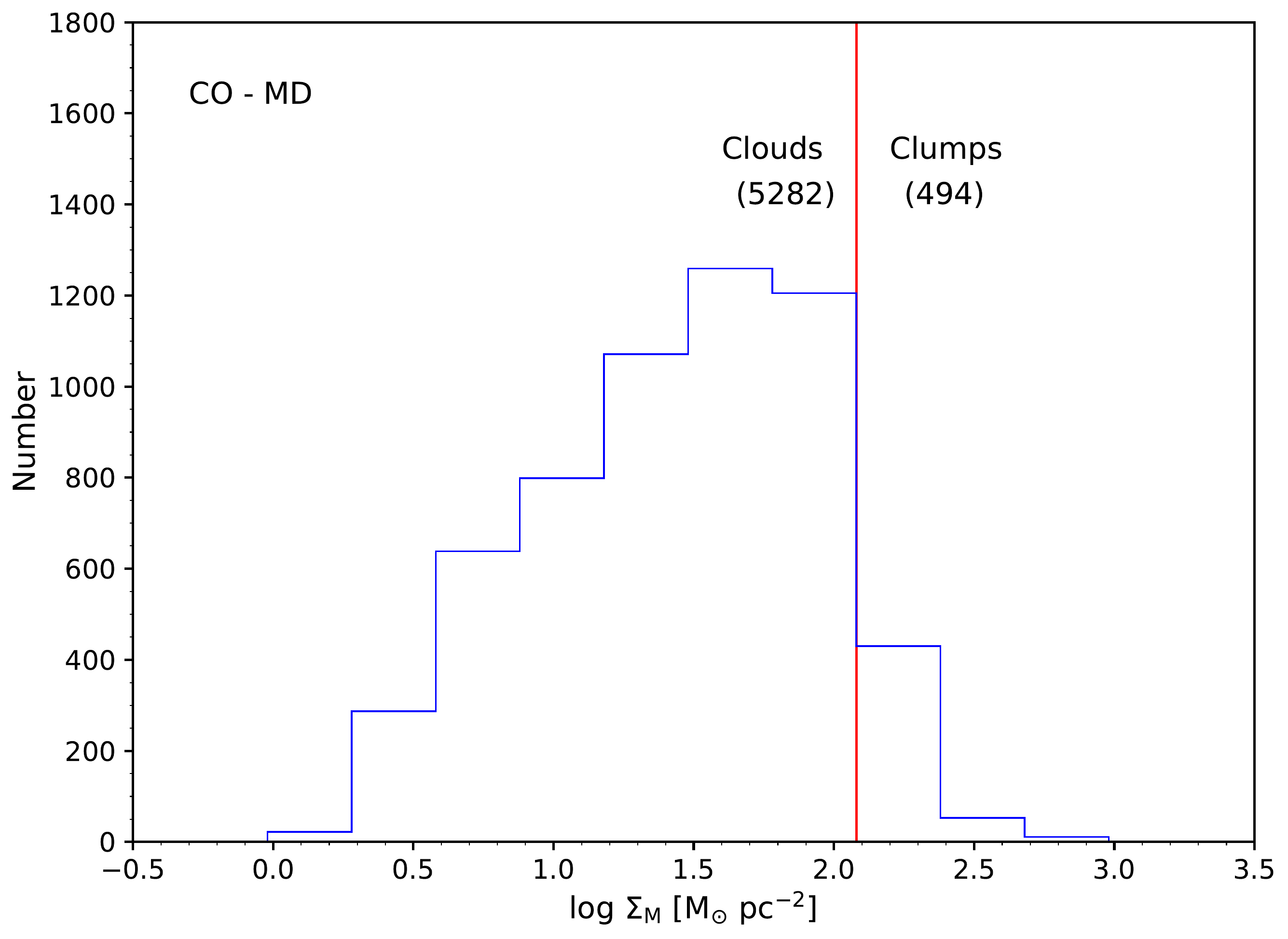}
\caption{
(Upper Left)
The mean and standard deviation of the virial ratio is plotted versus the logarithm of the clump mass, for the catalog of 
\citet{2017ApJ...834...57M}
after selecting only those catalog entries with $\alphavir < 100$ and
$\mco > 1$ \msun.
Median values are plotted in a magenta dashed line.
(Upper Right) The same quantities are plotted versus the mass surface density.
The horizontal line at $\alphavir = 2$ demarcates nominally unbound clouds above the
line from nominally bound clouds below the line. The vertical line in the right panel
indicates $\sigmam = 120$ \msunpc.
(Lower left) The histogram of values of log \alphavir, with a vertical red line
at $\alphavir = 2.0$. 
The number of structures in each category are given in parentheses.
(Lower right) The histogram of values of log \sigmam, with a vertical red line at
$\sigmam = 120$ \msunpc.
}
\label{md1}
\end{figure*}

The  catalog of
\citet{2017ApJ...834...57M},
as noted in \S \ref{samples},
contains some outliers with very high nominal \alphavir, so we
exclude those greater than a maxiumum value, \amax.
Entries in Table \ref{tabstats} are given for two choices, 
$\amax = 100$ and $\amax = 20$.
The mean and standard deviation are plotted versus mass in Figure
\ref{md1}. As we saw with the catalog of 
\citet{2016ApJ...822...52R}, 
the $\mean{\alphavir}$ values decline with \mco. In this case, they reach
$\mean{\alphavir} = 2$ only for the most massive bin. 
The mean and median values of \alphavir\ decline with \sigmam, but never
drop below $\alphavir = 2$. The histograms indicate that the vast majority
of structures are unbound clouds. The fraction of the
mass that is bound is relatively insensitive to the choice of \amax, at
about 0.19 for all structures.
The fact that these results differ subtantially from those of
\citet{2016ApJ...822...52R}
will be discussed in \S \ref{disc:method}.

\begin{figure*}
\center
\includegraphics[scale=0.3]{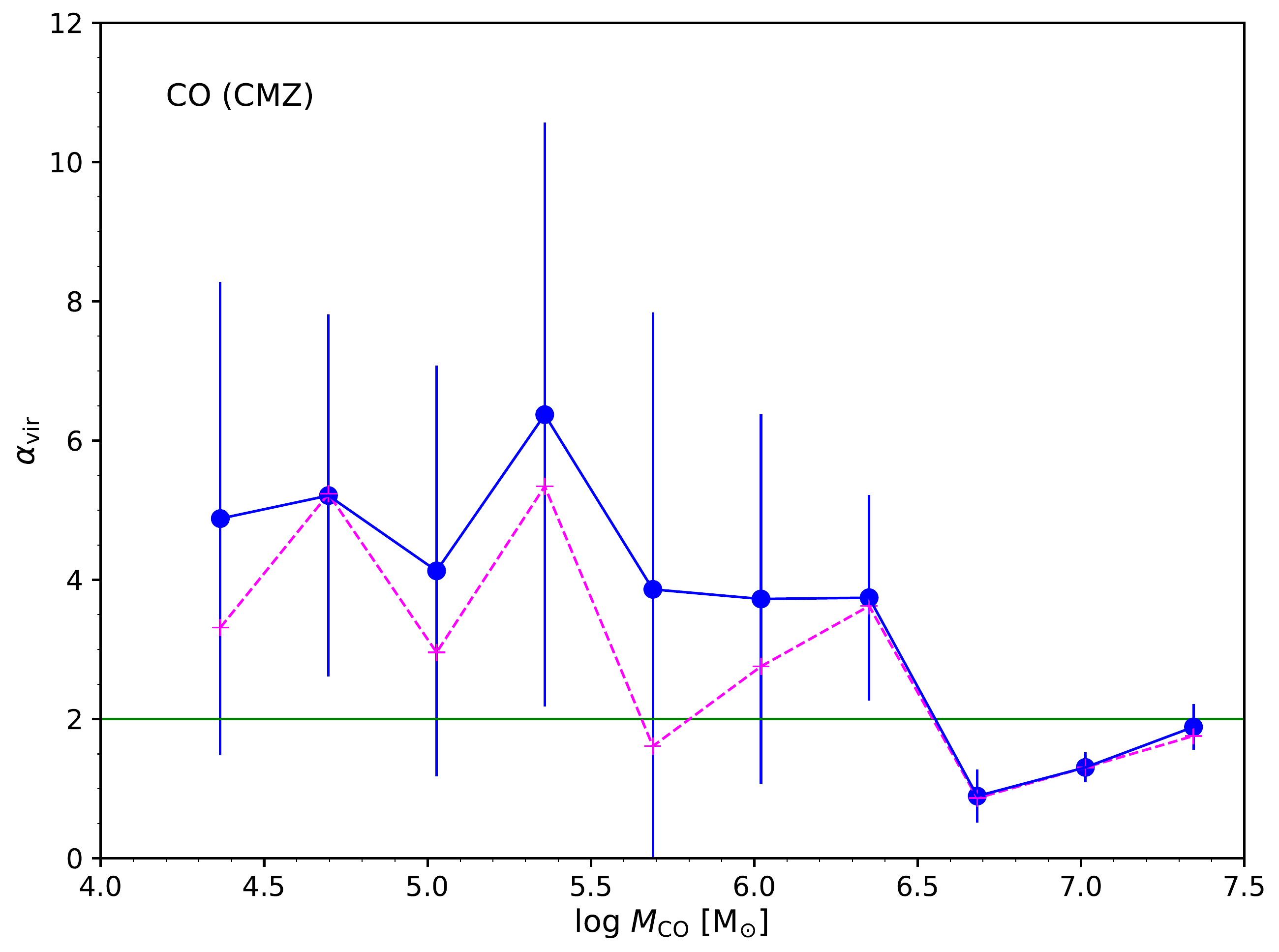}
\includegraphics[scale=0.3]{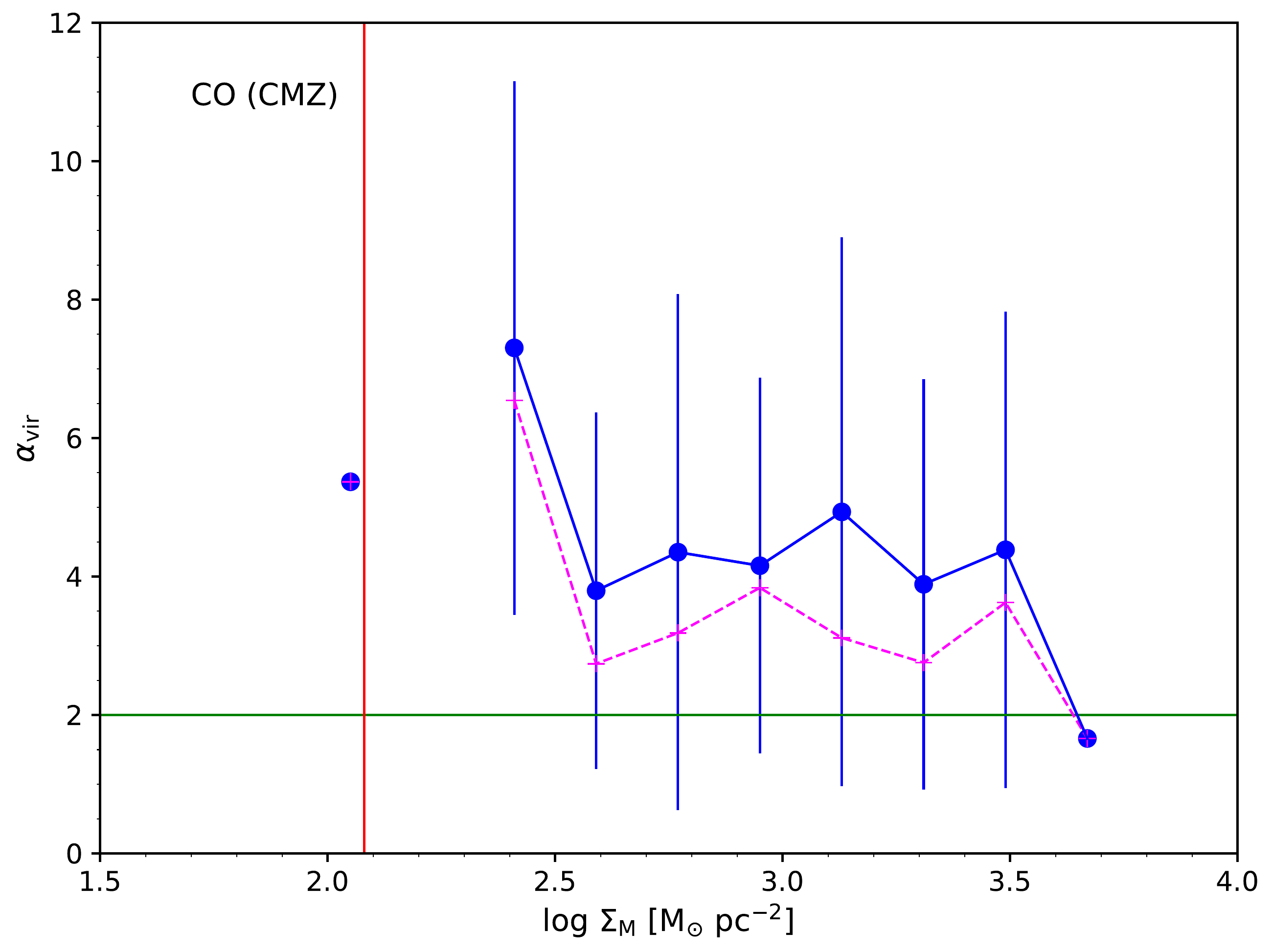}
\includegraphics[scale=0.3]{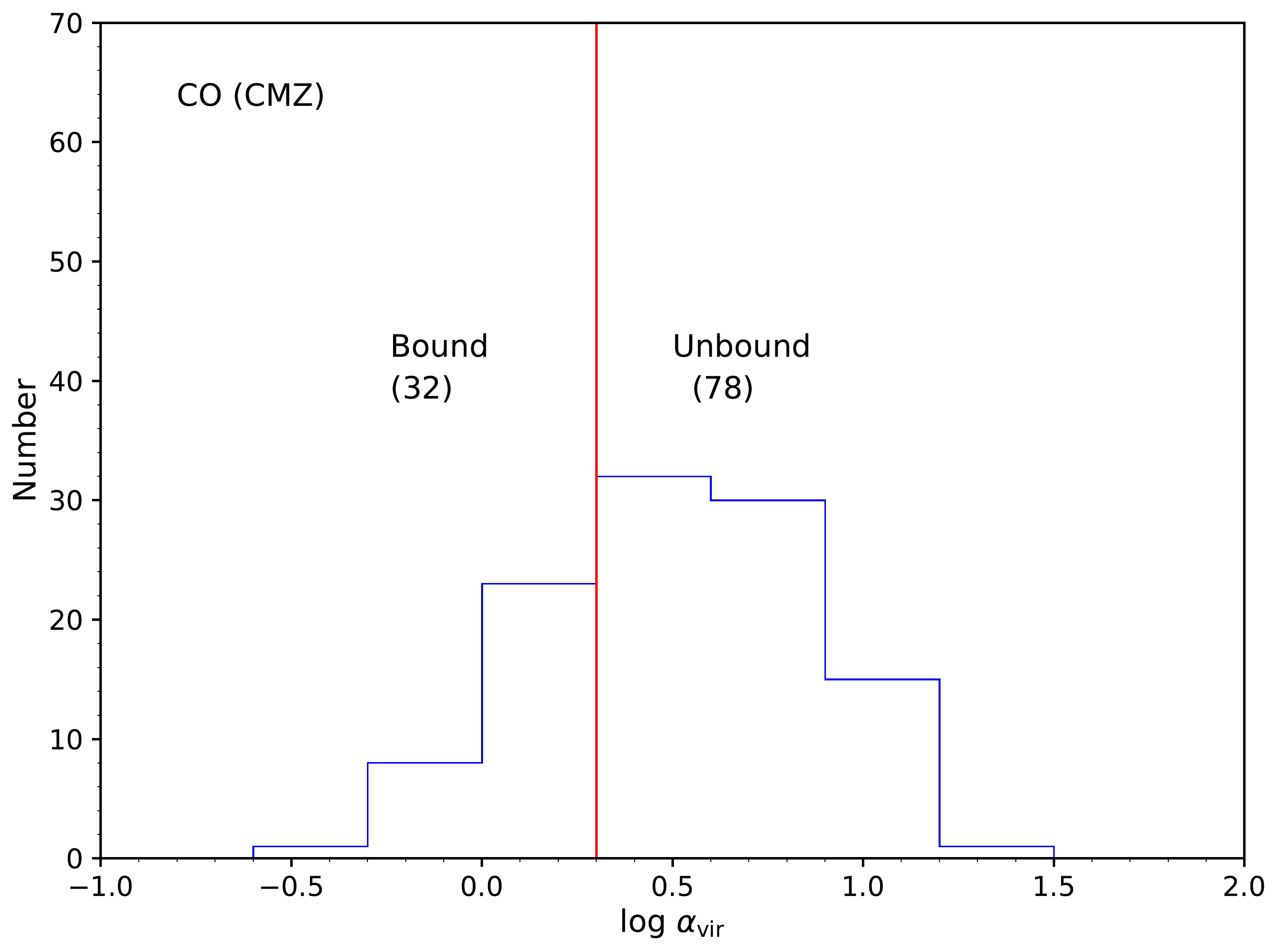}
\includegraphics[scale=0.3]{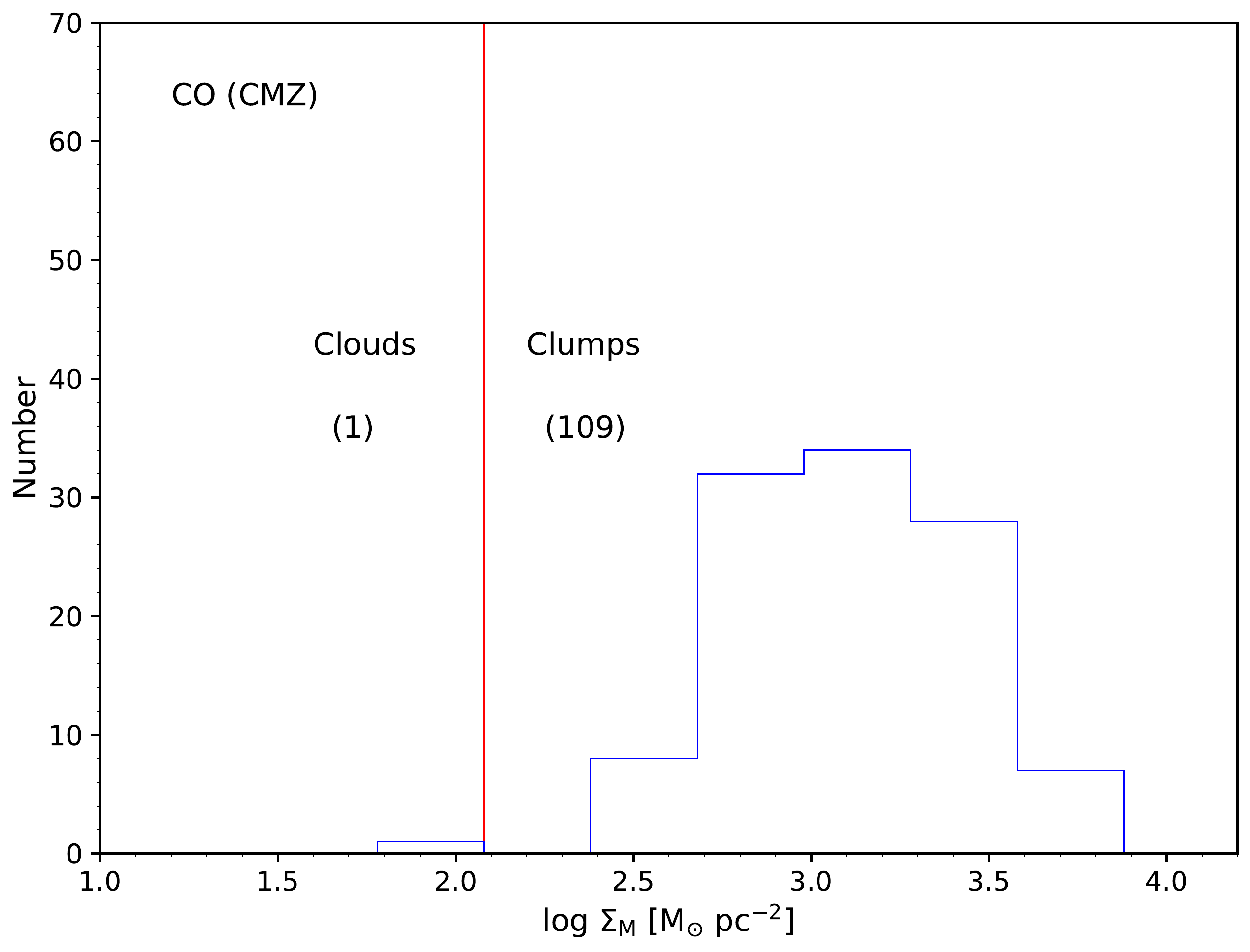}
\caption{
The catalog of 
\citet{2001ApJ...562..348O} is plotted.
(Upper Left)
The mean and standard deviation of the virial ratio is plotted versus the logarithm of the clump mass. Median values are plotted in a magenta dashed line.
(Upper Right) The same quantities are plotted versus the mass surface density.
The horizontal line at $\alphavir = 2$ demarcates nominally unbound clouds above the
line from nominally bound clouds below the line. The vertical line 
indicates $\sigmam = 120$ \msunpc.
(Lower left) The histogram of values of log \alphavir, with a vertical red line
at $\alphavir = 2.0$. 
The number of structures in each category are given in parentheses.
(Lower right) The histogram of values of log \sigmam, with a vertical red line at
$\sigmam = 120$ \msunpc.}
\label{oka}
\end{figure*}

Figure \ref{oka} shows that
the structures in the CMZ in the catalog of 
\citet{2001ApJ...562..348O}
are generally unbound even though they almost all
 have $\sigmam > 120$ \msunpc, satisfying
the local criterion for star formation. 
Clearly, the definition of a clump based on the local surface density
criterion for active star formation does not work in the CMZ.
While most structures are unbound, the tendency for higher 
\replaced{masss}{mass structures} to be bound
results in a relatively high $\fmass = 0.71$, with the caveat that these masses
may well be overestimated (\S \ref{samples}).

\begin{figure*}
\center
\includegraphics[scale=0.3]{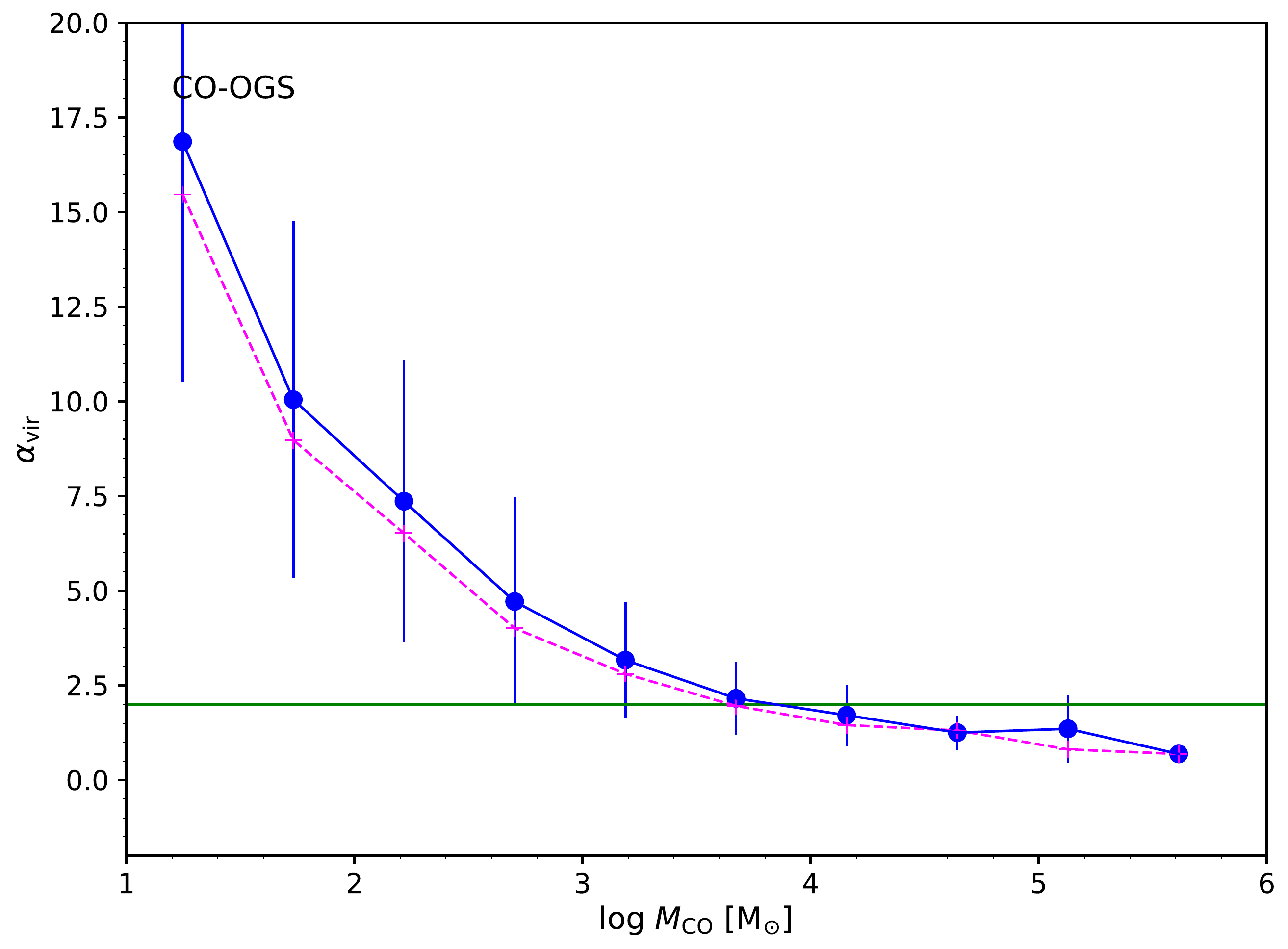}
\includegraphics[scale=0.3]{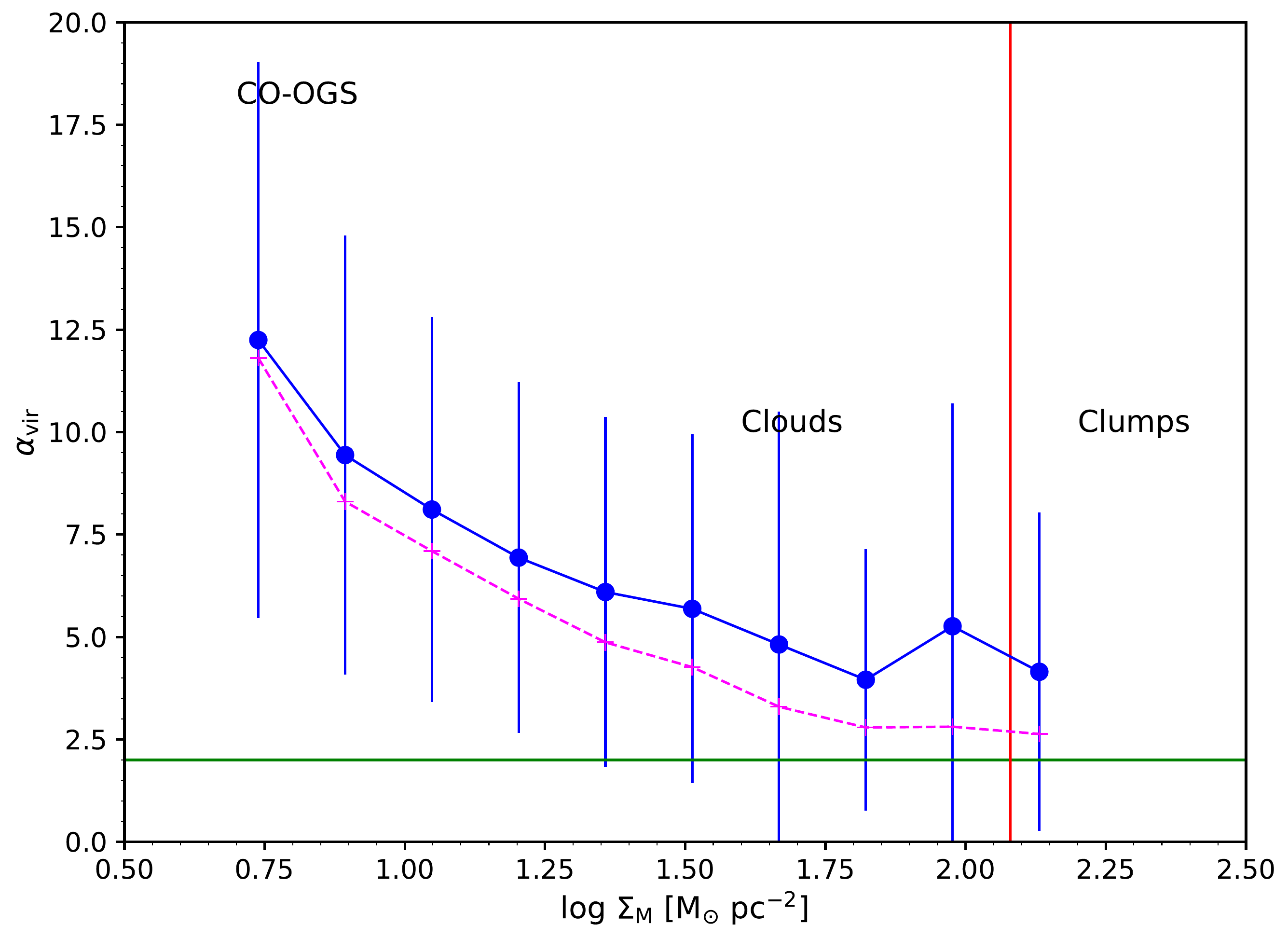}
\includegraphics[scale=0.3]{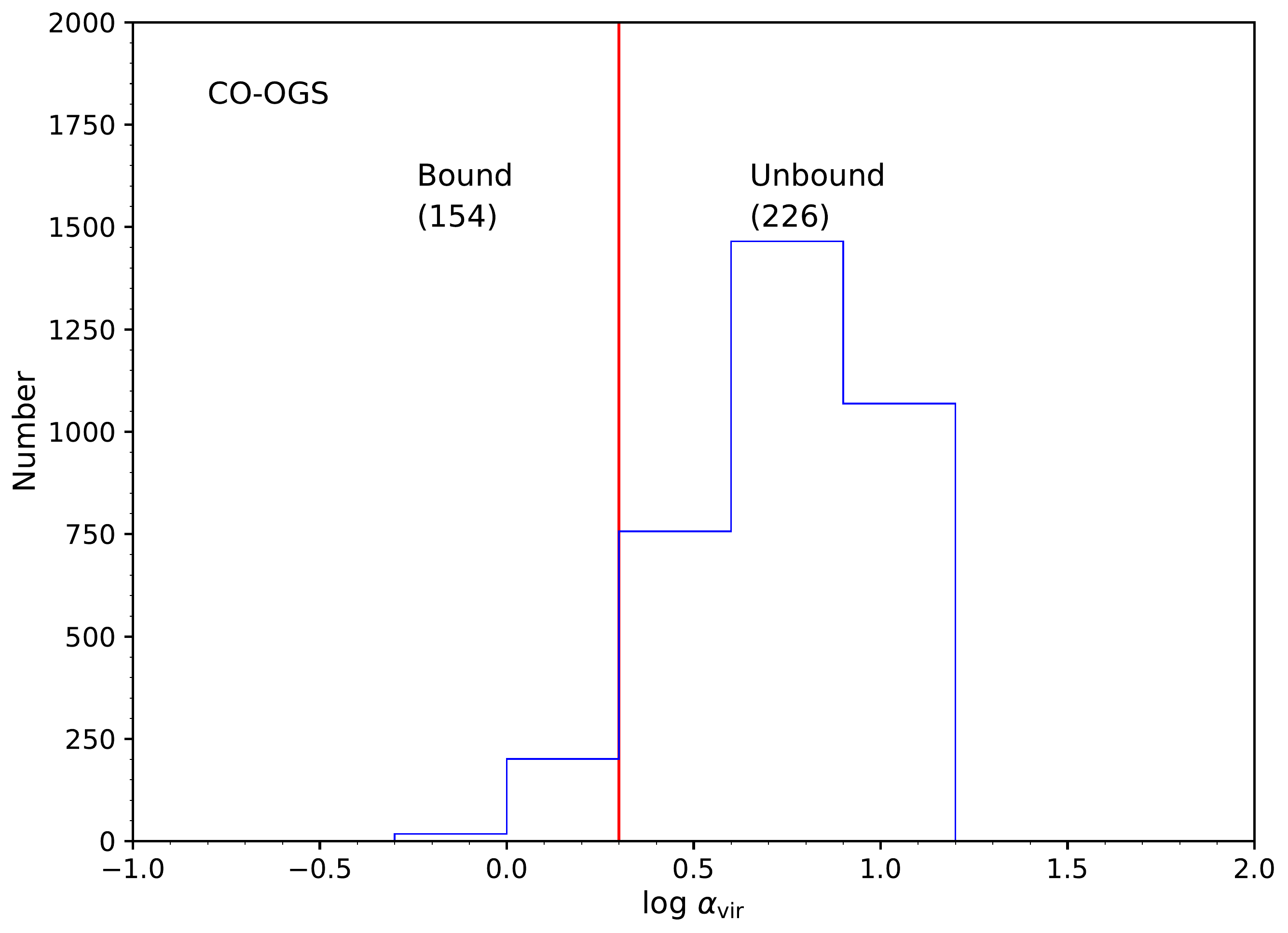}
\includegraphics[scale=0.3]{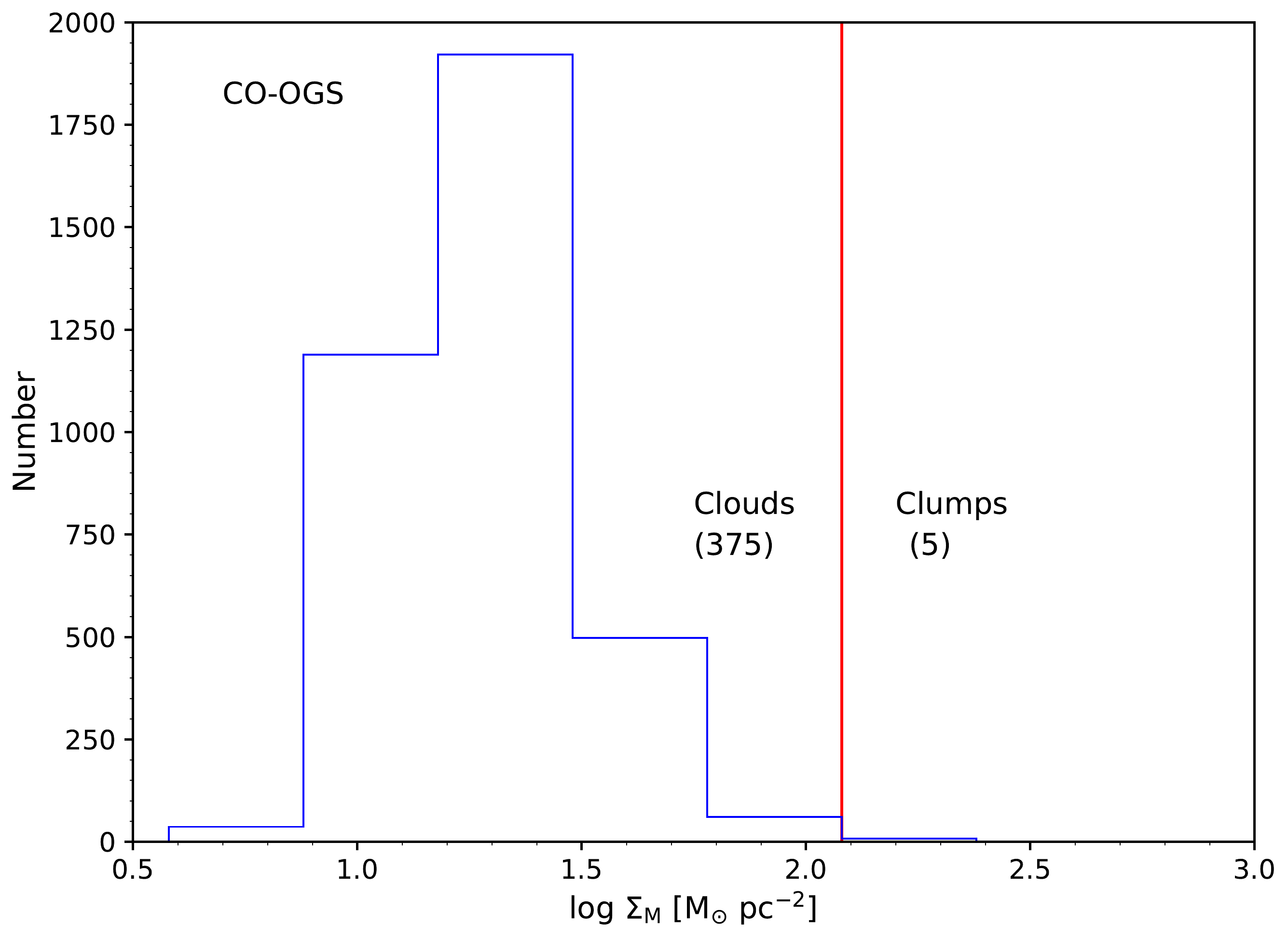}
\caption{
The catalog of 
\citet{2001ApJ...551..852H} is plotted.
The masses were limited to $\mco > 2.5\ee3$ \msun\  for completeness and $\vlsr < -25$ \kms\ to avoid
distance ambiguity.
(Upper Left)
The mean and standard deviation of the virial ratio is plotted versus the logarithm of the clump mass. Median values are plotted in a magenta dashed line.
(Upper Right) The same quantities are plotted versus the mass surface density.
The horizontal line at $\alphavir = 2$ demarcates nominally unbound clouds above the
line from nominally bound clouds below the line. The vertical line 
indicates $\sigmam = 120$ \msunpc.
(Lower left) The histogram of values of log \alphavir, with a vertical red line
at $\alphavir = 2.0$. 
The number of structures in each category are given in parentheses.
(Lower right) The histogram of values of log \sigmam, with a vertical red line at
$\sigmam = 120$ \msunpc.}
\label{omw}
\end{figure*}

The outer Galaxy sample (OGS) from the catalog of
\citet{2001ApJ...551..852H}
shows \replaced{the opposite}{a similar} behavior. Most are \replaced{bound}{unbound, but more than half the mass is in bound structures} despite having
$\sigmam < 120$ \msunpc, with $\fmass = 0.72$ (Table \ref{tabstats}).
 To some extent, this result depends on the requirement that
$\mco > 2.5\ee3$ \msun\ imposed by
\cite{2001ApJ...551..852H} to ensure completeness. Since completenesss
is less important for our purposes, we relaxed that criterion to
$\mco > 10$ \msun\ to see the effect. While \fn\ dropped to 0.06,
$\fmass = 0.58$, still substantial because most of the bound mass was 
in the higher mass structures.

\subsection{Structures Defined by CO \jj21\ Emission in Other Galaxies}
 
\begin{figure*}
\center
\includegraphics[scale=0.3]{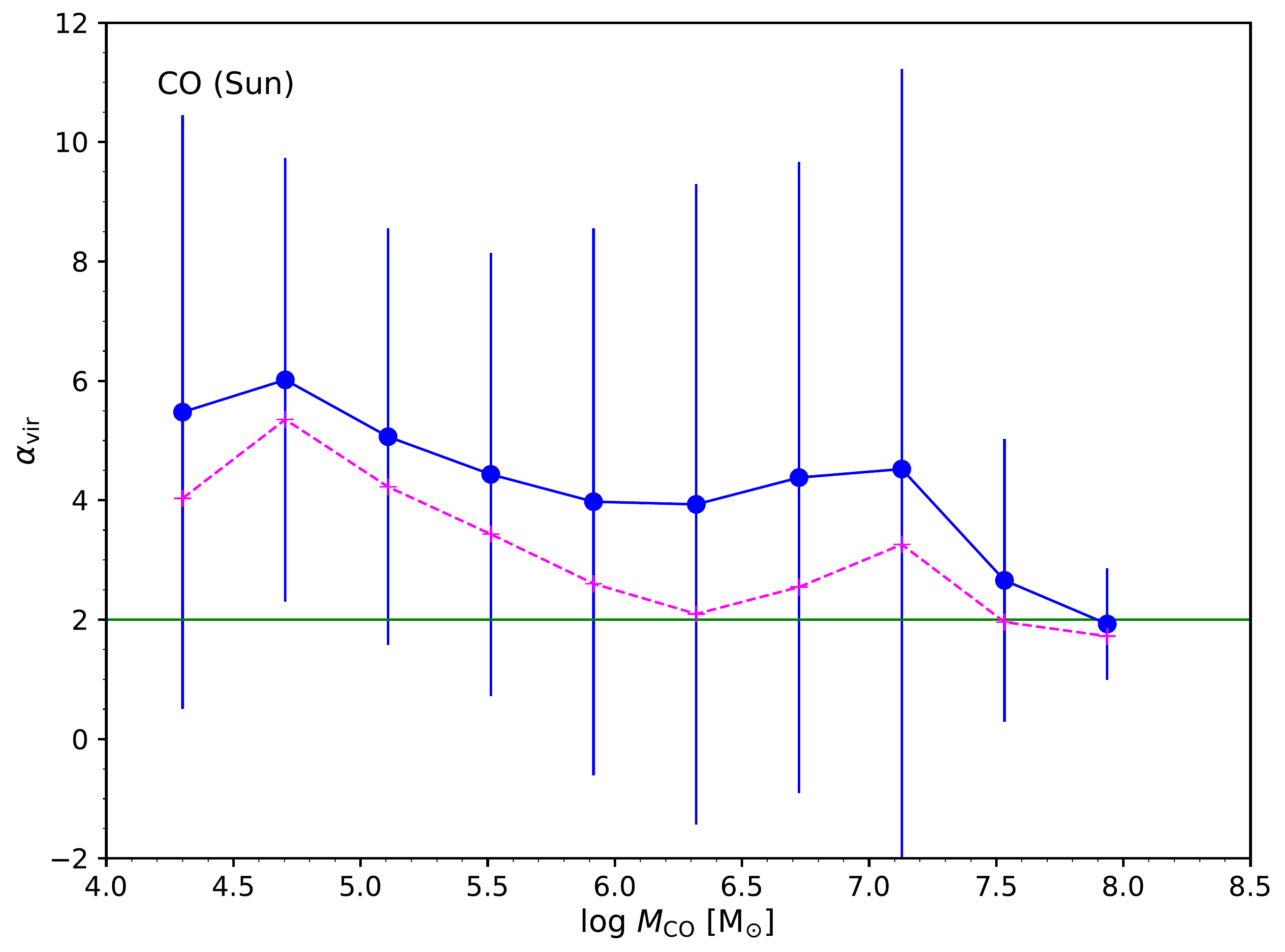}
\includegraphics[scale=0.3]{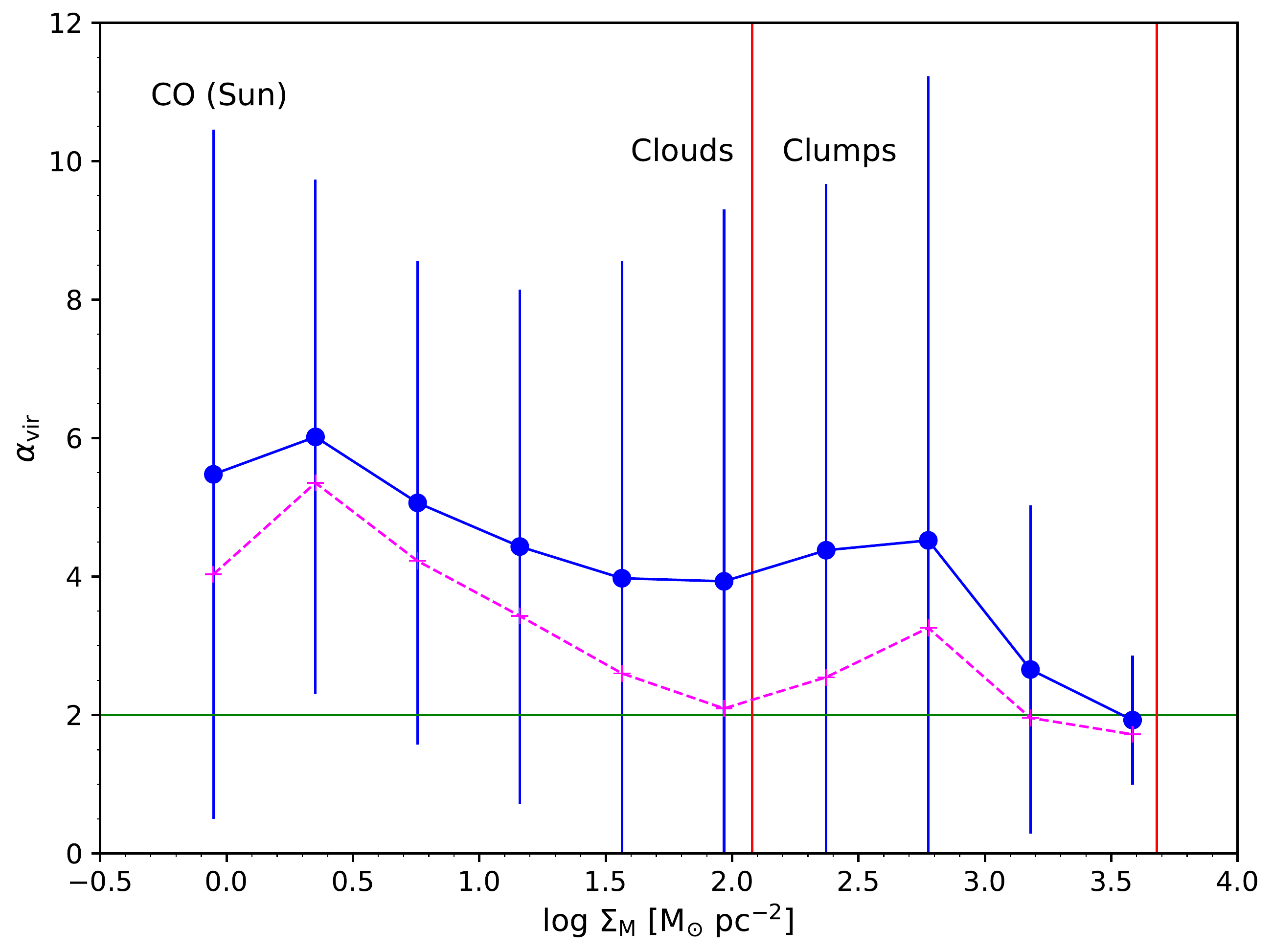}
\includegraphics[scale=0.3]{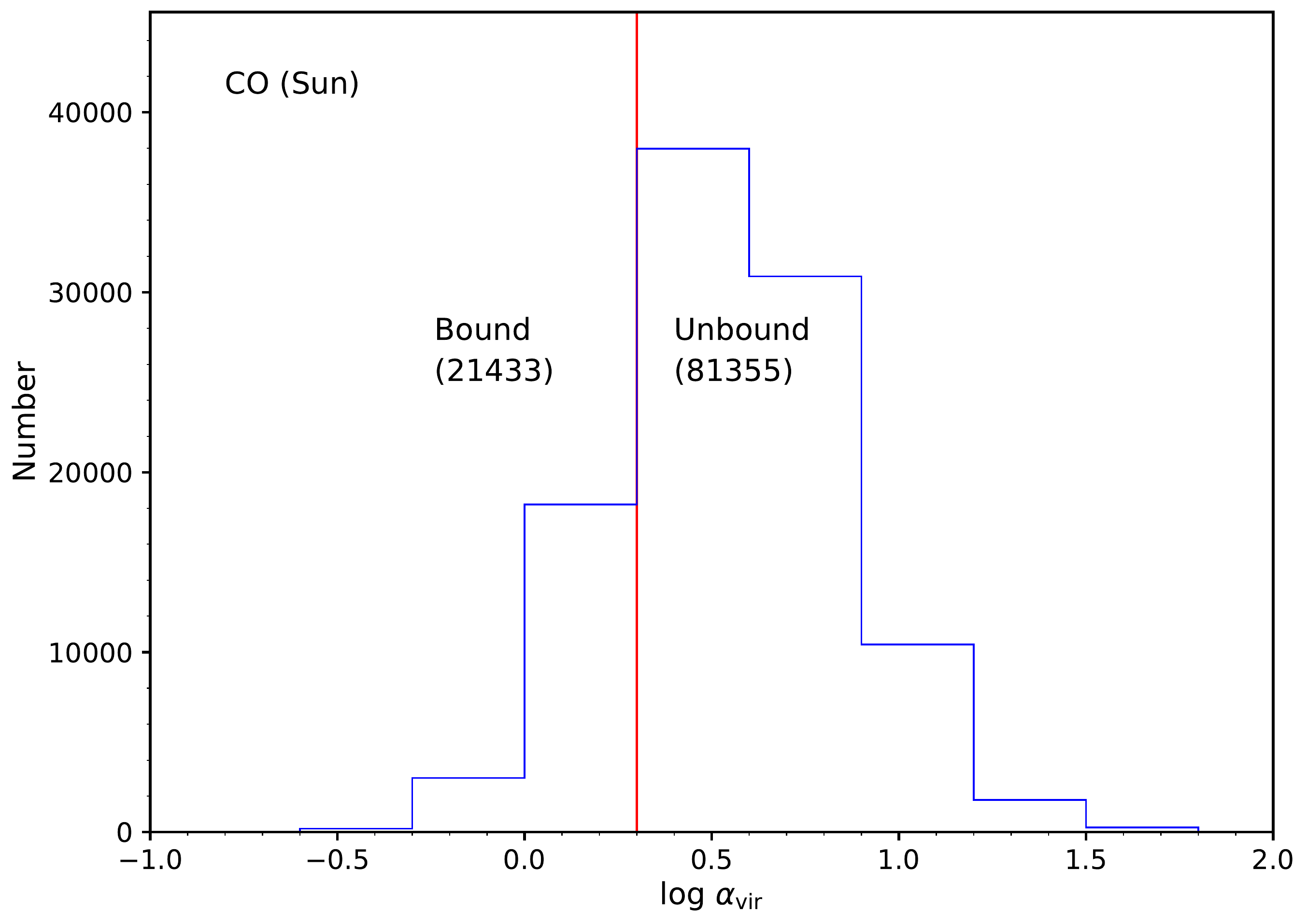}
\includegraphics[scale=0.3]{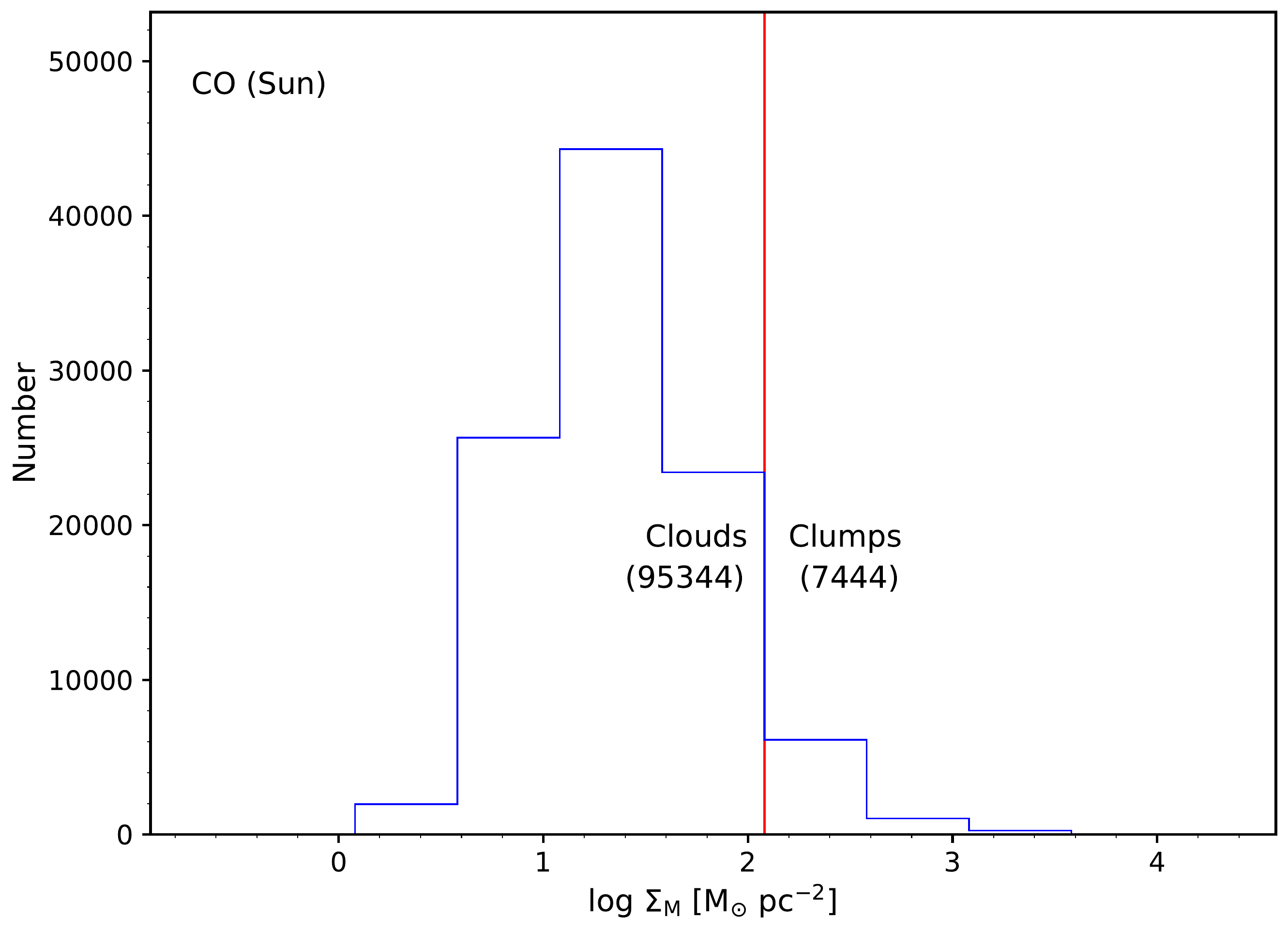}
\caption{
The mean and standard deviation  of the virial ratio is plotted in bins versus the logarithm of the clump mass, for 
The catalog of 
\citet{2020ApJ...901L...8S} is plotted.
(Upper Left)
The mean and standard deviation of the virial ratio is plotted versus the logarithm of the clump mass. Median values are plotted in a magenta dashed line.
(Upper Right) The same quantities are plotted versus the mass surface density.
The horizontal line at $\alphavir = 2$ demarcates nominally unbound clouds above the
line from nominally bound clouds below the line. Vertical lines 
indicate $\sigmam = 120$ \msunpc\ and $\sigmam = 1$ g \cmc.
(Lower left) The histogram of values of log \alphavir, with a vertical red line
at $\alphavir = 2.0$. 
The number of structures in each category are given in parentheses.
(Lower right) The histogram of values of log \sigmam, with a vertical red line at
$\sigmam = 120$ \msunpc.}
\label{sun20}
\end{figure*}

We were able to use all the catalog entries \added{with 150 pc resolution}
from table B1 of
\citet{2020ApJ...901L...8S}, 
save one entry with a negative value for \alphavir.
\citet{2020ApJ...901L...8S} themselves
found an area-weighted median $\alphavir = 3.5$ with a 1-$\sigma$\ spread
of 0.6 dex, based on over \eten5\ pixels of size 150 pc in 66 galaxies.
The mass-weighted median was smaller at 2.7. 
All the caveats about neglect of tidal
effects, etc. apply quite strongly to studies with 150 pc resolution.
Independent analysis of the data in their table B1 confirms the
median value and finds a mean value of $\mean{\alphavir} = 4.7 \pm 4.0$.
The averages of \alphavir\ in Figure \ref{sun20}
all lie above $\alphavir = 2$, but they approach that value for the most
massive data points. A similar pattern applies to the plot of \alphavir\
versus \sigmam. 
While many points have very high \alphavir, the fraction with
$\alphavir < 2$ is \replaced{$\fn = 0.19$}{$\fn = 0.21$}
and the fraction of the total mass in those structures is 
\replaced{$\fmass = 0.32$}{$\fmass = 0.35$},
reflecting the usual trend for lower $\alphavir$ in more massive clouds.
The median and mean surface densities
are \replaced{24}{22} and 50 \msunpc, but with a maximum at $\sigmam = 9.7\ee3$ \msunpc.
The last value is clearly more characteristic of dense clumps, not clouds. 
All points above
log $\mco = 6.43$ \msun\  correspond to $\sigmam > 120$ \msunpc\
or $\av > 8$ mag.
Vertical lines on Figure \ref{sun20} indicate surface
densities of 120 \msunpc\ and 1 g \cmc. The second of these corresponds
to the criterion for massive star formation 
\citep{2008Natur.451.1082K}. 
The fact
that \alphavir\ drops to values near two as the surface density approaches
this value is interesting. To have such high surface density averaged over
a region of 150 pc is quite remarkable. If \sigmam\ is restricted to be less than 120 \msunpc,
the fraction of the total mass with $\alphavir < 2$ becomes 0.15.

\subsection{Structures Defined by \coo\ \jj10\ Emission}

To use  data from all three surveys and to compare methods for
structure identification, we have reanalyzed the three data sets, using
the CO clouds defined by 
\citet{2017ApJ...834...57M} to constrain the \coo\ voxels included.
The detailed method is explained in the Appendix.
However, we begin with  the original catalog, but with updated
abundance assumptions. These results will allow a comparison between different
cloud identification methods.

\subsubsection{GRS: Original Sample}

\begin{figure*}
\center
\includegraphics[scale=0.3]{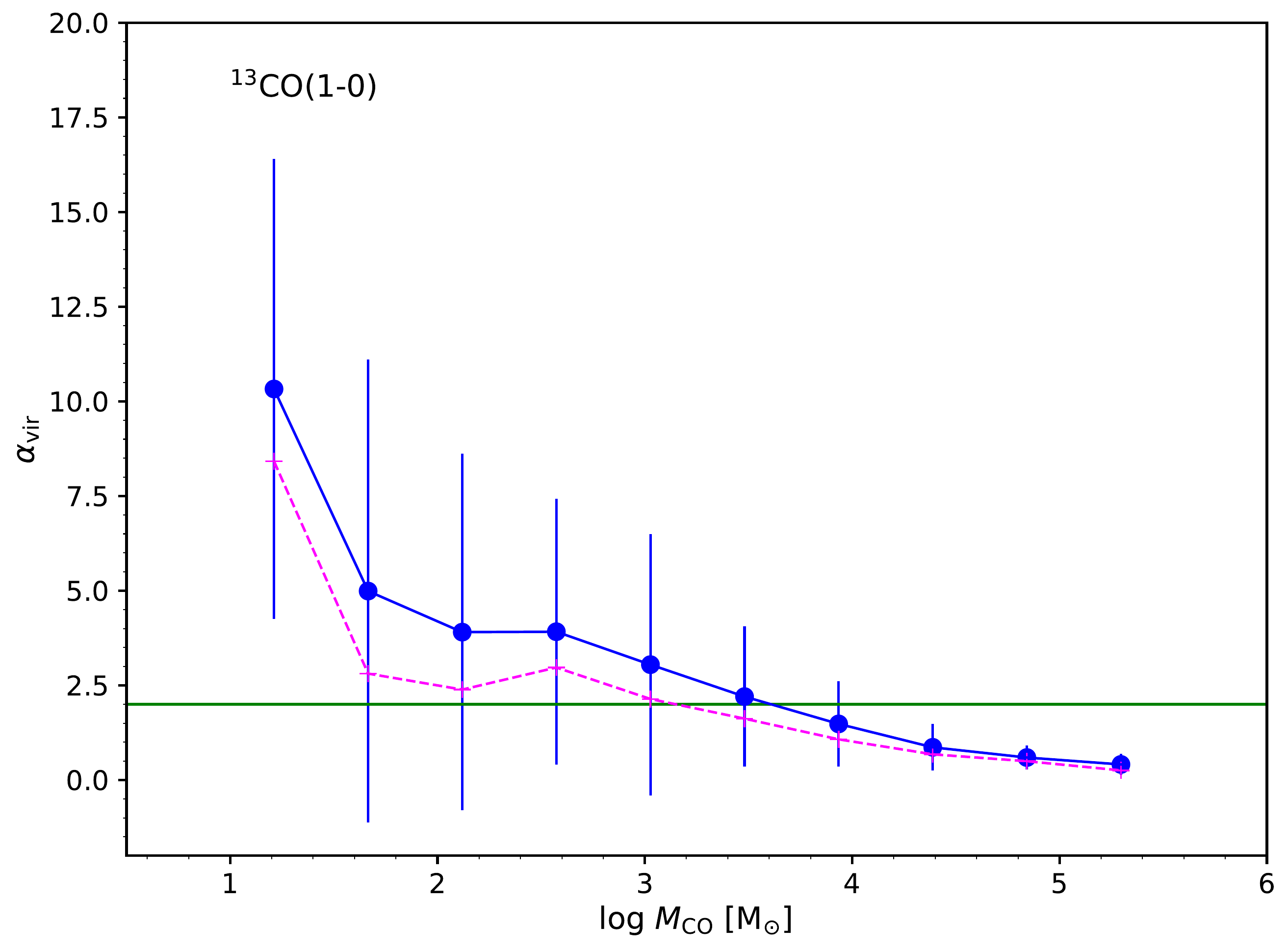}
\includegraphics[scale=0.3]{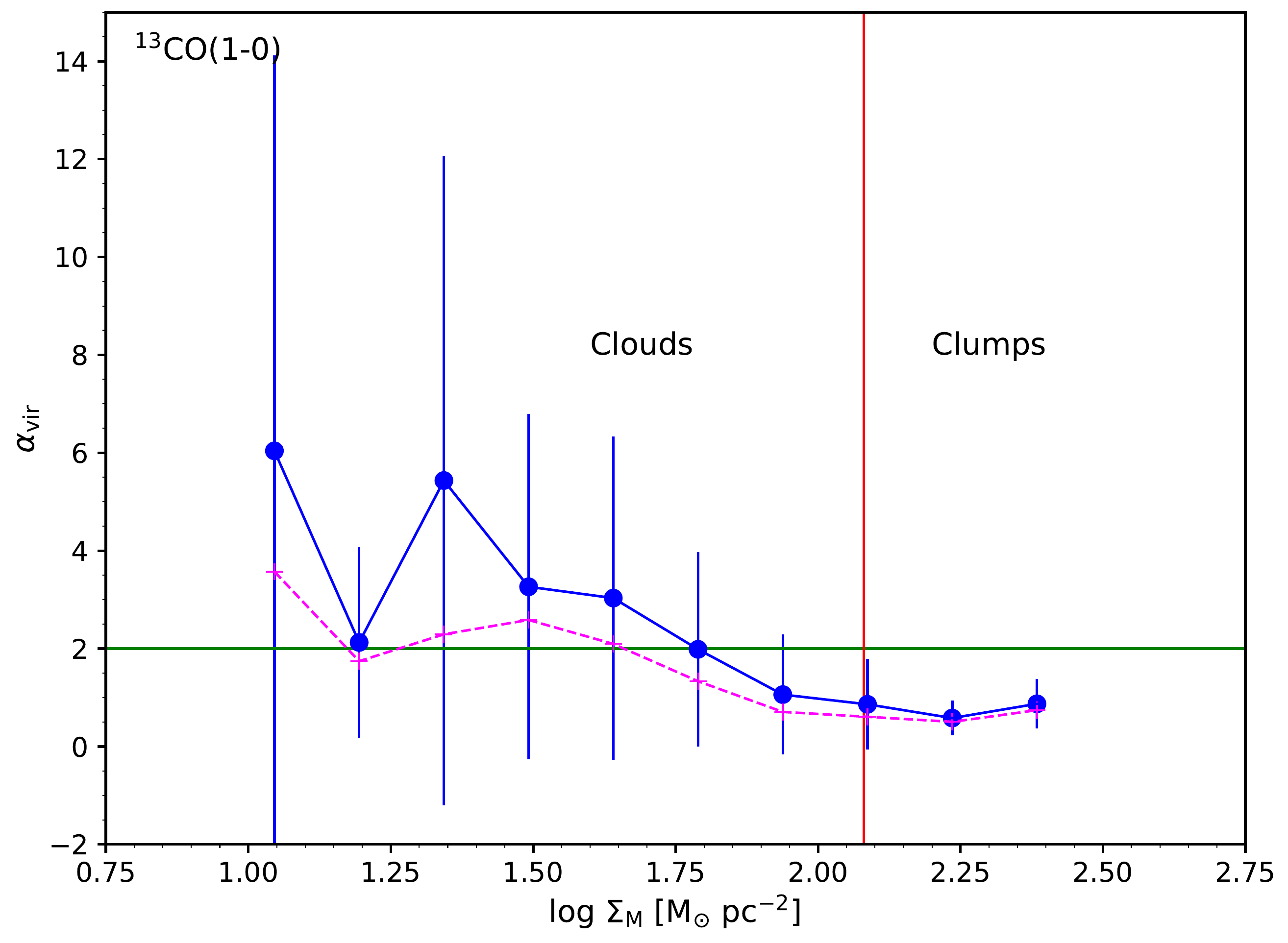}
\includegraphics[scale=0.3]{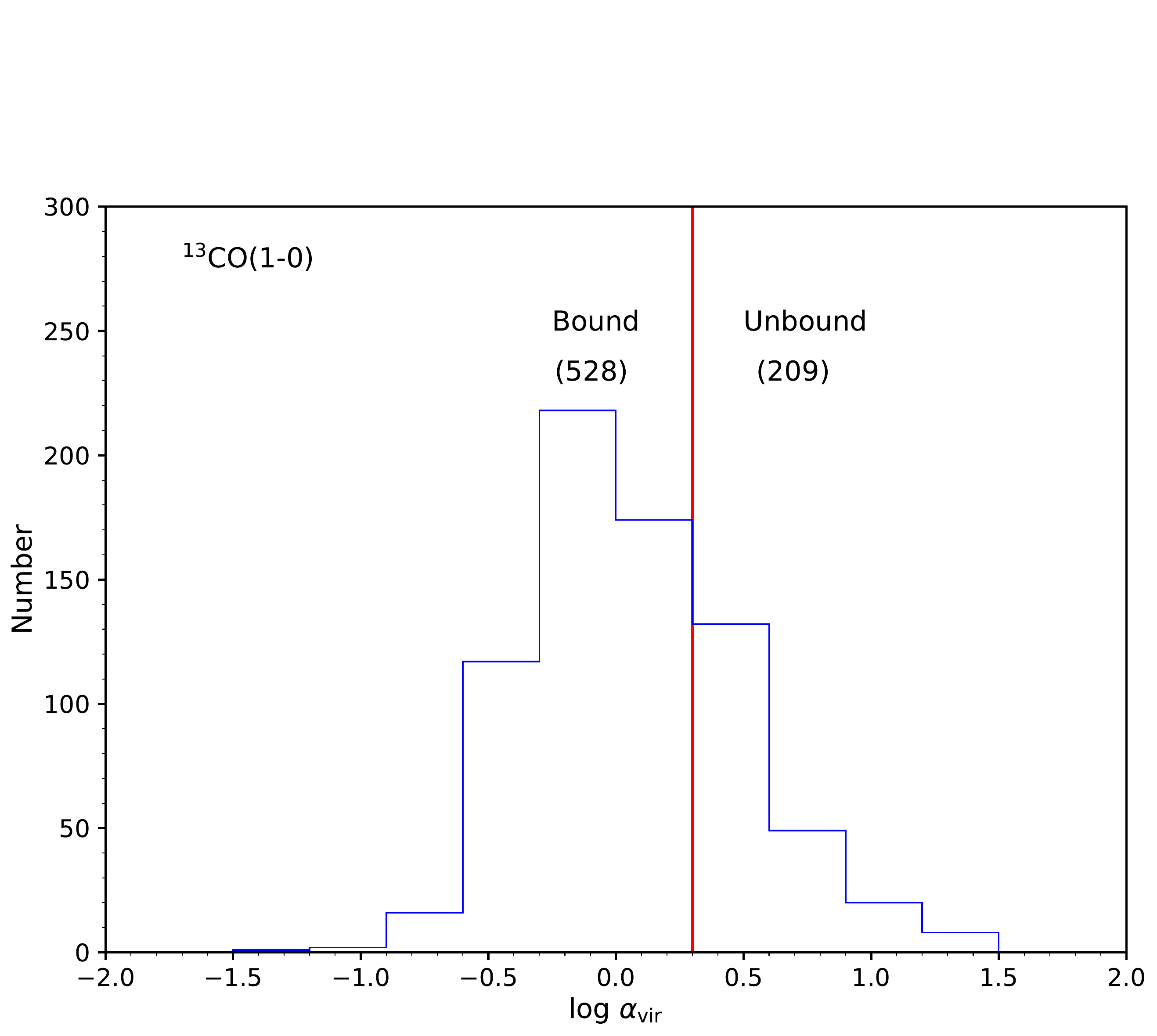}
\includegraphics[scale=0.3]{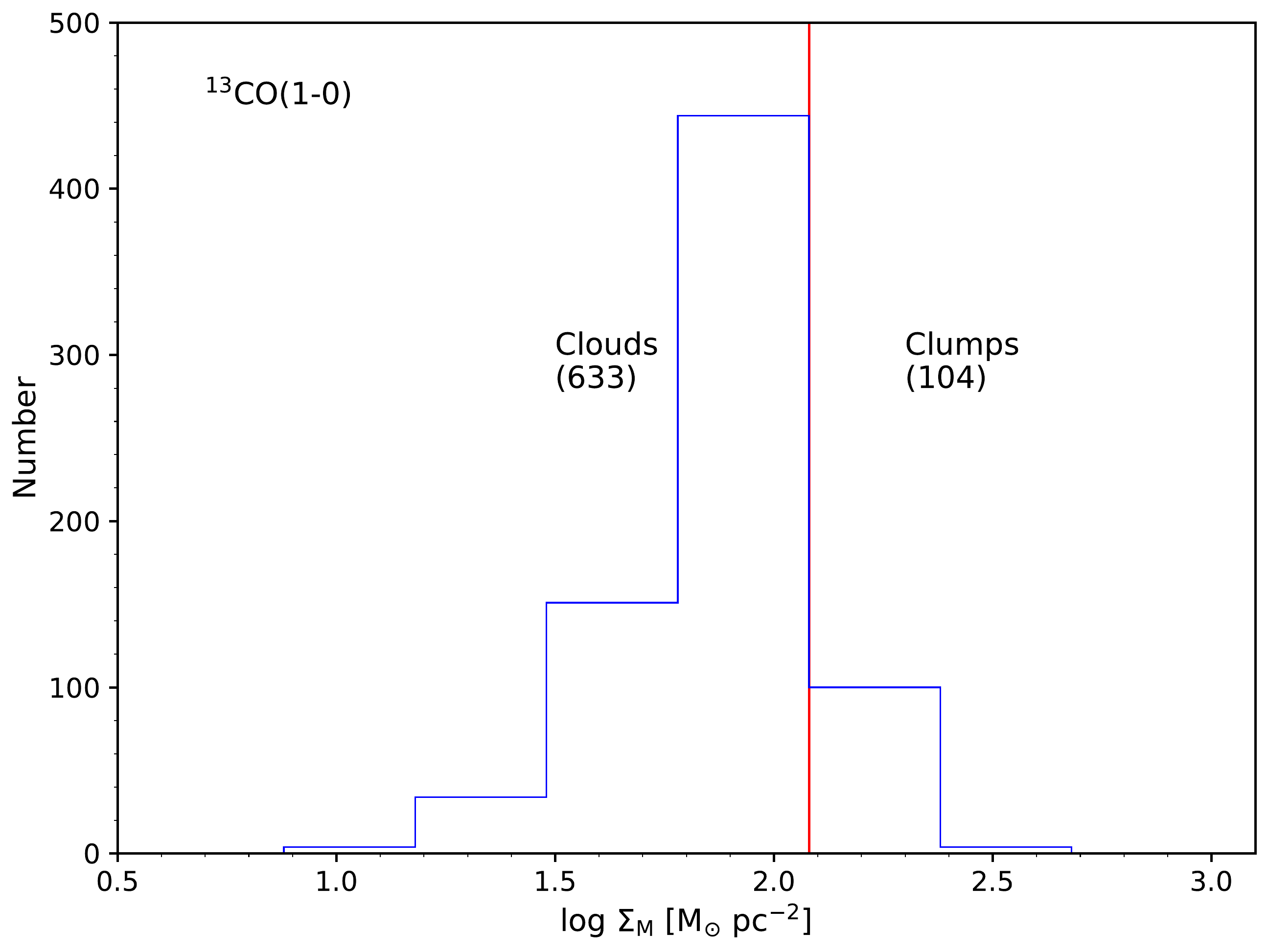}
\caption{
(Upper Left)
The mean and standard deviation of the virial ratio is plotted versus the logarithm of the clump mass, for the catalog of 
\citet{2010ApJ...723..492R}.
after selecting only those catalog entries with $\sigmam > 10$ and
$\mco > 1$ \msun, and $\alphavir < 20$.
Median values are plotted in a magenta dashed line.
(Upper Right) The same quantities are plotted versus the mass surface density.
The horizontal line at $\alphavir = 2$ demarcates nominally unbound clouds above the
line from nominally bound clouds below the line. The vertical line in the right panel
indicates $\sigmam = 120$ \msunpc.
(Lower left) The histogram of values of log \alphavir, with a vertical red line
at $\alphavir = 2.0$. 
The number of structures in each category are given in parentheses.
(Lower right) The histogram of values of log \sigmam, with a vertical red line at
$\sigmam = 120$ \msunpc.
}
\label{rd1}
\end{figure*}

The tabulated masses and surface densities for the \coo\ GRS data from
\citet{2010ApJ...723..492R}
originally used values of \hh/CO of 12500 and $\muhh = 2.37$.
As discussed in \S \ref{intro}, we instead use the current best values for
\hh/CO and \muhh, resulting in masses and surface densities lower by a factor
of 0.567 and virial parameters higher by a factor of 1.76. For a first analysis,
(entry 8 in Table \ref{tabstats}), we retain the assumption from 
\citet{2010ApJ...723..492R}
of a constant CO/\coo\ of 45.
The usual quantities 
are plotted in Figure \ref{rd1} and the results are in Table \ref{tabstats}.
The majority of the sample would be classified as clouds, with $\sigmam < 120$
\msunpc, but about 1/6 are denser.
The fraction of mass in bound structures, $\fmass = 0.95$, but the
fraction for clouds is much less, $\fcmass = 0.55$.

\subsubsection{GRS: New Analysis}

From now on, we use the CO-defined structures and the \coo-derived properties,
with the abundance assumptions
described at the end of \S \ref{intro}, including the variation of
\isorat\ with \rgal.
We begin with a re-analysis of the GRS survey in \coo\
\citep{2010ApJ...723..492R}.
The structures are defined in the catalog of
\citet{2017ApJ...834...57M},
but only for a sub-sample.
The sub-sample of CO clouds is somewhat biased towards
higher \sigmam\ and lower \alphavir\ than the full catalog:
$\mean{\sigmam} = 95 \pm 64$ \msunpc, nearly twice that for the full catalog;
$\mean{\alphavir} = 5.62 \pm 5.28$, 0.81 times that for the full catalog.

\begin{figure*}
\center
\includegraphics[scale=0.3]{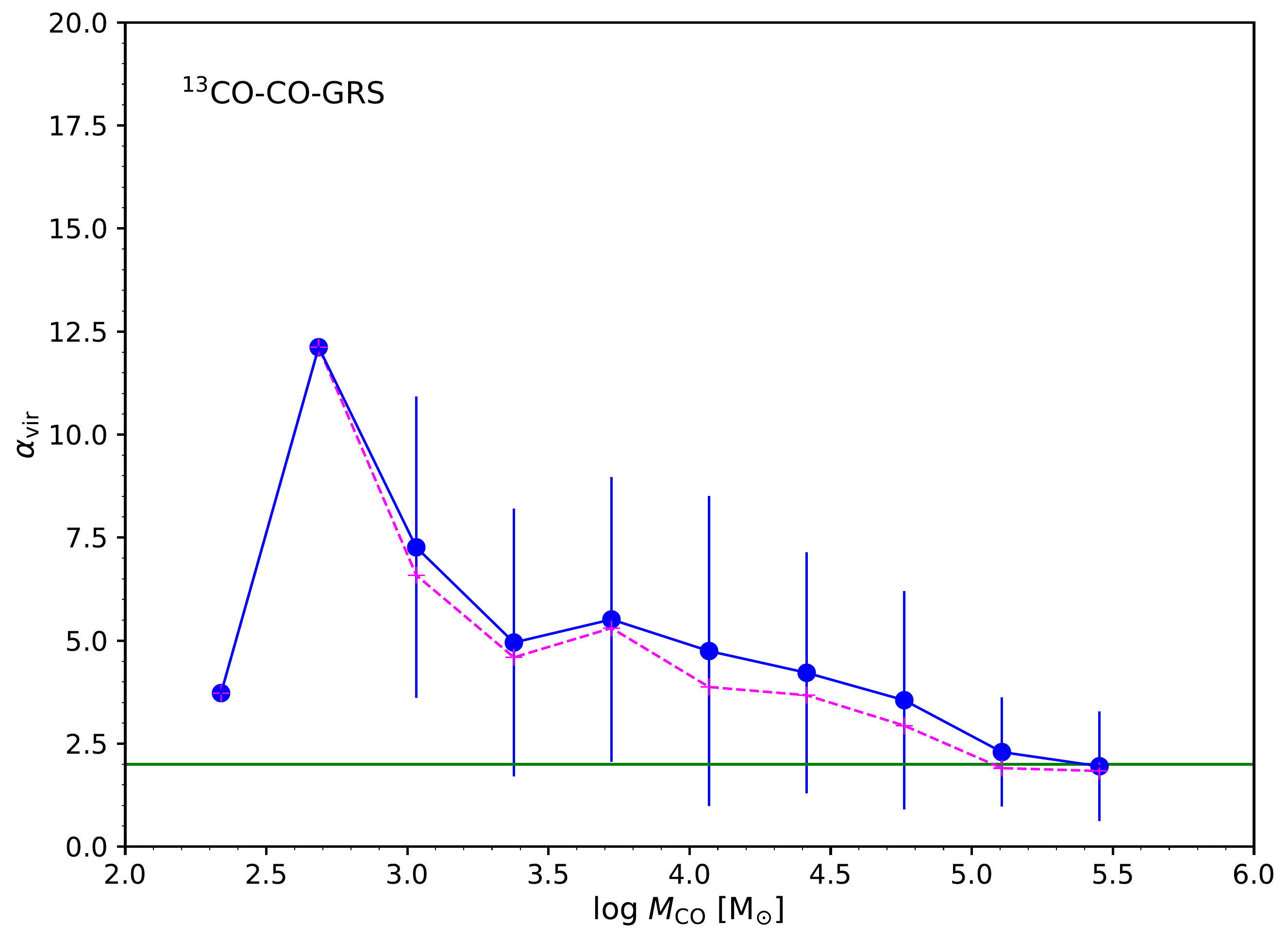}
\includegraphics[scale=0.3]{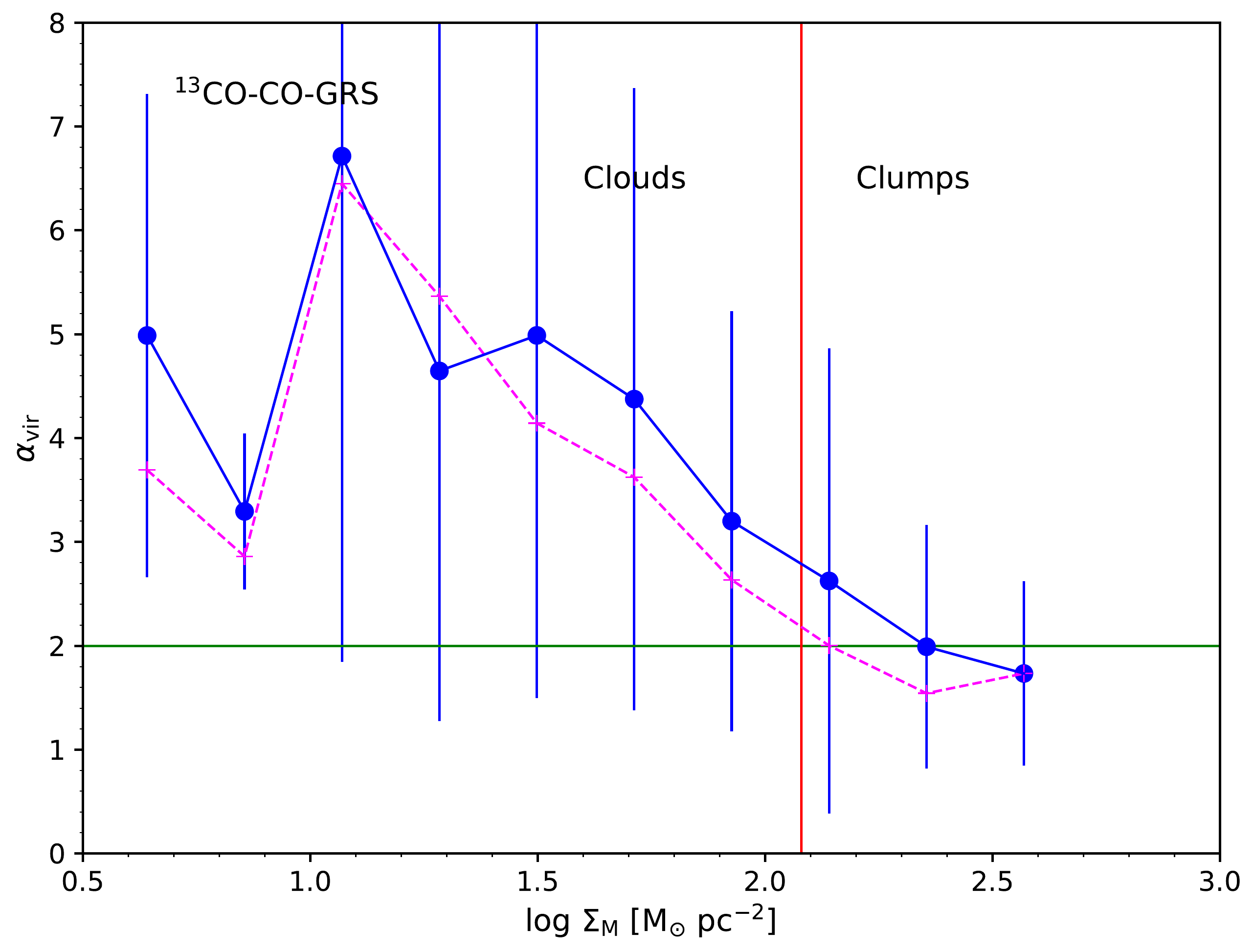}
\includegraphics[scale=0.3]{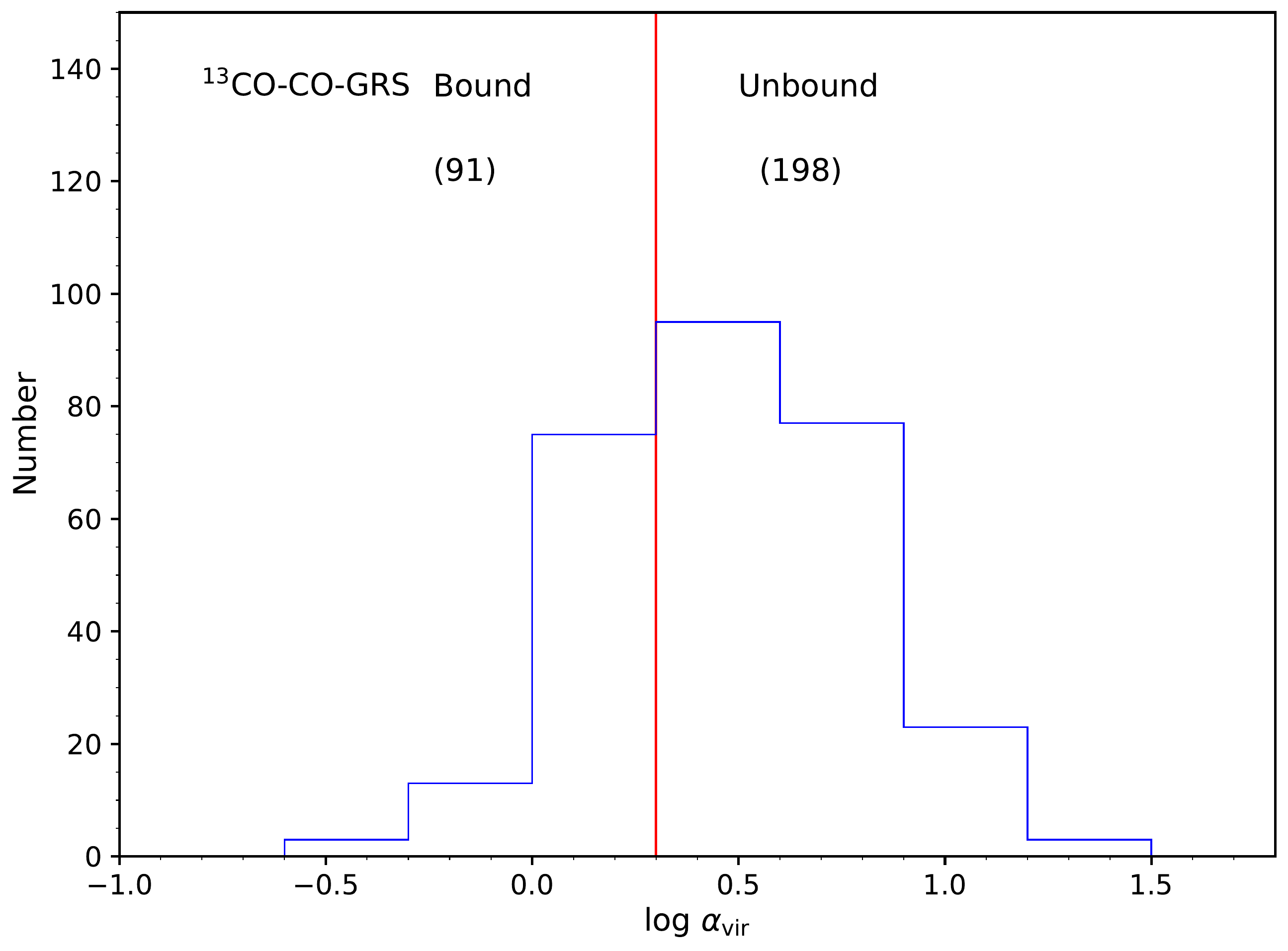}
\includegraphics[scale=0.3]{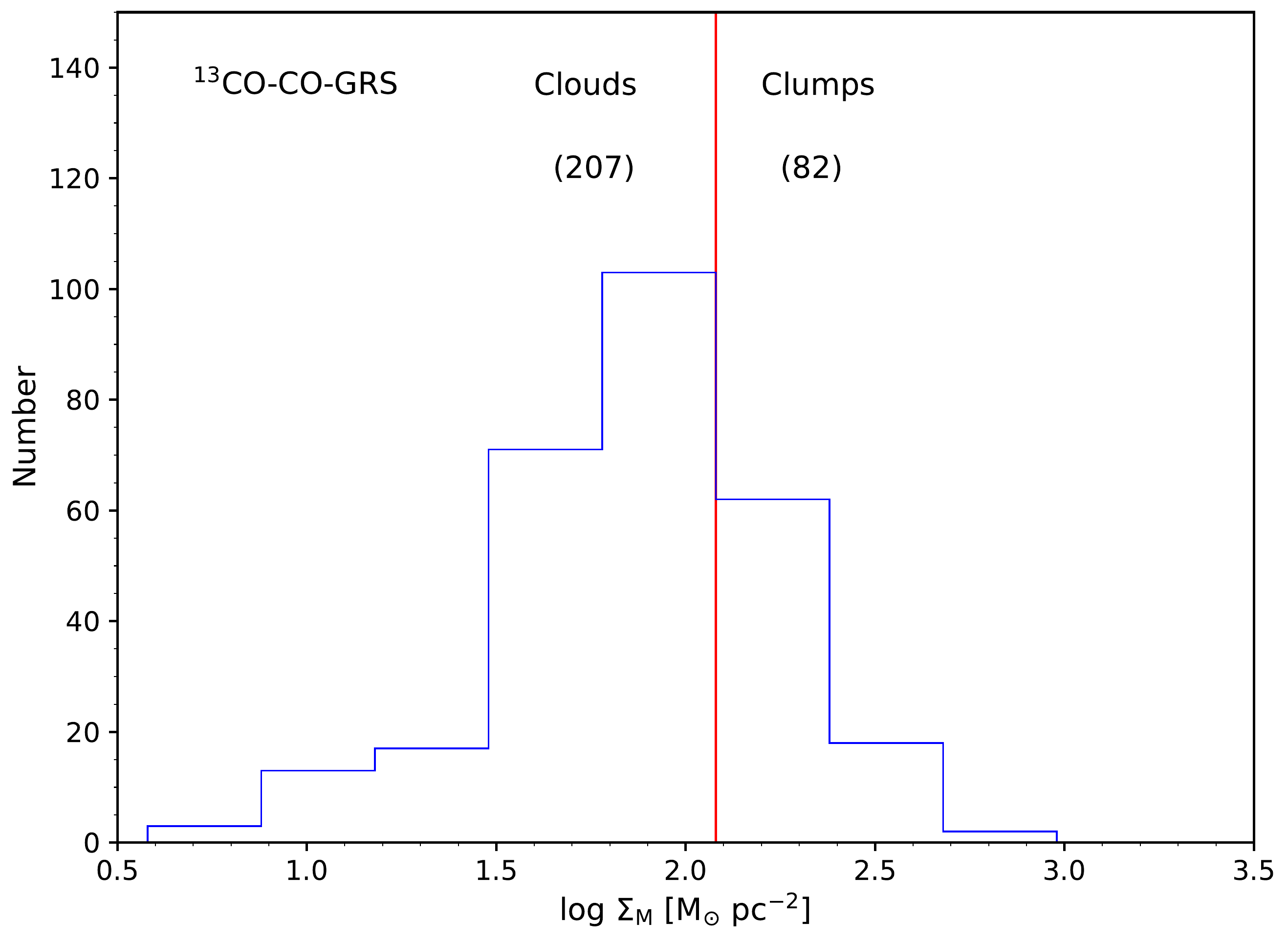}
\caption{
(Upper Left) The mean and standard deviation of the virial ratio is plotted versus the logarithm of the clump mass, for the catalog produced for the GRS in combination with CO.
Median values are plotted in a magenta dashed line.
(Upper Right) The same quantities are plotted versus the mass surface density.
The horizontal line at $\alphavir = 2$ demarcates nominally unbound clouds above the line from nominally bound clouds below the line. 
The vertical line in the right panel indicates $\sigmam = 120$ \msunpc.
The mass and surface density are the total (gas+dust)  determined from 
submm continuum emission.
(Lower left) The histogram of values of log \alphavir, with a vertical red line
at $\alphavir = 2.0$. 
The number of structures in each category are given in parentheses.
(Lower right) The histogram of values of log \sigmam, with a vertical red line at
$\sigmam = 120$ \msunpc.
}
\label{heyer20grs}
\end{figure*}

The analysis of the GRS data with the method described in the
Appendix is listed as entry 9 of Table \ref{tabstats} and the usual plots are in
Figure \ref{heyer20grs}. The main difference from entry 8 is that the clouds
were identified in the CO analysis and the \coo\ data were taken only within
the boundaries of those clouds. The change in 
cloud definition makes the mean \alphavir\ larger and \fmass\ smaller by
substantial amounts ($\fmass = 0.46$).

The fraction of the mass of the CO-defined structure contained in the
\coo\ emission (\mrat) is a valuable guide to the fraction of
 possibly bound structures within unbound CO clouds.
For the GRS sample, $\mean{\mrat} = 0.22 \pm 0.08$ with a median
$\mrat = 0.21$.

\subsubsection{EXFC55-100}

\begin{figure*}
\center
\includegraphics[scale=0.3]{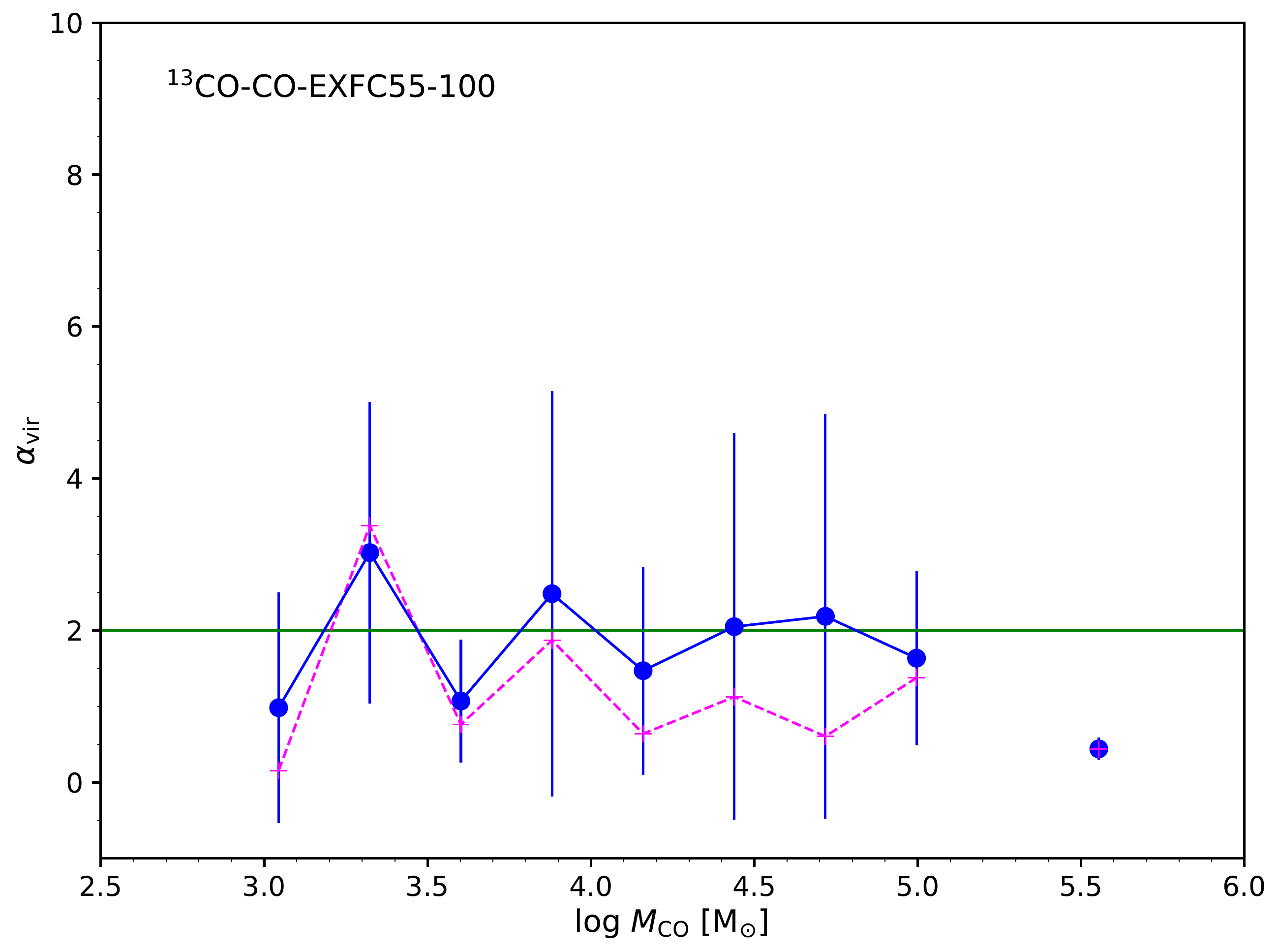}
\includegraphics[scale=0.3]{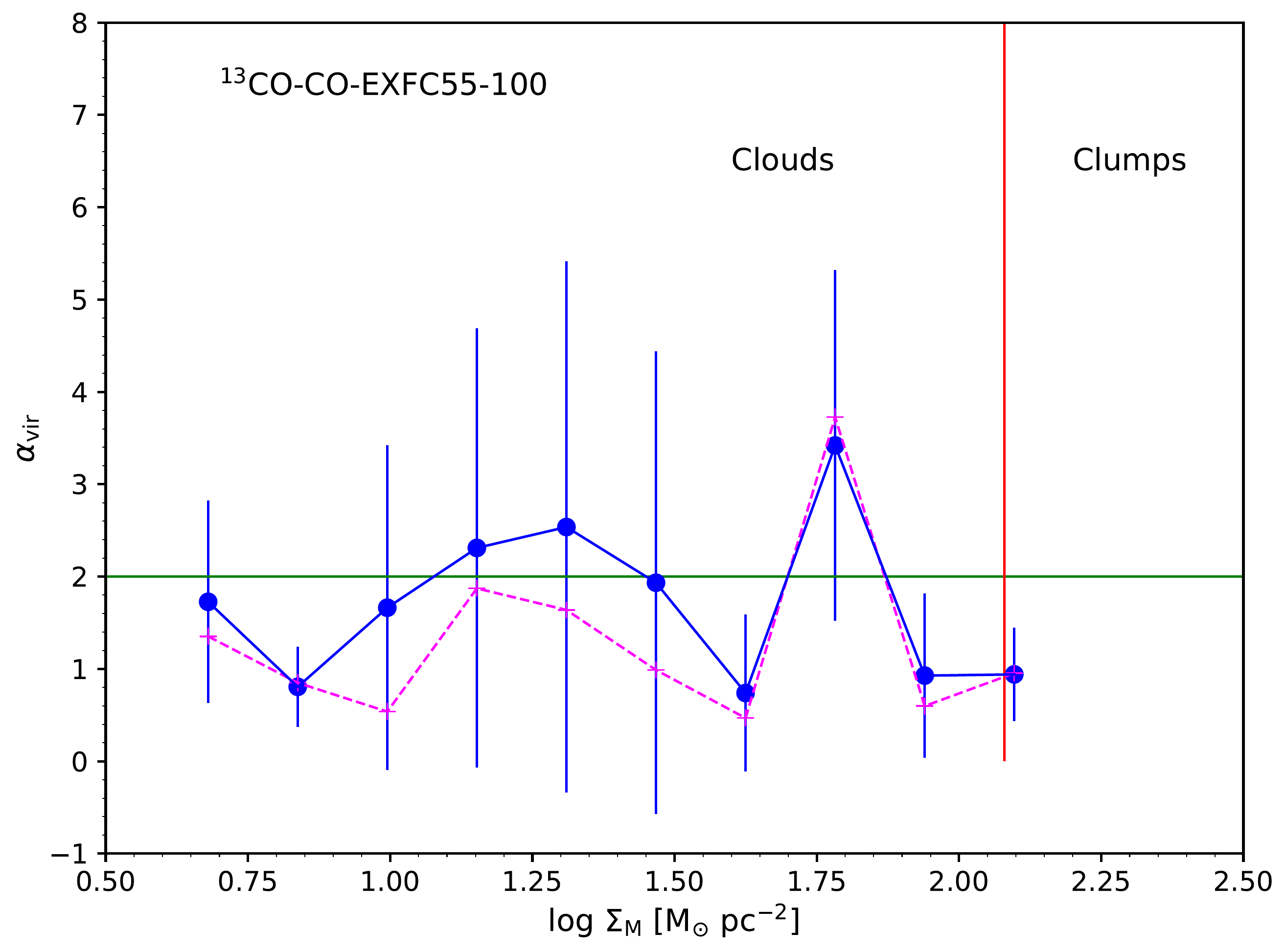}
\includegraphics[scale=0.3]{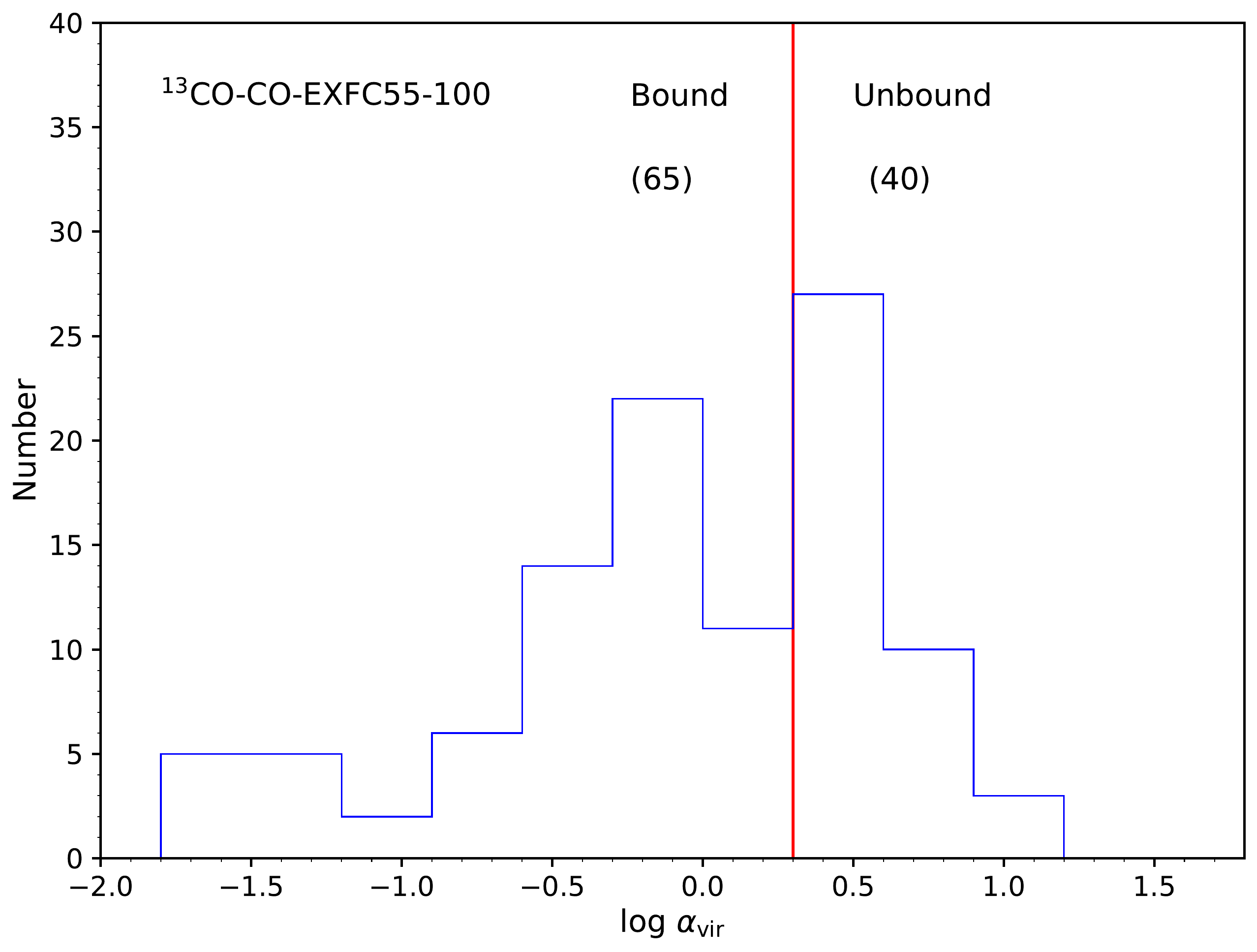}
\includegraphics[scale=0.3]{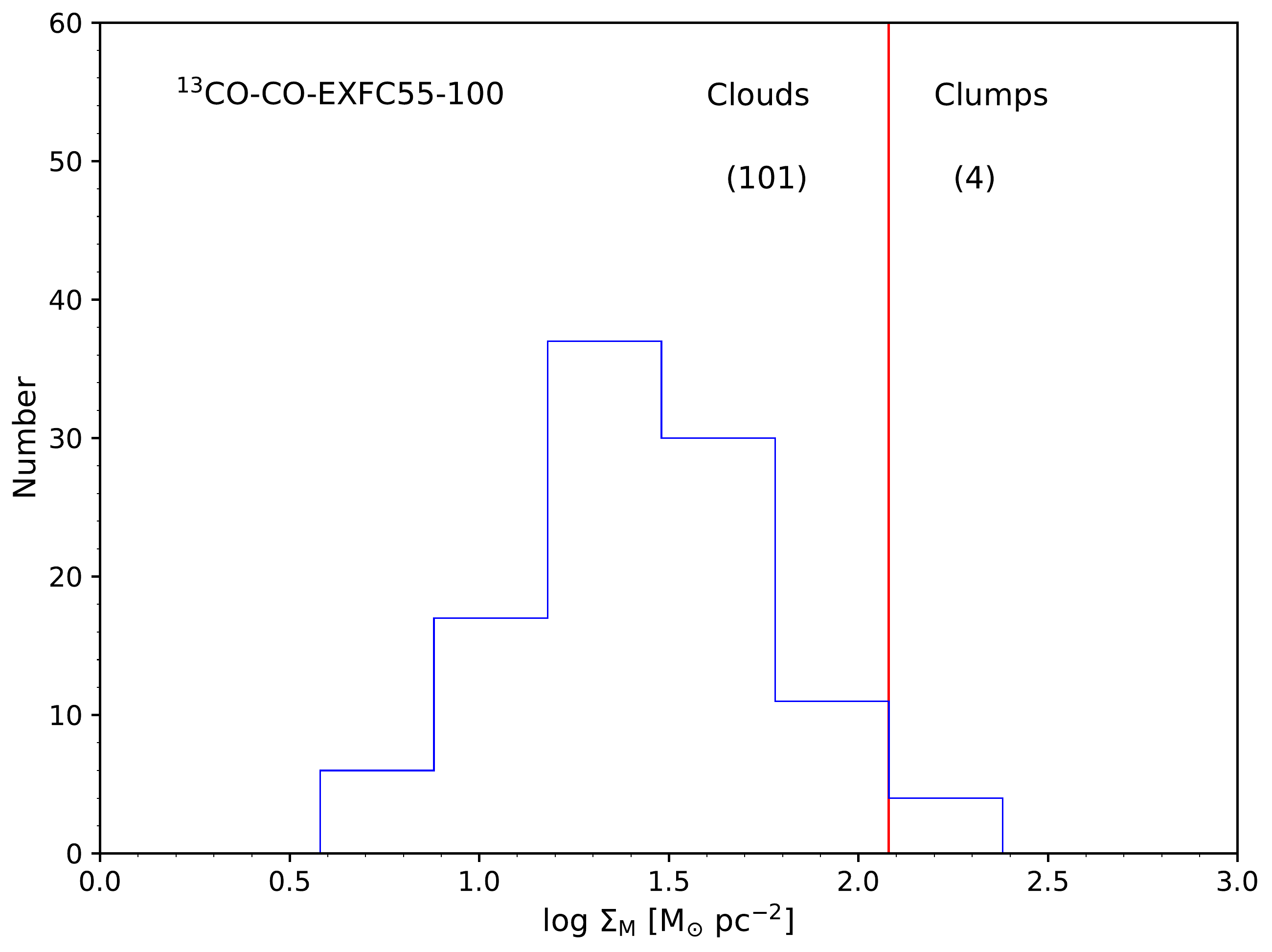}
\caption{
(Upper Left) The mean and standard deviation of the virial ratio is plotted versus the logarithm of the clump mass, for the catalog produced for the EXFC55-100 sample in combination with CO and using new abundances.
Median values are plotted in a magenta dashed line.
(Upper Right) The same quantities are plotted versus the mass surface density.
The horizontal line at $\alphavir = 2$ demarcates nominally unbound clouds above the line from nominally bound clouds below the line. 
The vertical line in the right panel indicates $\sigmam = 120$ \msunpc.
The mass and surface density are the total (gas+dust)  determined from 
submm continuum emission.
(Lower left) The histogram of values of log \alphavir, with a vertical red line
at $\alphavir = 2.0$. 
The number of structures in each category are given in parentheses.
(Lower right) The histogram of values of log \sigmam, with a vertical red line at
$\sigmam = 120$ \msunpc.
}
\label{heyer20local}
\end{figure*}

For this sample, most  structures are identified as clouds, but
they are mostly bound, despite having low \sigmam. The usual plots are in
Figure \ref{heyer20local} and the statistics are given as entry 10 in Table \ref{tabstats}.
The fraction of the CO mass recovered by the \coo\ emission, $\mean{\mrat} = 0.79 \pm 0.80$ with a median
$\mrat = 0.61$. These values are much higher than those in the GRS.

\subsubsection{EXFC135-195}

\begin{figure*}
\center
\includegraphics[scale=0.3]{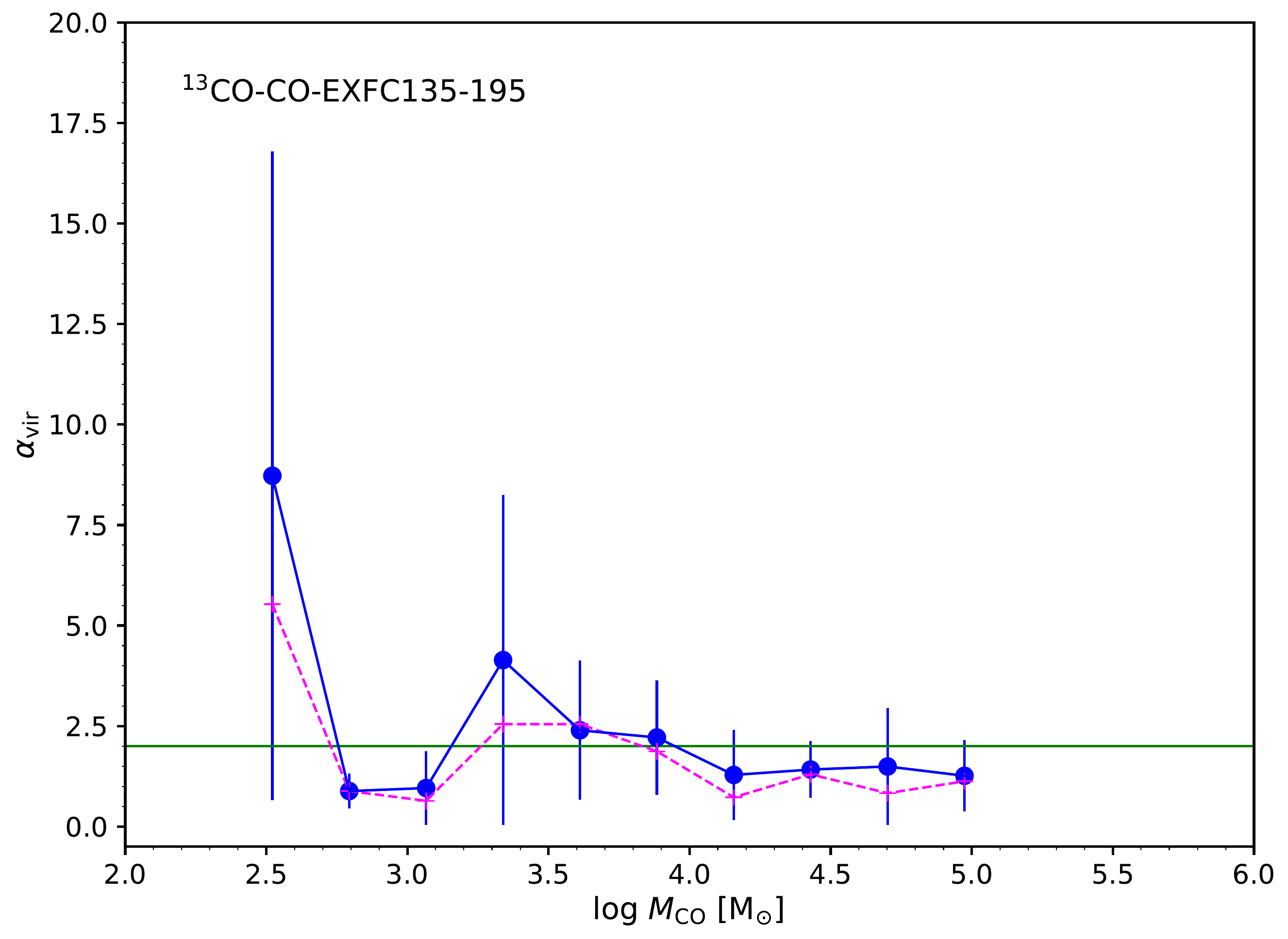}
\includegraphics[scale=0.3]{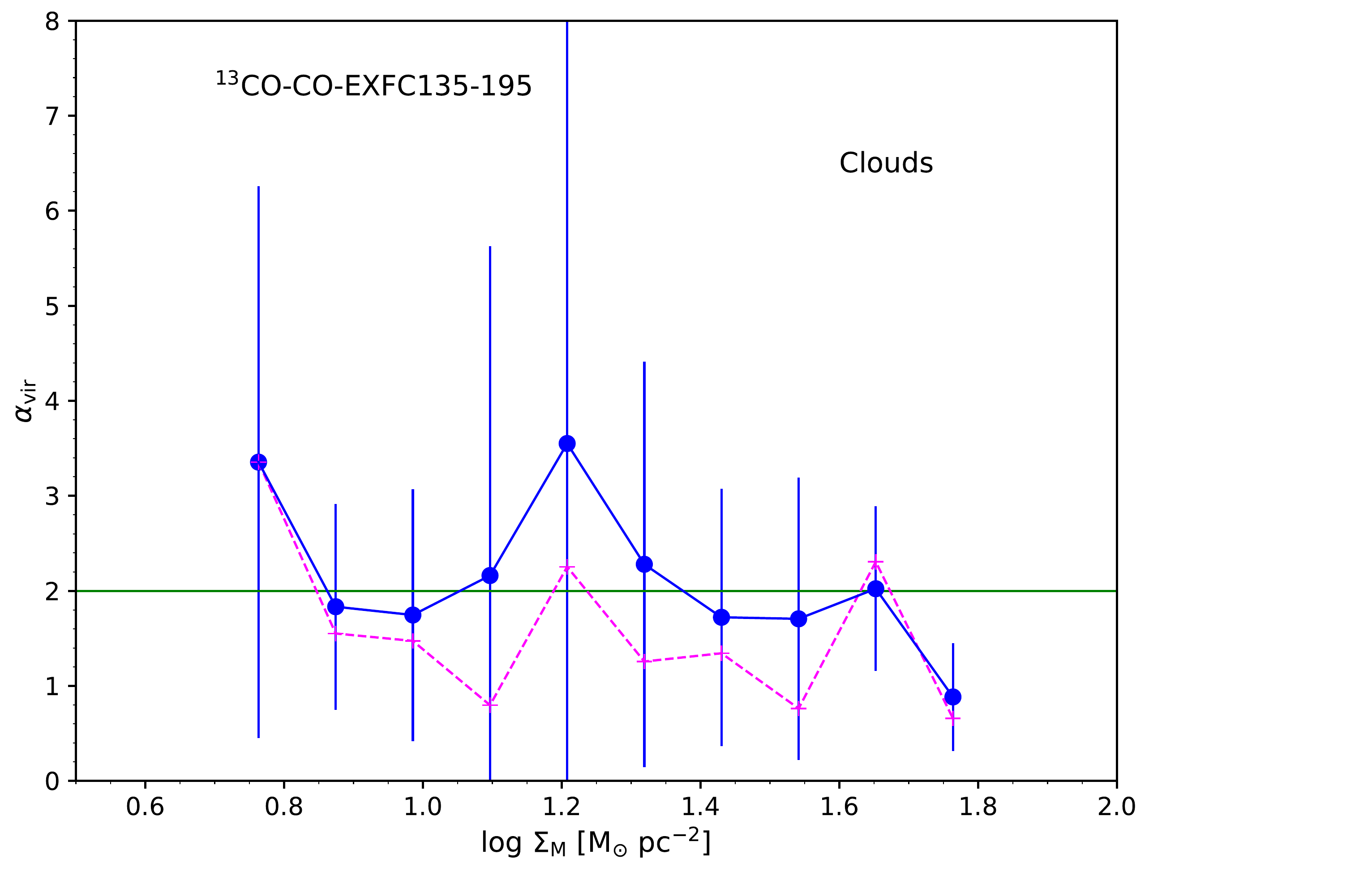}
\includegraphics[scale=0.3]{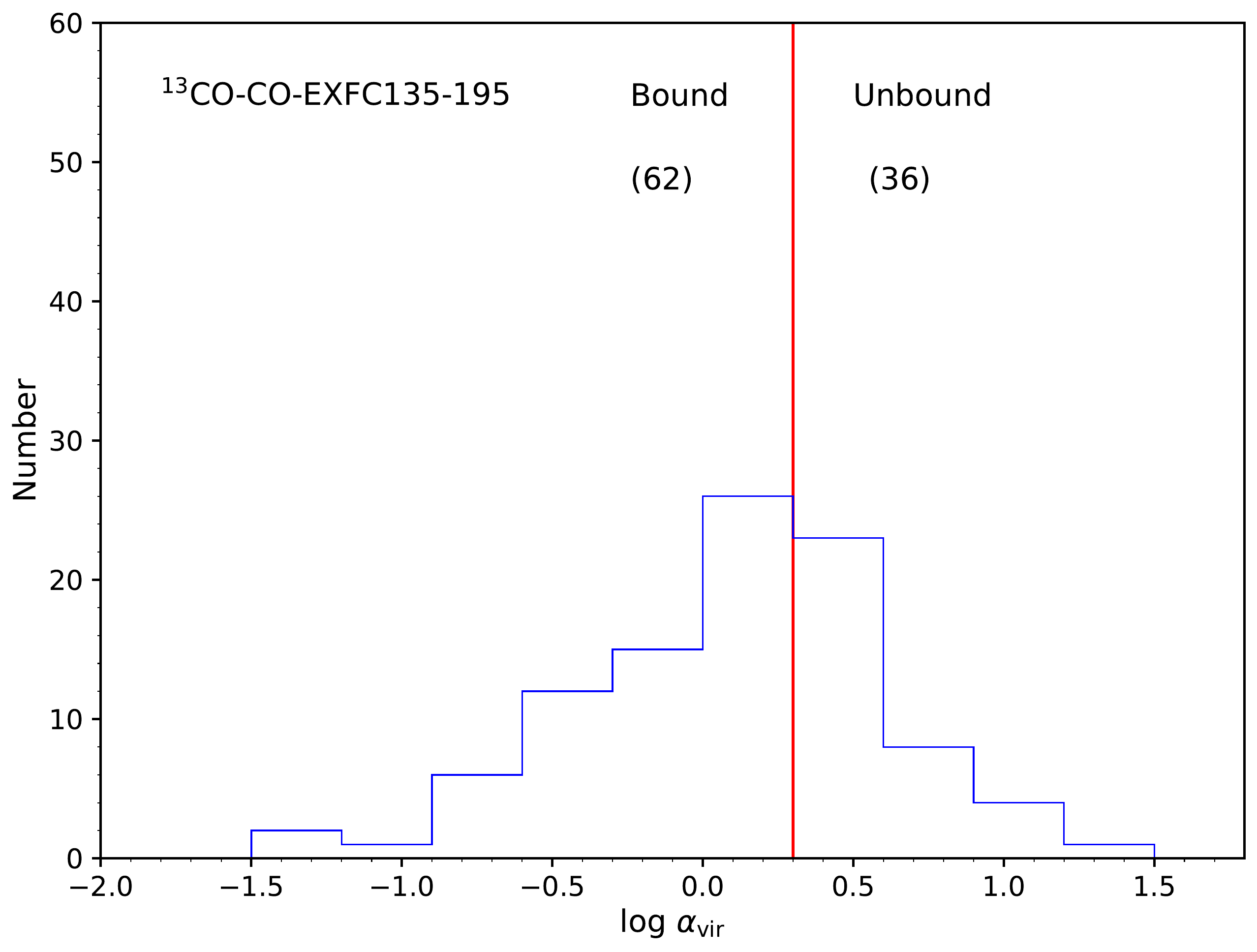}
\includegraphics[scale=0.3]{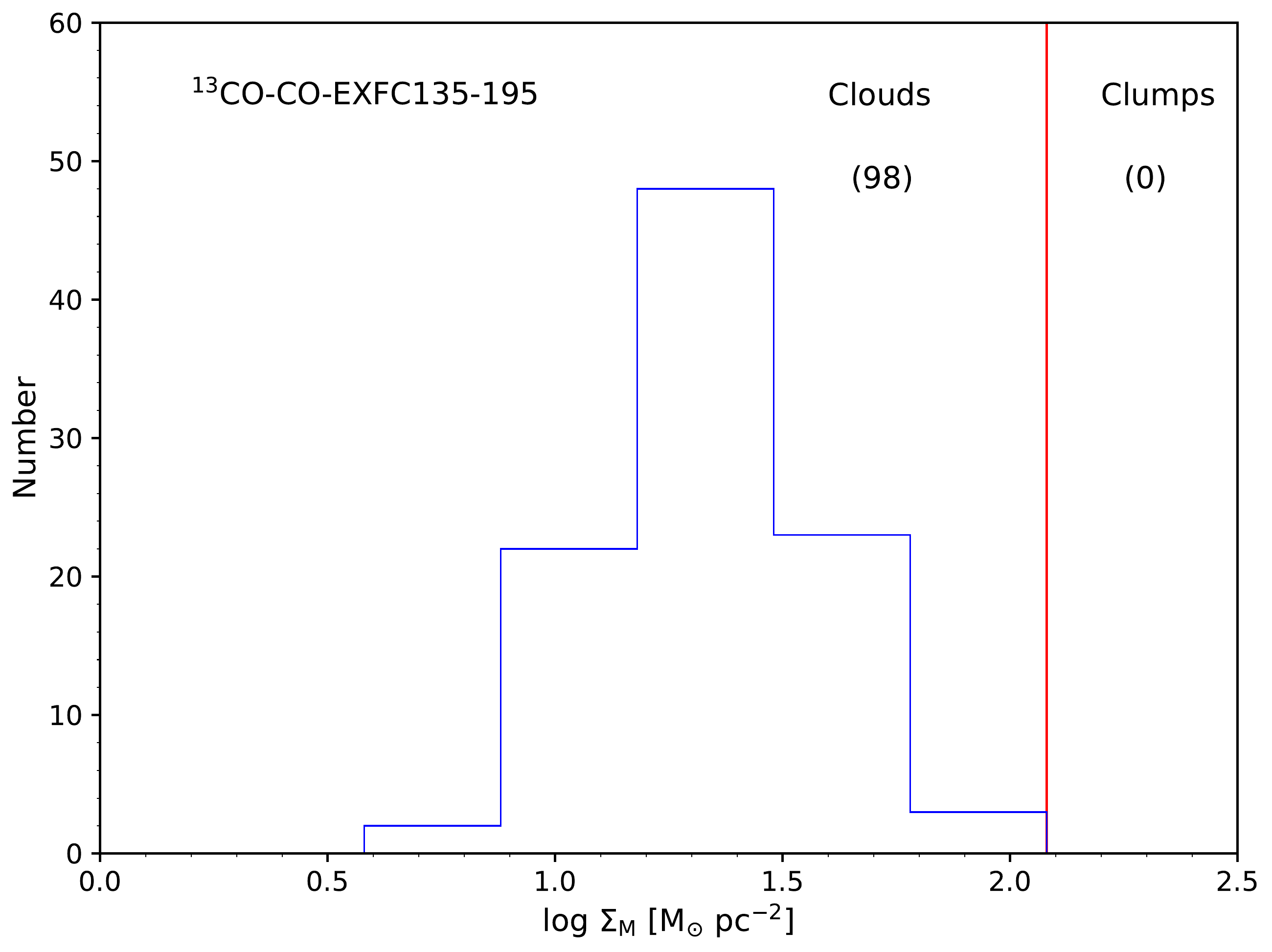}
\caption{
(Upper Left) The mean and standard deviation of the virial ratio is plotted versus the logarithm of the clump mass, for the catalog produced for the EXFC135-195 sample in combination with CO and using new abundances.
Median values are plotted in a magenta dashed line.
(Upper Right) The same quantities are plotted versus the mass surface density.
The horizontal line at $\alphavir = 2$ demarcates nominally unbound clouds above the line from nominally bound clouds below the line. 
The vertical line in the right panel indicates $\sigmam = 120$ \msunpc.
The mass and surface density are the total (gas+dust)  determined from 
submm continuum emission.
(Lower left) The histogram of values of log \alphavir, with a vertical red line
at $\alphavir = 2.0$. 
The number of structures in each category are given in parentheses.
(Lower right) The histogram of values of log \sigmam, with a vertical red line at
$\sigmam = 120$ \msunpc.
}
\label{heyer20outer}
\end{figure*}

As for EXFC55-100, most structures are identified as clouds, but
they are mostly bound, despite having very low \sigmam. The usual plots are in
Figure \ref{heyer20outer} and the statistics are given as entry 11 in Table \ref{tabstats}.
For the EXFC135-195 sample, $\mean{\mrat} = 0.76 \pm 0.34$ with a median
$\mrat = 0.66$. These values are much higher than those in the GRS and similar
to those of the EXFC55-100 sample.

\subsubsection{Summary of Consequences of New Analysis}

The main effect of the different method of structure identification is to decrease the fraction of
mass in bound structures. In the inner Galaxy, most become
unbound, but 46\% of the mass is still in bound structures. 
For larger \rgal, the
 structures become less dense but more bound. For the outer Galaxy (EXFC135-195)
sample, $\fmass = 0.75$ even though $\mean{\sigmam} = 24$ \msunpc. This trend
reflects a strong decrease in velocity dispersion with \rgal. The fraction of
the mass recovered by \coo\ (\mrat) is low in the inner Galaxy,
but quite high in the other two samples.

\subsection{ \coo\ \jj21}

\begin{figure*}
\center
\includegraphics[scale=0.3]{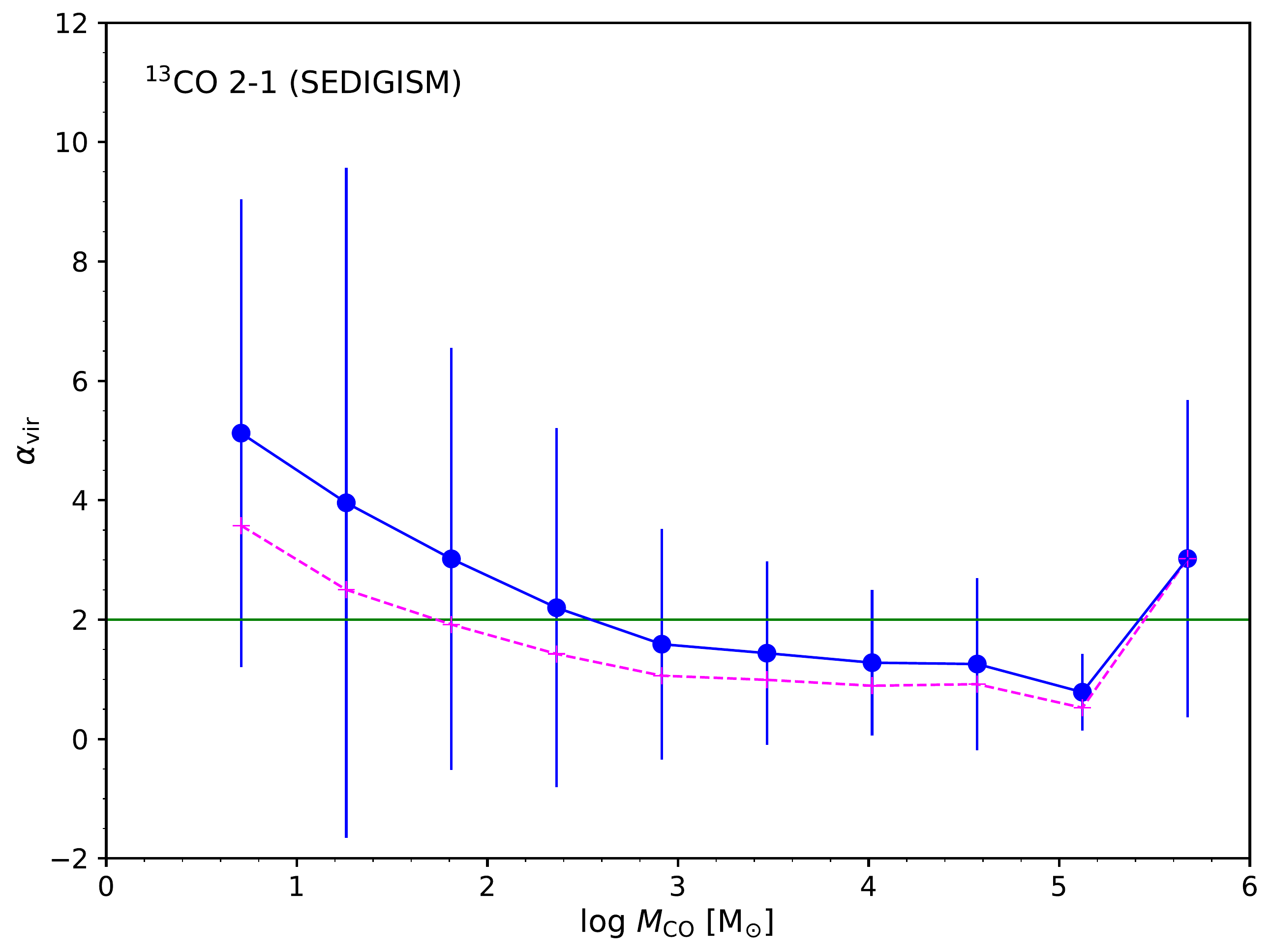}
\includegraphics[scale=0.3]{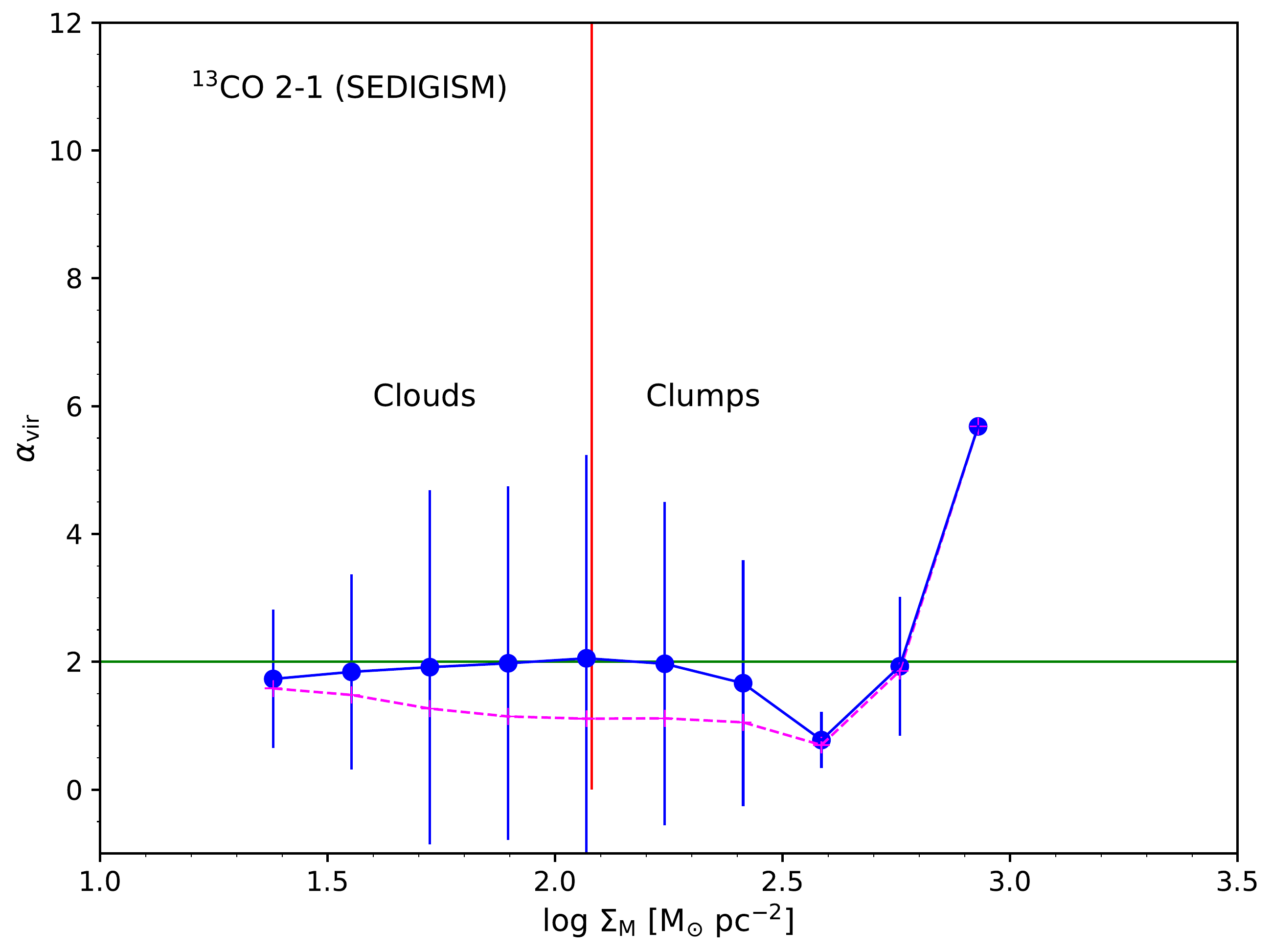}
\includegraphics[scale=0.3]{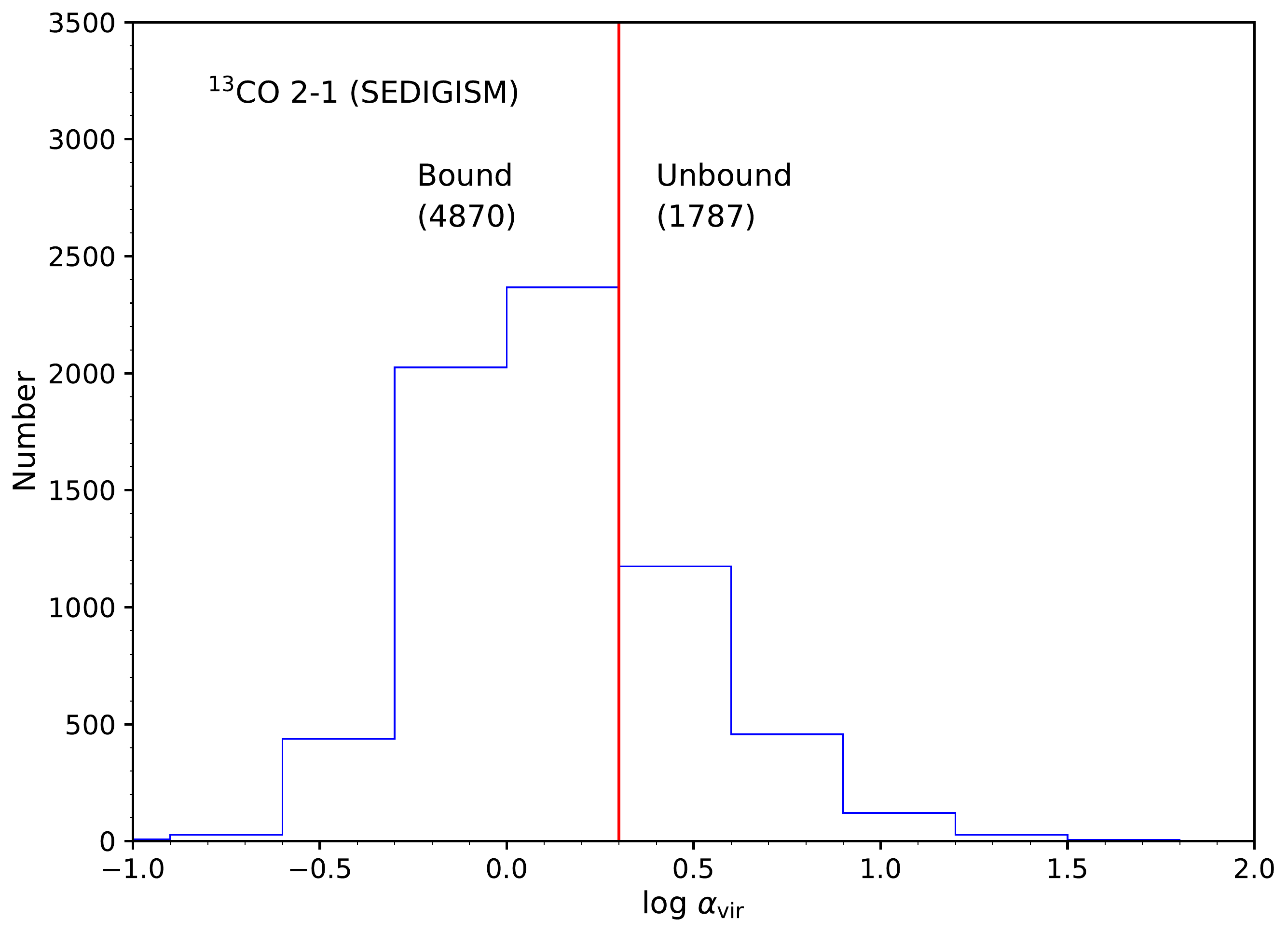}
\includegraphics[scale=0.3]{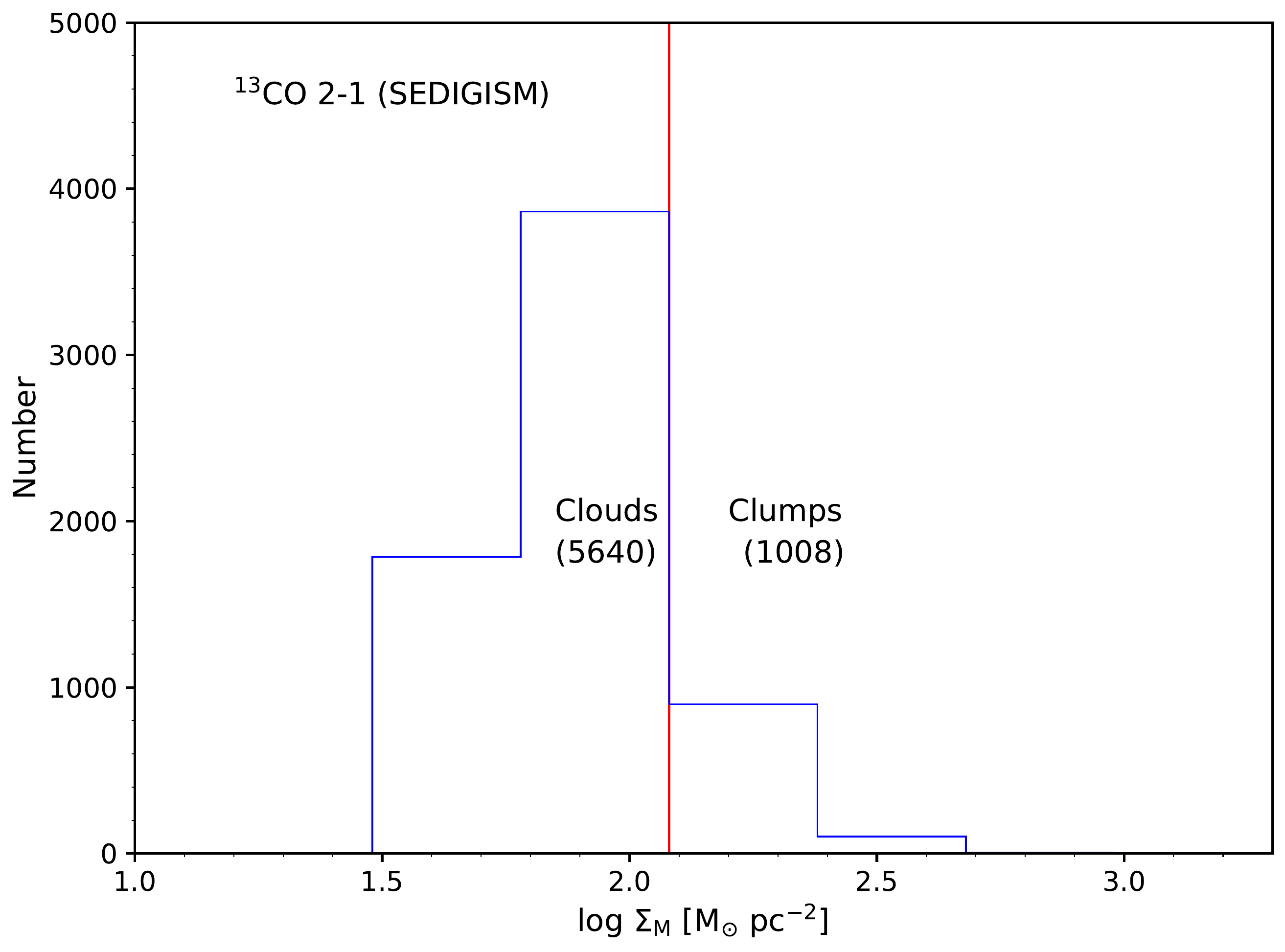}
\caption{
(Upper Left)
The mean and standard deviation of the virial ratio is plotted versus the logarithm of the clump mass, for the catalog of 
\citet{2021MNRAS.500.3027D},
after selecting only those catalog entries with good distances, not on edges and at
least 114 pixels. Median values are plotted in a magenta dashed line.
(Upper Right) The same quantities are plotted versus the mass surface density.
The horizontal line at $\alphavir = 2$ demarcates nominally unbound clouds above the
line from nominally bound clouds below the line. The vertical line in the right panel
indicates $\sigmam = 120$ \msunpc.
(Lower left) The histogram of values of log \alphavir, with a vertical red line
at $\alphavir = 2.0$. 
The number of structures in each category are given in parentheses.
(Lower right) The histogram of values of log \sigmam, with a vertical red line at
$\sigmam = 120$ \msunpc.
}
\label{sed}
\end{figure*}

The situation for the \jj21\ transition of \coo\
\citep{2021MNRAS.500.3027D},
shown in Figure \ref{sed}, is different. Most clouds defined in this
sample are nominally bound, even for relatively low mass and \sigmam.
However, \mean{\alphavir} is lower and \fmass\ is higher than for the
\jj10\ in the GRS sample with the new analysis, but comparable for the other \jj10\ samples.

\subsection{Stuctures Defined by Herschel and \ammonia}

\begin{figure*}
\center
\includegraphics[scale=0.3]{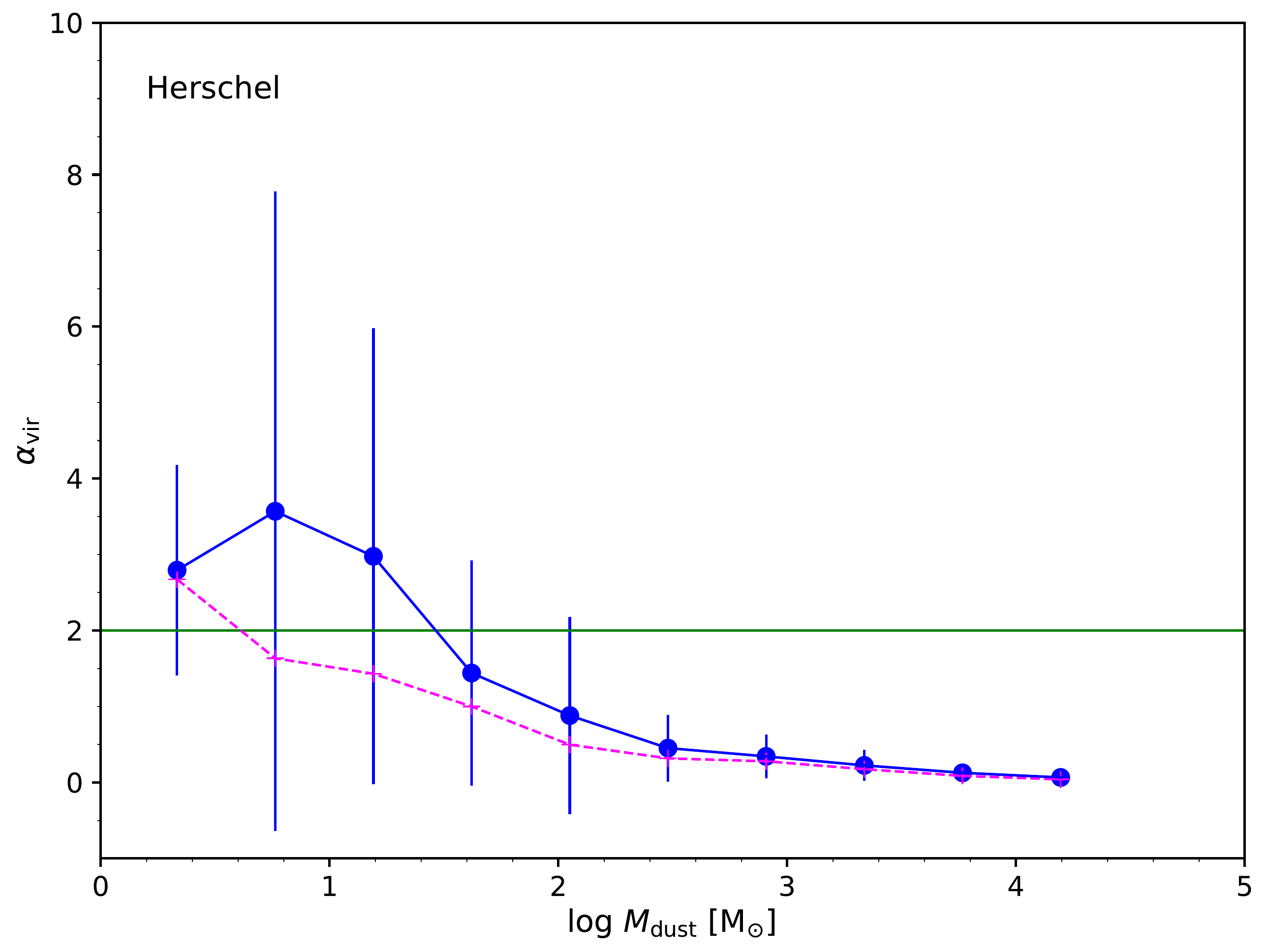}
\includegraphics[scale=0.3]{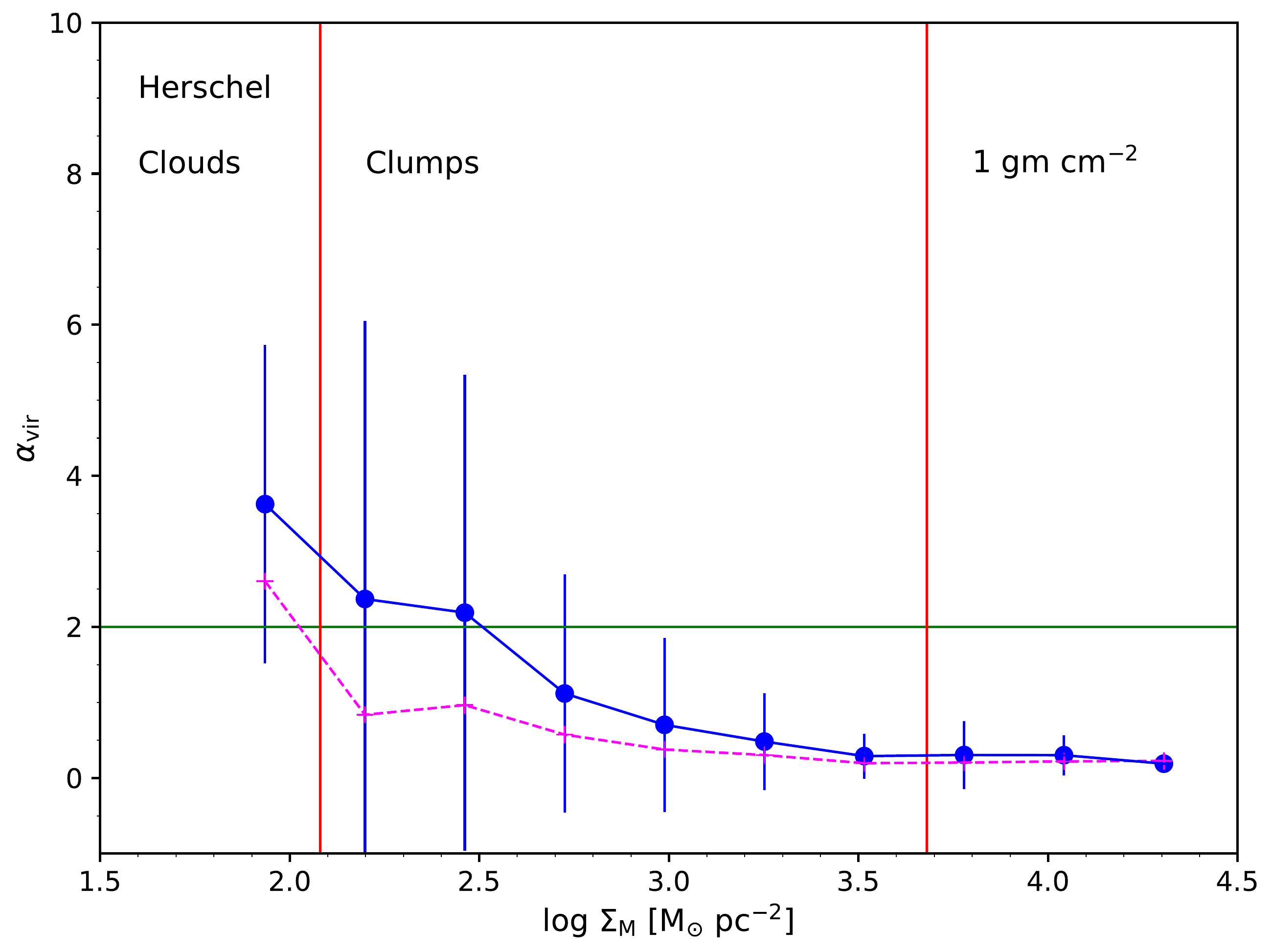}
\includegraphics[scale=0.3]{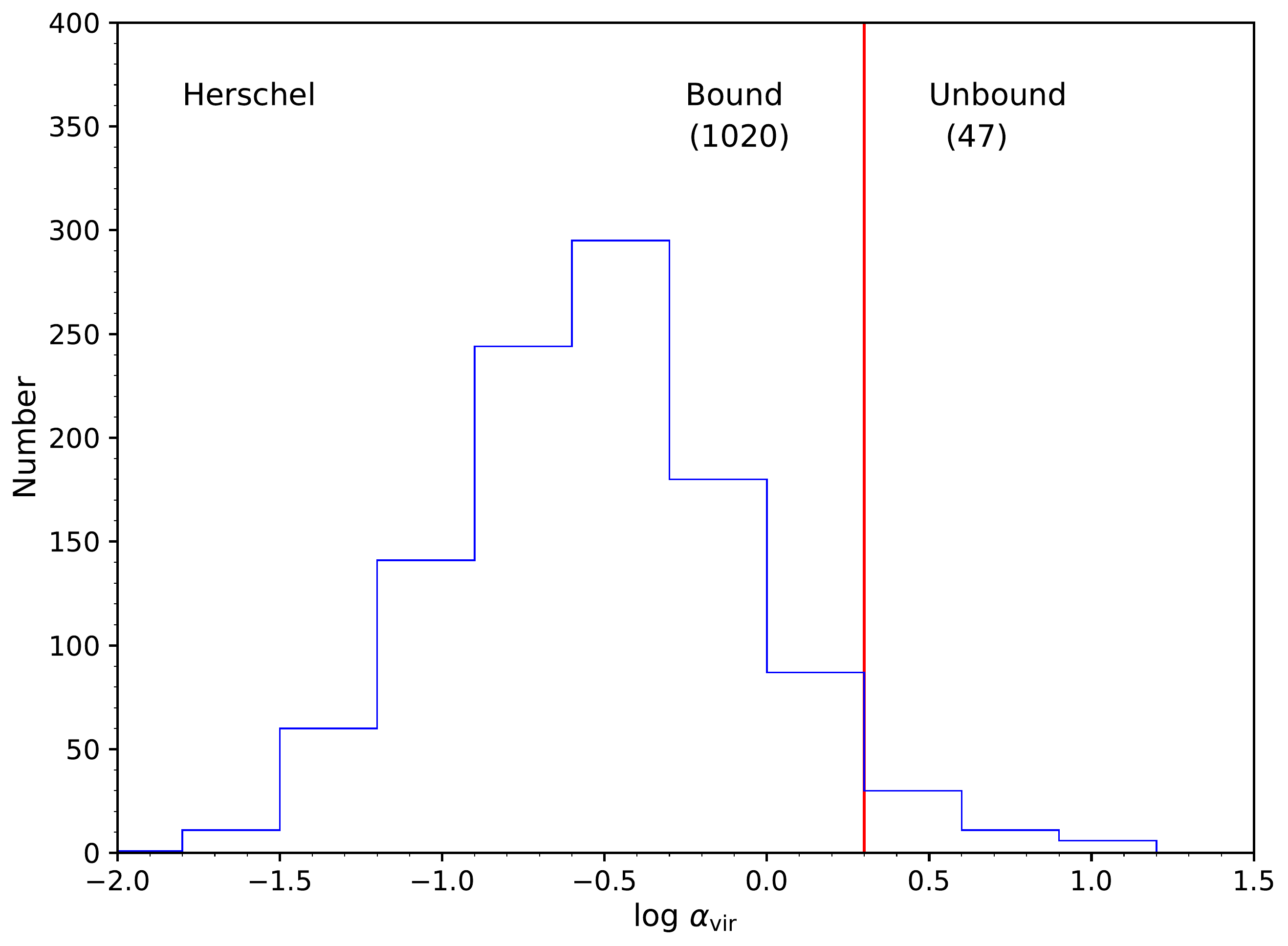}
\includegraphics[scale=0.3]{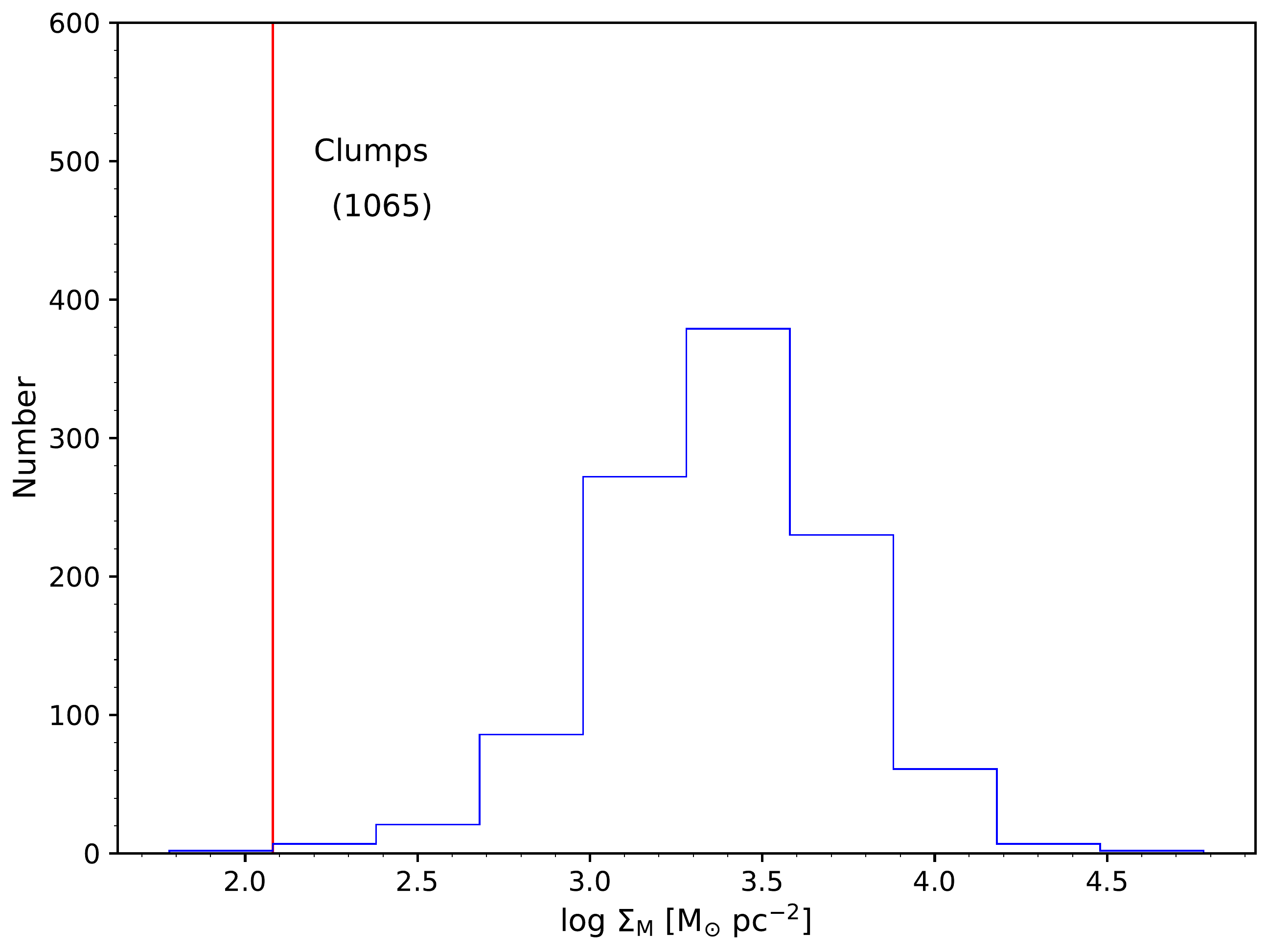}
\caption{
(Upper Left) The mean and standard deviation of the virial ratio is plotted versus the logarithm of the clump mass, for the catalog of 
\citet{2019MNRAS.483.5355M}.
Median values are plotted in a magenta dashed line.
(Upper Right) The same quantities are plotted versus the mass surface density.
The horizontal line at $\alphavir = 2$ demarcates nominally unbound clouds above the line from nominally bound clouds below the line. 
The vertical lines in the right panel indicate $\sigmam = 120$ \msunpc\
and $\sigmam = 1$ g \cmv.
The mass and surface density are the total (gas+dust)  determined from 
submm continuum emission.
(Lower left) The histogram of values of log \alphavir, with a vertical red line
at $\alphavir = 2.0$. 
The number of structures in each category are given in parentheses.
(Lower right) The histogram of values of log \sigmam, with a vertical red line at
$\sigmam = 120$ \msunpc.
}
\label{merello}
\end{figure*}

Compact structures identified by their submm emission with Herschel have
very low virial parameters, strongly indicative of being gravitationally
dominated and $\fmass = 1.0$. 
Figure \ref{merello} plots the virial parameter versus
the logarithm of clump mass from an analysis by
\citet{2019MNRAS.483.5355M}.
Only the clumps with the lowest mass and \sigmam\ are unbound.
While most of these contain star formation, even the 157 ``prestellar'' clumps
are strongly gravitationally dominated, with 
$\mean{\alphavir} = 0.57$, almost identical to the value for the full
sample. The values for \fcn\ and \fcmass\ of zero reflect the fact that
the very few structures in the sample with surface densities small enough
to be considered clouds are all unbound.


\section{Discussion}\label{discussion}

In this section, we will explore how the boundedness of structures depends
on methods and properties. We include sub-sections on the method used to identify
structures, the tracer used to define the structures and measure the mass,
the location in the Galaxy, mass of the structure, and the surface density
of the structure. 
 
\subsection{Dependence on Method Used to Identify Structure}\label{disc:method}

The first issue to consider is the fact that the results from two different
decompositions of the CO \jj10\ survey by 
\citet{2001ApJ...547..792D}
show quite different results. The difference is clear from the Figures,
but it can be encapsulated in the values of \fmass:
the structures identified by
\citet{2016ApJ...822...52R}
have $\fmass = 0.73$
while those of 
\citet{2017ApJ...834...57M}
have $\fmass = 0.19$.

Clearly, the choice of method to identify structures plays a major role
in the different results. We can borrow the terminology of taxonomists
to distinguish lumpers and splitters. In our context, lumpers would
aggregate emission into larger structures, while splitters would break
emission into smaller structures.
\citet{2016ApJ...822...52R}
used a dendrogram analysis to identify 1064 clouds, accounting for
about 25\% of the molecular mass in the Galaxy. They note that their
method avoids merging unrelated clouds into pseudo-clouds. In this
sense, they are splitters. Comparing their catalog to that of 
\citet{1986ApJ...305..892D}
(based on an early version of the first quadrant survey),
they recover all of the same clouds, but some
are split and less massive. They use the size-linewidth relation and
latitude distribution to break the kinematic distance ambiguity.
In contrast,
\citet{2017ApJ...834...57M}
might reasonably be called lumpers. They
performed a Gaussian decomposition of each spectrum and then
lumped them together by a clustering procedure. This process
led to larger linewidths in the clouds (median $\sigmav = 3.6$ \kms)
than in the individual Gaussian components (median $\sigmav = 1.65$ \kms).
Interestingly, this method allowed them to identify 98\% of the CO
emission with 8107 clouds. They found that the size-linewidth relation
has too much scatter, so they use the relation $\sigmav \propto (R \sigmam)^{0.5}$, 
where $R$ is the cloud size, to resolve the kinematic distance ambiguity.
While the mass also increased by this process, the net effect was
probably to produce larger \alphavir.
Were they lumpers or splitters? In the sense that they aggregated
velocity components, they were lumpers, but they decomposed four times
as much of the CO emission into 8 times as many clouds as Rice, making
them splitters. Lumping in velocity and splitting in number of clouds
is likely to lead to larger \alphavir. However, the fact that the scaling
relation mentioned above holds indicates that unrelated gas is not being
lumped in pseudo-clouds.
\added{Despite these differences, the scaling relations for the two
catalogs are similar \citep{2020ApJ...898....3L}.
}

Another difference between the results of the two methods is apparent in the
x-axis values for the plots of \alphavir\ versus \mco. The plot for
\citet{2016ApJ...822...52R}
starts at $\log \mco = 3.5$, while that of
\citet{2017ApJ...834...57M}
starts at $\log \mco = 0.5$.
The latter catalog contains many more clouds of lower mass.
However, this does not explain the difference in \fmass\ because
the total mass is still dominated by the massive clouds.

Probably the main factor in the difference is the completeness in
accounting for the CO emission of the Galaxy. Because
\citet{2016ApJ...822...52R}
account for only 25\% of the CO emission, their catalog is less
suitable for addressing the question of slow star formation on the Galactic
scale, where the problem was defined by taking the total mass of
molecular gas from the CO survey and dividing by a free-fall time.

In this context, the data from
\citet{2020ApJ...901L...8S}
are interesting as no method is used to define structures.
Instead only the mass within a 150 pc region is tabulated.
The fact that \replaced{$\fmass = 0.32$}{$\fmass = 0.35$} and 
\replaced{$\fcmass = 0.14$}{$\fcmass = 0.15$}, 
averaged over many galaxies and
galactocentric radii, even with some very high values for \sigmam,
suggests that the most likely value for \fmass\ is relatively small
for structures measured by CO, regardless of which of the two transitions
is used.

The importance of the identification method is also apparent in the
difference between the results from the original structures in the
\citet{2010ApJ...723..492R}
GRS 
sample (entry 8 in Table \ref{tabstats}) and the new analysis, where structures were defined by the CO
catalog of
\citet{2017ApJ...834...57M}
(entry 9 in Table \ref{tabstats}).

These differences suggest that procedures like CLUMPFIND or dendrogram
analyses tend to split structures or minimize low level, extended emission.
For addressing the global issue of slow star formation, procedures that
account for all the emission are necessary.

\subsection{Dependence on the Tracer Used to Define Structure}

We begin with the extremes: the samples of
\citet{2017ApJ...834...57M}
and
\citet{2020ApJ...901L...8S} 
showing low values for \fmass\ for mass measured from CO emission,
versus the sample of
\citet{2019MNRAS.483.5355M}
showing $\fmass = 1.0$ for structures traced by Herschel peaks and
\ammonia\ linewidths. The distinction between clouds, primarily traced
by CO, and dense clumps, primarily traced by peaks in \smm\ emission and lines
of species rarer than CO, has been based on this difference, but the
analysis of \alphavir\ strongly supports this distinction. The much higher
mean and median \sigmam\ for the Herschel-\ammonia\ data also indicate that
these are clearly distinct structures.

What about the structures probed by \coo? They clearly lie between the extremes
of clouds and dense clumps. For the GRS sample  with our new analysis,
$\fmass = 0.46$ and $\mean{\alphavir} = 3.83$ and $\mean{\sigmam} = 104$. 
For the sample defined by \coo\ \jj21, from
\citep{2021MNRAS.500.3027D},
$\fmass = 0.79$ and $\mean{\alphavir} = 1.94$, even though
$\mean{\sigmam} = 87.7$ \msunpc, lower than that of \coo\ \jj10\ in the GRS,
but higher than that of \coo\ \jj10\ in the other two surveys.

\begin{figure*}
\center
\includegraphics[scale=0.3]{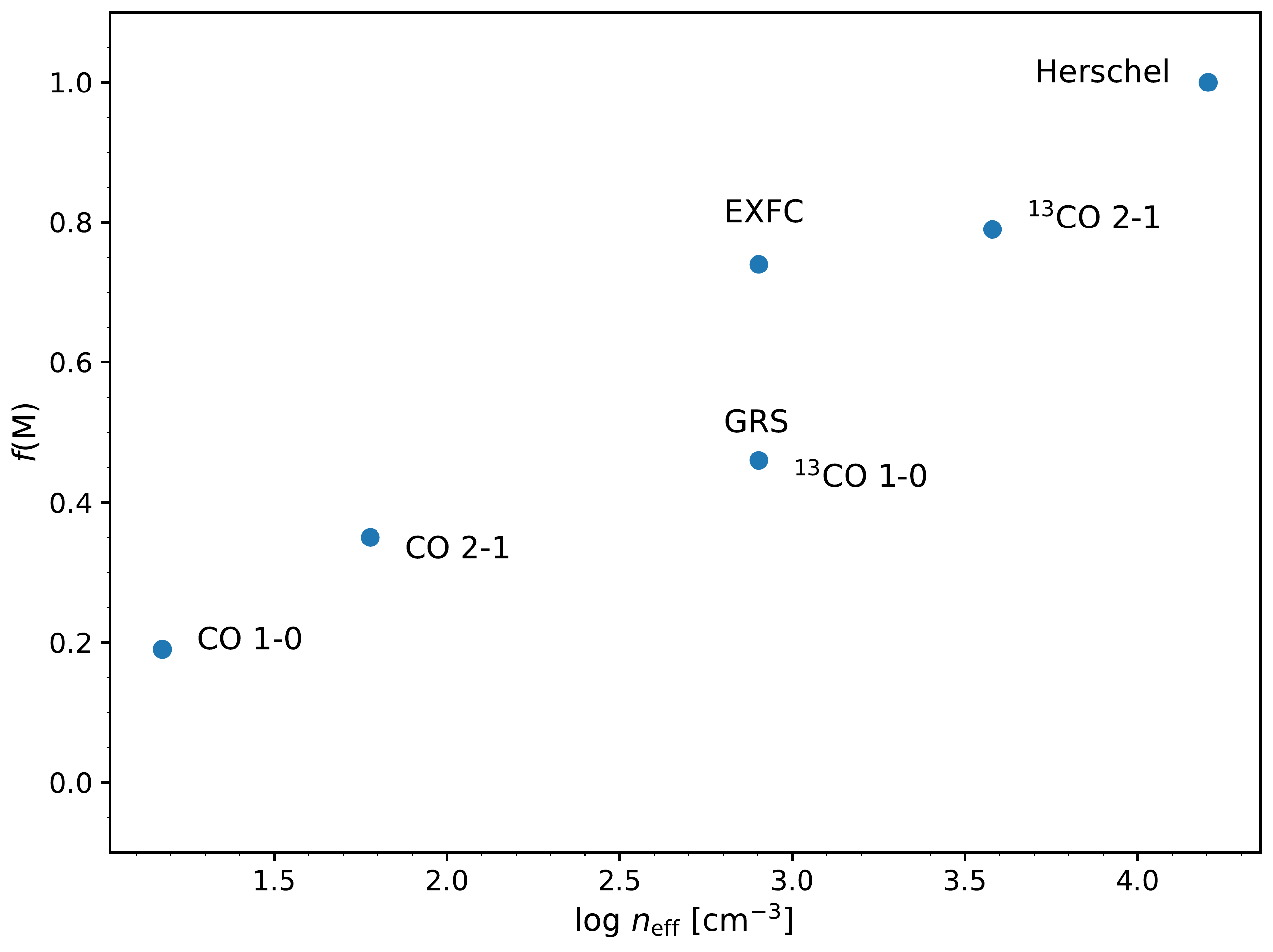}
\caption{
The value of \fmass\ versus \neff\ of the tracer. The higher point for \coo\ \jj10\ is for the average of the two EXFC surveys, while the
lower point is from the analysis in this paper for the GRS survey. Both points for \coo\ \jj10\ use the structures in the CO catalog of \citet{2017ApJ...834...57M}.
}
\label{fmvsneff}
\end{figure*}

The trend with tracer is that \alphavir\ decreases (structures are more
likely to be bound) as the tracer changes from CO to \coo\ \jj10\ to  \coo\
\jj21\ 
to Herschel-\ammonia. This can be related to a trend in the
characteristic density of the material being traced.
The characteristic density probed by each tracer can be crudely estimated from 
the effective density of the tracer used to define the structure. The
effective density (\neff), defined by 
\citet{1999ARA&A..37..311E}
and developed further by
\citet{2015PASP..127..299S},
is the density of particles needed to produce a 1 \kkms\ observed
line for an assumed column density and kinetic temperature.
Because  effective densities for CO and \coo\ were not calculated
by those references, we calculated them using the on-line tool RADEX
\citep{2007A&A...468..627V}
assuming $\tk = 10$ K and column density corresponding to $\av = 1$ mag, so
$N({\rm CO}) = 1\ee{17}$ \cmc\ and $N(\coo ) = 1.7\ee{15}$ \cmc.
The resulting effective densities are 15, 60, 800, 3800 \cmv\
for CO \jj10, CO \jj21, \coo\ \jj10, \coo\ \jj21, respectively.
The comparable density for the Herschel sample is less clear.
The structures were identified from dust continuum emission, but
\ammonia\ observations of both (1,1) and (2,2) inversion transitions
were required in order to estimate the kinetic temperature, \tk. The effective density for
the (2,2) line at $\tk = 15$ K is $\neff = 1.6\ee4$ \cmv\
\citep{2015PASP..127..299S}, 
similar to the density of about 2\ee4 \cmv\ needed to make \tk\
close to the dust temperature, \td, as found by
\citet{2019MNRAS.483.5355M}.
We take
$\neff = 1.6\ee4$ for the Herschel sample.

The values for \fmass\ are plotted versus these effective densities
in Figure \ref{fmvsneff}.
We use only the results for entries 3, 7, 9, 12, and 13, along with the
average of 10 and 11 (the EXFC samples), from Table
\ref{tabstats} as most representative of structure selection by
each tracer.
Clearly the fraction of the mass in bound structures increases rapidly
with \neff, reaching values about 0.8 for densities near the effective
density of \coo\ \jj21\ and unity for the Herschel sample. 
The data roughly follow a line in the semilog plot, but the two points
for \coo\ \jj10\ suggest that location in the Galaxy also matters.

\subsection{Dependence on Location in the Galaxy}

The contrast between the inner Galaxy (CMZ)
\citep{2001ApJ...562..348O}
and the outer Galaxy (OGS)
\citep{2001ApJ...551..852H}
is perhaps the most striking.
The structures identified in the CMZ have very high surface densities
($\mean{\sigmam} = 1700$ \msunpc), but $\fmass = 0.71$. 
In contrast the structures in the OGS have very low surface densities
($\mean{\sigmam} = 17.7$ \msunpc), but $\fmass = 0.58$ for entry 6,
with the more inclusive lower limit on cloud mass.
With the criterion of $\sigmam < 120$ \msunpc, none of the structures
in the CMZ defined by CO are bound (both \fcn\ and \fcmass\ are zero),
while $\fcmass = 0.46$ or 0.56 in the OGS, depending on the mass limits used.
Clearly, the local criterion of $\sigmam = 120$ \msunpc\ for boundedness,
based on a star formation threshold, is inappropriate for these parts of
the Galaxy. In the CMZ, a clump would need to be defined with a much
higher \sigmam, higher than 3000 \msunpc. In the outer Galaxy, the opposite
applies. 

A difficulty with this comparison is that very different criteria were used
to define structures in the CMZ compared to the disk or outer Galaxy.
A high threshold was needed to isolate structures in the highly confused
CMZ. 
Using the same procedure for all \rgal, 
\citet{2017ApJ...834...57M}
did not find a drop in median of \alphavir\ at large \rgal, but did
see a rise inside a few kpc.
The sample of other galaxies
\citep{2020ApJ...901L...8S}
also uses a common method for all radii.
There are \replaced{3358}{1562} data points in their catalog with $\rgal < 0.5$ kpc, similar to the
definition of the CMZ in the Milky Way. In that restricted sample,
the median, mean, and standard deviation of \sigmam\ are 
\replaced{103, 312, 570}{130, 375, 670} \msunpc, much higher than the full sample, but
still less than those for the CMZ.
The median, mean, and standard deviation of \alphavir\ are
\replaced{6.01,  8.43,  8.14}{5.45, 8,28, 9.03}, about twice those for the full sample, and
larger than those for the CMZ, while the fractions in bound
structures are \replaced{$\fn = 0.08$}{$\fn = 0.11$} and 
\replaced{$\fmass = 0.22$}{$\fmass = 0.27$}, substantially
smaller than for the full sample or for the CMZ of the Milky Way.
The data from other galaxies supports the idea that a much higher
surface density is needed to produce bound structures in the inner
regions of galaxies and suggests that most of the molecular gas there
is unbound despite very high surface densities.

Comparing the GRS and EXFC samples in \coo\ \jj10\ reveals a similar pattern.
More mass is in bound structures despite the lower \sigmam\ in the outer 
parts of the Galaxy.

\subsection{Dependence on Mass}

A trend common to all the samples is a decrease in \alphavir\ with 
increasing mass. At some level, this trend is a result of a trend 
of decreasing \alphavir\ with increasing \sigmam. This is most obvious
for the sample of 
\citet{2020ApJ...901L...8S},
where the size of the region considered is constant, so the trend
with mass is really a trend with surface density.

Contrary evidence that the most massive clouds may be unbound can be seen
in Figure 8 from
\citet{1986ApJ...305..892D}
who consider the 26 most massive complexes in the first quadrant of the
Milky Way. About half have linewidths larger than those expected from
virial equilibrium.
A more recent analysis of the most massive complexes by
\citet{2016ApJ...833...23N}
found that most were unbound. While some are engaged in what they called
``mini-starbursts'' which could be unbinding the cloud, those sources do not have a very different distribution of virial
parameters.
It is these very large complexes that will dominate studies of CO 
in other galaxies. They also dominate the total mass of molecular gas
in our Galaxy because the cloud mass function slope is less than 2, 
unlike the case for stars. However, the mass function peaks at much
lower masses in the outer Galaxy, as shown in Figure 7 of
\citet{2017ApJ...834...57M}.
A similar trend is seen for molecular clouds in M33
\citep{2005PASP..117.1403R}.

\subsection{Dependence on Mass Surface Density}

\begin{figure*}
\center
\includegraphics[scale=0.3]{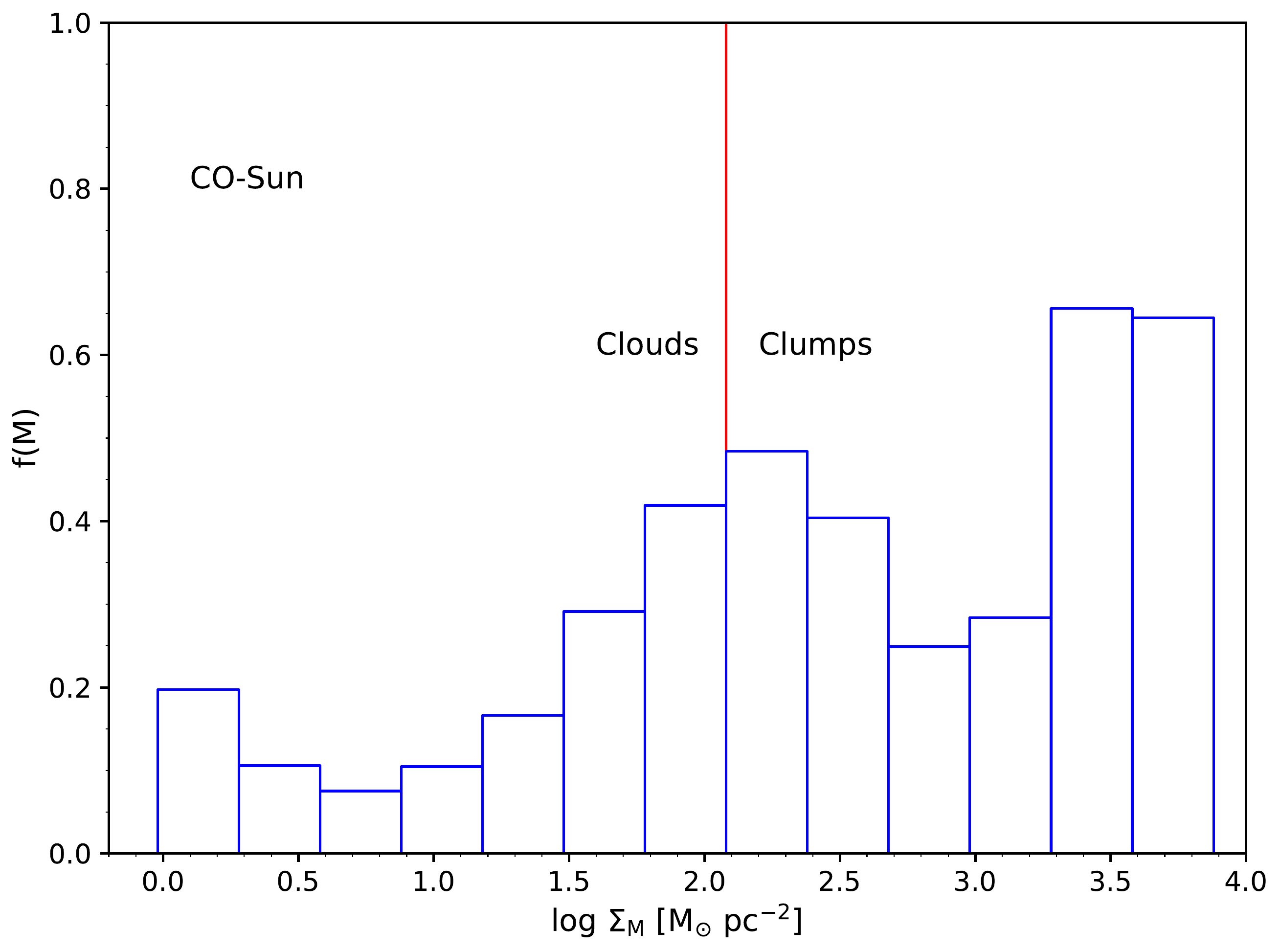}
\includegraphics[scale=0.3]{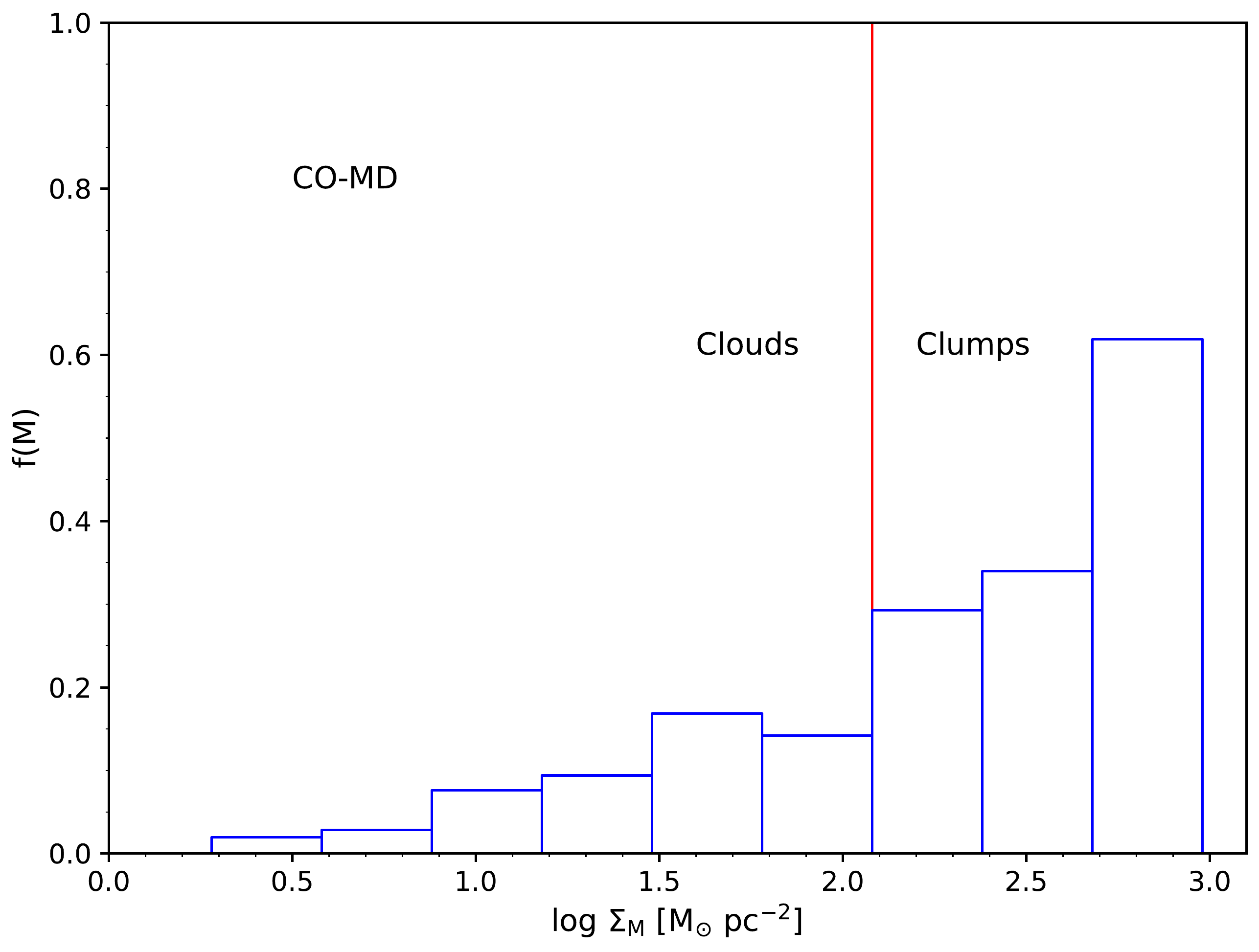}
\includegraphics[scale=0.3]{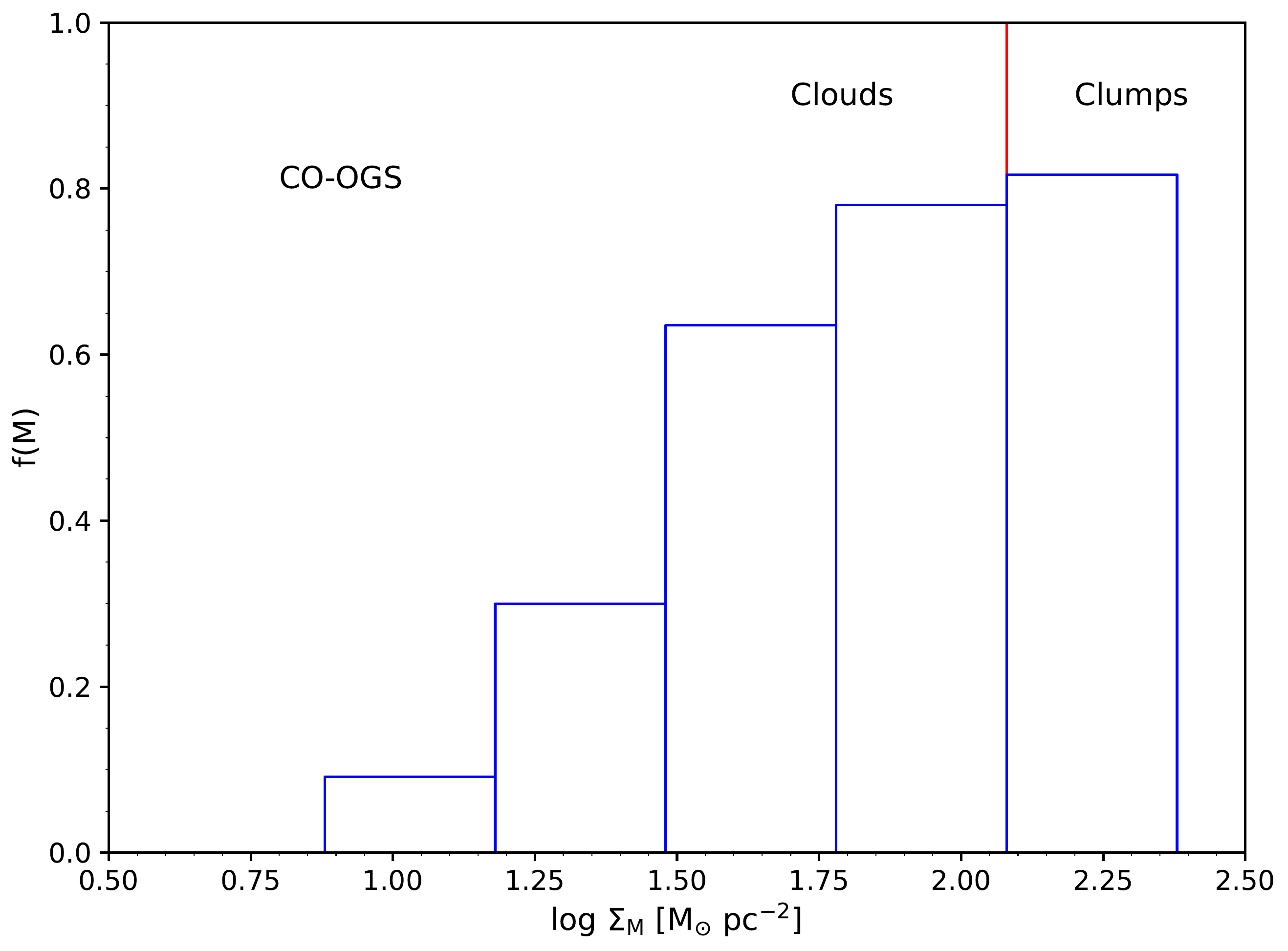}
\includegraphics[scale=0.3]{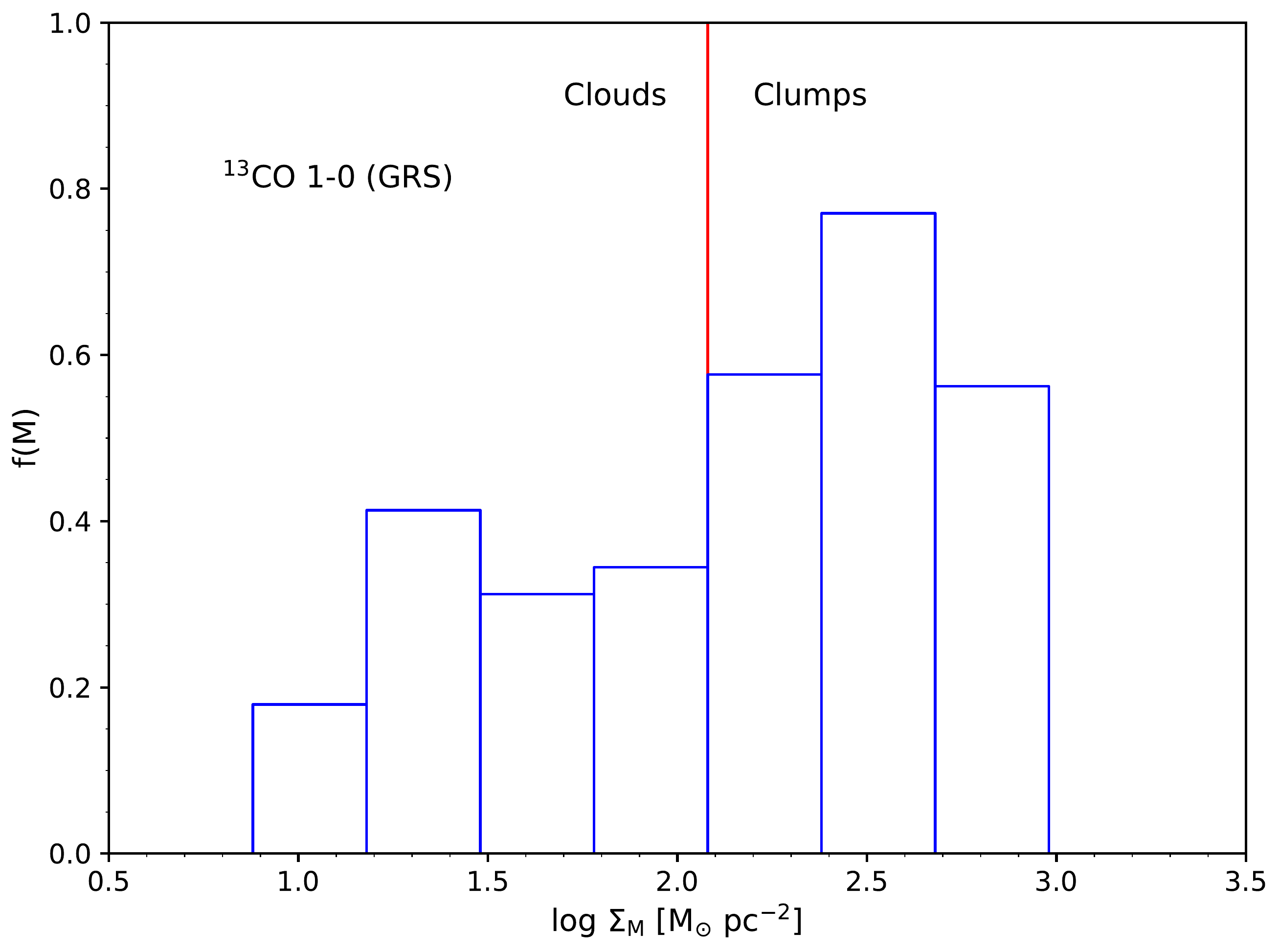}
\caption{
The value of \fmass\ is plotted versus log \sigmam\ for the CO samples of
\citet{2020ApJ...901L...8S},
\citet{2017ApJ...834...57M},
\citet{2001ApJ...551..852H},
and the \coo\ \jj10\ sample for the GRS, with structures identified by CO
(entry 9 in Table \ref{tabstats}).
}
\label{fvssigma}
\end{figure*}

The decrease in median and mean \alphavir\ with increasing \sigmam\
is clear from the relevant plots for most samples that cover the
relevant range of \sigmam. The distinction between clouds and clumps
adopted for discussion in this paper is $\sigmam = 120$ \msunpc.
To explore this connection further, we have calculated the mass fraction
in bound structures in bins of 0.3 in log \sigmam\ for samples that
cover the transition well. As shown in Figure \ref{fvssigma}, these all
show a strong increase in the fraction of mass that is bound as
\sigmam\ increases.

The dependence on surface density is also reflected in the values of
\fcn\ and \fcmass\ in Table \ref{tabstats}. These are generally smaller
than the \fn\ and \fmass\ computed for all structures. Only a small fraction
of the gas in structures we have called clouds is bound. This statement
applies both to Milky Way structures and to those in other galaxies.

The decline of virial parameter with mass may in part
be an artifact of observational biases in the way that the
linewidth is measured
\citep{2018A&A...619L...7T}.
They point out that the tracers used for determining the velocity 
dispersion may be weighted more heavily to denser regions, with 
smaller dispersions that do not reflect the full dispersion of the
entire structure. This effect would be most noticeable in samples
for which the tracers of mass and velocity dispersion are different.
The effect might contribute to the very low values of \alphavir\ and
the strong trend with mass in the Herschel sample.

\subsection{Caveats and Uncertainties}\label{caveats}

As discussed extensively in \S \ref{intro}, there are substantial
uncertainties in computing the virial parameter. \deleted{ or the virial mass.}
For any given structure, the uncertainty in
\alphavir\ is likely to be a factor of 2 or 3.
Analyses of the uncertainties in estimating cloud mass, velocity
dispersion, etc. can be found in
\citet{2013ApJ...777..173B}.
Masses from CO luminosities have uncertainties, as discussed
in the introduction, 
so that we expect uncertainties of at least a factor of 3, and these
may be systematic uncertainties, at least in the case of \mco.
\citet{2018ApJ...858...16G}
argue that X(CO) may be overestimated based on
``observing" simulations. They get a value of X(CO) about
half the standard value of 2\ee{20}. If that is correct,
masses from CO would be overestimated and virial parameters
underestimated. 
\added{\citet{2020ApJ...903..142G} has further examined this issue and
provided conversion factors that are shallower functions of 
metallicity ($Z$) than those used by 
\citet{2020ApJ...901L...8S}.
}
Finally, we re-emphasize the dependence on the method
used to define a cloud.

For \coo, our newer analysis accounts for the latest 
measurements of the isotopic ratio, \isorat, as a function of \rgal\
and the latest value of \hh/CO. 
On the other hand, we have made no correction for changes in the CO
abundance with metallicity, which is known to decrease with \rgal.
\citet{2000MNRAS.311..329D}
found a gradient in the oxygen abundance gradient over the
range of 5-15 kpc:
$ 12 + \log {\rm O/H} = (-3.95\pm 0.49)\ee{-2}\rgal + (8.82 \pm 0.05)$.
The effect of metallicity on CO emission is not simple, but extragalactic
observers commonly use a correction factor: 
$\alphaco = 4.35 Z^{-1.6}$ \msun (\kkms pc$^2$)$^{-1}$
(\citealt{2020ApJ...901L...8S} and references therein),
where $Z$ is the metallicity relative to solar.
If we applied the same correction factor, CO clouds at 3 kpc would 
have twice higher \alphavir, while those at 15 kpc would have about
one-third the value of \alphavir. Such a correction would strengthen the
already strong trend for clouds to be unbound in the inner Galaxy, where
most of the molecular mass resides, and to be bound in the outer Galaxy.
\added{However, the divergence of opinion on how \alphaco\ depends on
$Z$ is currently substantial
(see, e.g., \citealt{2017MNRAS.470.4750A, 2020ApJ...903..142G, 2020ApJ...898....3L}).
}

We reiterate that methods of structure identification remain one of the
most significant sources of uncertainty.

\section{Consequences for the Star Formation Rate Problem}\label{sfrredux}

The results of this analysis have consequences for the problem of the star 
formation rate, discussed in the introduction. If only a fraction
 of the
mass of molecular gas in the Galaxy is in bound structures, the
over-prediction of the star formation rate of 300 \msunyr, can be decreased substantially. Based on the catalog of
\citet{2017ApJ...834...57M},
\added{the only one that accounts for all the CO emission,}
only 19\% of structures defined by CO \jj10\ are bound, decreasing
the theoretical star formation rate to 57 \msunyr, using the same assumption
about the free-fall time as the value found in \S \ref{intro}. 

We can make a more accurate estimate by assessing the star formation
rate cloud by cloud, then adding those rates up.
If we assume that only clouds with $\alphavir \leq 2$ form stars in a
free fall time based on the mean density of that cloud, 
the predicted star formation rate for an individual cloud is 
\begin{equation}
\sfrth = M/\tff = 6.04\ee{-8} M_{\msun}^{1.5} r_{\rm pc}^{-1.5}
\end{equation}

Using the catalog of 
\citet{2017ApJ...834...57M}
because it accounts for all the molecular clouds defined by CO emission,
and summing the star formation rate only over bound clouds, we find a total
star formation rate for the Milky Way of 46.4 \msunyr, a factor of 6.5 less
than the value in the introduction. This is mostly due to the small fraction
of bound clouds, but there is also a contribution from a longer \tff\ because
the mean density of the clouds identified by
\citet{2017ApJ...834...57M}
 is less than 100 \cmv. The resulting depletion time of {\it bound} molecular gas is decreased to $\tdep = 1.0\ee8$ yr, closer to the free-fall time, but still considerably larger.

Of course, regions of clouds are likely to be bound even if the entire cloud 
is not. The fraction of bound mass could be assessed from comparison of masses indicated by different tracers. Ideally, we would do the same analysis for
a complete survey of the Galaxy in other tracers, especially \coo.
A similar calculation for the \coo\ GRS survey, using the full sample
(entry 8 in Table \ref{tabstats}) predicts a star formation
rate of 7.4 \msunyr.
However, the existing \coo\ surveys do not cover the whole Galaxy, and
the new analysis was possible only for a subset of the \coo\ data, so we
cannot do a direct calculation. Instead we calculate the fraction of the
mass of the unbound CO-defined structures that is found in the bound \coo-derived properties.
This value is $0.22$ for the GRS, which applies to the inner Galaxy, which 
contains most of the molecular mass. The fraction is higher (0.77-0.80) in
the outer parts of the Galaxy. So this material would increase the predicted star
formation rate. If all of the \coo-derived mass in the unbound CO-defined
clouds in the GRS sample form stars at the fiducial \tff, they would contribute
another 55 \msunyr\ to the Galactic  star formation rate.
Clearly, a full sky survey of \coo\ would be very valuable.

\replaced{Our results do not directly help the other problem of}{We have
focused on \epsff, rather than the final efficiency (\epssf)}
\added{, but the simulations do find lower final efficiencies (\epssf) for
higher \alphavir\ [a factor of 5-10 decrease as \alphavir\ increases from 1 to 5 \citep{2021ApJ...911..128K}].
}
Because the most massive clouds seem most likely to be bound (subject to the
caveats above), the difference between mass functions of clouds and clusters is,
if anything, increased. The resolution to this problem may lie in the inclusion
of tidal forces since most of these very massive clouds reside in the inner Galaxy.


\section{Conclusions}\label{conclusions}

We list the main conclusions before discussing them:

\begin{enumerate}

\item Clearly, the choice of method to identify clouds plays a major role
in the different results.

\item The tracer (both molecule and transition) that is used to define
structures has a strong effect on whether those structures are bound.

\item At a fixed \rgal, the fraction of the mass in bound structures increases rapidly
with the effective density of the tracer, suggesting that regions with
larger volume densities contribute more to bound gas.

\item Many clouds identified by CO are unbound and most of the mass
in those clouds is unbound.

\item Structures defined by dust continuum emission and linewidths from \ammonia\ emission are almost all bound.

\item Structures defined by \coo\ emission are more likely to be bound
than those defined by CO emission. However, their boundedness also depends
on the method used to identify them and on assumptions about isotopic and elemental abundances.

\item Boundedness correlates strongly with surface density.

\item Structures within 0.5 kpc of a galaxy center have much higher
surface density, but much of the mass is unbound because of higher
turbulence. The opposite is true in outer regions.

\item More generally, structures identified by CO or \coo\ have lower
surface density, but are more likely to be bound, at larger \rgal.

\item More massive clouds are more likely to be bound, but this trend
is partially confused with the trend with surface density. 

\item For the most complete \replaced{survey}{catalog} of structures traced by CO
\citep{2017ApJ...834...57M}, the fraction of mass in bound structures,
$\fmass = 0.19$, alleviating, but not eliminating, the
fundamental problem of slow  star formation in
the Milky Way. 

\end{enumerate}

The most important conclusion is that most of the mass traced by 
CO emission is in unbound structures. This result was anticipated by
theorists, notably
\citet{2011MNRAS.413.2935D},
whose title asked ``Why are most molecular clouds not gravitationally bound?".
Their answer involved cloud-cloud collisions and stellar feedback; shredding and
merging resulted in clouds not being well-defined entities with the same gas over
cloud ``lifetimes."
\citet{2017ApJ...840...48P}
argued that supernova feedback kept the interstellar medium sufficiently stirred
up that many structures were unbound even though their densities were such that
they were likely molecular. As mentioned in \S \ref{intro},
\citet{2021ApJ...911..128K} found that unbound clouds are needed to reach
the observational constraints on the star formation rate.

Observers have often found, but seldom emphasized, this result.
\citet{2016ApJ...818..144R} suggested that gas detected in CO, but not \coo\
might be ``diffuse, non-star-forming gas," but did not explicitly suggest that
this gas was unbound. 
Figure 16 of
\citet{2017ApJ...834...57M}
already showed that almost all the structures in their catalog are 
unbound, but this result was not particularly emphasized.
\citet{2016ApJ...833...23N}
 showed that mini-starburst complexes, which have enhanced star formation rate density, are actively forming massive stars but are not necessarily bound, suggesting that being bound is not the major
factor regulating their star formation activity.
The idea that structures identified by molecular emission are well-defined,
bound structures, unlike the more diffuse atomic interstellar medium,
is long-standing and creates cognitive dissonance with contrary factual
evidence. Uncertainties are often invoked to argue that clouds are bound even
when the evidence suggests otherwise (see \citealt{2011MNRAS.413.2935D} for
discussion of this tendency.).
However, there is no good reason to believe that the atomic-molecular
transition is identical to the unbound-bound transition. This would be plausible
if most of the support were thermal, but it is turbulent 
\citep{1974ApJ...192L.149Z}.
The fact that these transitions are {\bf close} can be coincidental
(see pg. 36 of
\citealt{1985prpl.conf...33E})
and may have led us astray.

\begin{acknowledgments}
We acknowledge stimulating discussions with many colleagues, but especially with E. Ostriker and J-G. Kim. NJE acknowledges C. Dobbs, A. Burkert, and J. Pringle for
their paper that opened his eyes to the possibility that molecular clouds are unbound.
The authors acknowledge Paris-Saclay University's Institut Pascal program ``The Self-Organized Star Formation Process" and the Interstellar Institute for hosting discussions that nourished the development of the ideas behind this work.
\end{acknowledgments}

\software{astropy
\citep{2013A&A...558A..33A,2018AJ....156..123A},
GILDAS
\citep{2005sf2a.conf..721P,2013ascl.soft05010G}
}




\appendix

\begin{figure*}
\center
\includegraphics[scale=0.6]{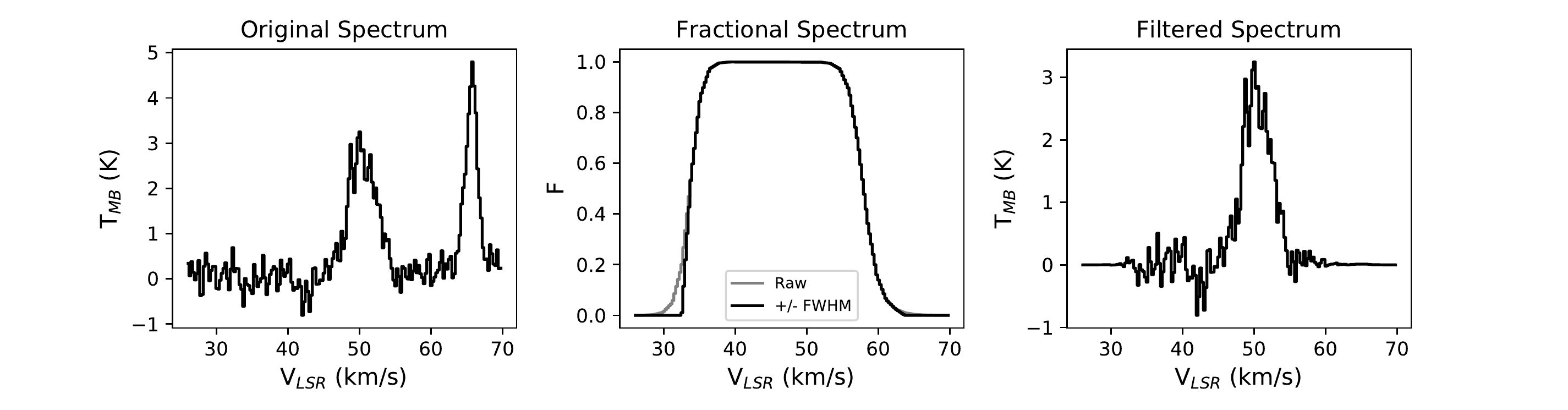}
\caption{
The method used to filter unrelated signal from \coo. The spectrum on the left is 
the original data. The center panel shows the function that isolates the velocity
range that is consistent with that of the CO-defined structure. The right panel
shows the filtered \coo\ spectrum that contributes to the \coo-defined mass and
other properties.
}
\label{filter}
\end{figure*}

We have selected a subset of clouds from the \citet{2017ApJ...834...57M} catalog which fit into the 
coverage of the \coo\ \jj10\ surveys from the Five College Radio Astronomy Observatory in order to investigate the 
boundedness of these clouds using a mostly optically thin tracer of \hh\ column density. 
To connect the \coo\ emission to the structures  defined by the sum of Gaussian components
\citep{2017ApJ...834...57M}, we developed the 
following procedure.  For a given cloud, we calculate the fraction of the observed CO emission that is
recovered by the summed Gaussian for each position and velocity channel, $F(X,Y,V)$. A typical profile of $F$
is flat-topped with values ranging from 0.9-1 in the core of the line (see Figure \ref{filter}). The fractional values rapidly decrease towards the 
wings but remain non-zero for the full extent of the spectral axis.  To exclude spectral channels beyond the 
core and wing of a cloud, we set $F(X,Y,|v-v_{cen}| > dv)=0$ 
where $v_{cen}$ and $dv$ are the velocity centroid and full width half maximum from the  catalog of
\citet{2017ApJ...834...57M}. 
A \coo\ data cube, $T_{13}(x,y,v)$,  
is constructed from the FCRAO surveys 
at the 
native 48\arcsec\ angular resolution and 22\arcsec\ sampling interval 
to match the CO coverage of the cloud.   
The spectral axis of the \coo\ data is smoothed and resampled to 0.25 \kms\ 
wide channels.   At this stage the \coo\ data cube may contain emission features not associated with the CO-defined cloud at
velocities outside of the core of the CO emission 
(see e.g., the left panel of Figure \ref{filter}).   
To further isolate the \coo\ emission, we first resample 
the spectral axis of the fraction cube, $F$, to  be aligned with the \coo\ spectral axis and then multiply the 
\coo\ spectral axis with the corresponding fractional spectrum, 
\begin{equation}
T_{13}(x,y,v)'= F(X,Y,v) T_{13}(x,y,v) 
\end{equation}
where $X-dX/2 < x < X+dX/2$,  $Y-dY/2 < y < Y+dY/2$, and  $dX$, $dY$ are the 0.125 degree pixel sizes of the CO data cube. 
The effect of this multiplication  is 
to suppress any contribution to the integrated \coo\ emission from signal outside of the CO velocity interval, as shown in Figure \ref{filter}.

  The cloud mass is  computed from $N(\hh)$ found from  equation 10 in 
\S \ref{intro} as follows:
\begin{equation}
M(\coo) = \muhh m_{\rm H} D^2 \int d\Omega  N(\hh)(l,b)
\end{equation}
where $\muhh = 2.8$, $m_{\rm H}$ is the atomic hydrogen mass, and $D$ is the distance to the cloud.
For each cloud, the velocity dispersion, $\sigma_v$(\coo) and cloud radius are calculated following 
\citet{2017ApJ...834...57M}.
Both $\sigma_v$(\coo) and radius are deconvolved from the native spectral and angular resolutions
of the data.  We also deconvolved the CO velocity dispersions and cloud sizes.
Clouds are excluded from any subsequent analyses if either the cloud size or velocity dispersion is 
not resolved. 
Finally, the \coo-derived virial mass is derived from Equation 8.
Accounting for these selection criteria, there are 289 clouds from the GRS, 105 clouds from the EXFC55-100 
survey, and 98 clouds from the EXFC135-195 survey.

\bibliographystyle{aasjournal}

\bibliography{more}

\end{document}